\documentclass[pdftex,twocolumn,epjc3]{svjour3}          % twocolumn

\RequirePackage[T1]{fontenc}

\smartqed  % flush right qed marks, e.g. at end of proof

\RequirePackage{graphicx}
\RequirePackage{mathptmx}      % use Times fonts if available on your TeX system
\RequirePackage{amsmath,amssymb}
\RequirePackage{tabularx}
\RequirePackage{booktabs}
\RequirePackage{flushend}
\RequirePackage[numbers,sort&compress]{natbib}
\RequirePackage[colorlinks,citecolor=blue,urlcolor=blue,linkcolor=blue]{hyperref}

\journalname{Eur. Phys. J. C}

\usepackage{lineno}
\begin{document}

% Title & co still to be decided
\title{Design and performance of the ENUBET monitored neutrino beam}

%\subtitle{Overview of the baseline design and results}

\author{
F. Acerbi\thanksref{1} \and I. Angelis\thanksref{21} \and  L. Bomben\thanksref{2,3} \and  M. Bonesini\thanksref{3} \and  F. Bramati\thanksref{3,4} \and  A. Branca\thanksref{3,4} \and  C. Brizzolari\thanksref{3,4} \and  G. Brunetti\thanksref{3,4} \and  M. Calviani\thanksref{6} \and  S. Capelli\thanksref{2,3} \and  S. Carturan\thanksref{7} \and  M.G. Catanesi\thanksref{8} \and  S. Cecchini\thanksref{9} \and  N. Charitonidis\thanksref{6} \and  F. Cindolo\thanksref{9} \and  G. Cogo\thanksref{10} \and  G. Collazuol\thanksref{5,10} \and  F. Dal Corso\thanksref{5} \and  C. Delogu\thanksref{5,10} \and  G. De Rosa\thanksref{11} \and  A. Falcone\thanksref{3,4} \and  B. Goddard\thanksref{6} \and  A. Gola\thanksref{1} \and  D. Guffanti\thanksref{3,4} \and  L. Halić\thanksref{20} \and  F. Iacob\thanksref{5,10} \and  C. Jollet\thanksref{16} \and  V. Kain\thanksref{6} \and  A. Kallitsopoulou\thanksref{24} \and  B. Kliček\thanksref{20} \and  Y. Kudenko\thanksref{13} \and  Ch. Lampoudis\thanksref{21} \and  M. Laveder\thanksref{5,10} \and  P. Legou\thanksref{24} \and  A. Longhin\thanksref{e1,5,10} \and  L. Ludovici\thanksref{15} \and  E. Lutsenko\thanksref{2,3} \and  L. Magaletti\thanksref{8,14} \and  G. Mandrioli\thanksref{9} \and  S. Marangoni\thanksref{3,4} \and  A. Margotti\thanksref{9} \and  V. Mascagna\thanksref{22,23} \and  N. Mauri\thanksref{9,18} \and  J. McElwee\thanksref{16} \and  L. Meazza\thanksref{3,4} \and  A. Meregaglia\thanksref{16} \and  M. Mezzetto\thanksref{5} \and  M. Nessi\thanksref{6} \and  A. Paoloni\thanksref{17} \and  M. Pari\thanksref{5,10} \and  T. Papaevangelou\thanksref{24} \and  E.G. Parozzi\thanksref{4} \and  L. Pasqualini\thanksref{9,18} \and  G. Paternoster\thanksref{1} \and  L. Patrizii\thanksref{9} \and  M. Pozzato\thanksref{9} \and  M. Prest\thanksref{2,3} \and  F. Pupilli\thanksref{5} \and  E. Radicioni\thanksref{8} \and  A.C. Ruggeri\thanksref{11} \and  G. Saibene\thanksref{2,3} \and  D. Sampsonidis\thanksref{21} \and  C. Scian\thanksref{10} \and  G. Sirri\thanksref{9} \and  M. Stipčević\thanksref{20} \and  M. Tenti\thanksref{9} \and  F. Terranova\thanksref{3,4} \and  M. Torti\thanksref{3,4} \and  S.E. Tzamarias\thanksref{21} \and  E. Vallazza\thanksref{3} \and  F. Velotti\thanksref{6} \and  L. Votano\thanksref{17}
}

\thankstext{e1}{e-mail: andrea.longhin@pd.infn.it}

\institute{Fondazione Bruno Kessler (FBK) and INFN TIFPA, Trento, Italy
\label{1}
\and
DiSAT, Università degli studi dell’Insubria, via Valleggio 11, Como, Italy \label{2}
\and
INFN, Sezione di Milano-Bicocca, piazza della Scienza 3, Milano, Italy
\label{3}
\and 
Phys. Dep. Università di Milano-Bicocca, piazza della Scienza 3, Milano, Italy
\label{4}
\and
INFN Sezione di Padova, via Marzolo 8, Padova, Italy
\label{5}
\and
CERN, Geneva, Switzerland
\label{6}
\and
INFN Laboratori Nazionali di Legnaro, Viale dell’Università, 2 - Legnaro (PD), Italy
\label{7}
\and
INFN Sezione di Bari, via Amendola 173, Bari, Italy
\label{8}
\and
INFN, Sezione di Bologna, viale Berti-Pichat 6/2, Bologna, Italy
\label{9}
\and
Phys. Dep. Università di Padova, via Marzolo 8, Padova, Italy
\label{10}
\and
INFN, Sezione di Napoli, via Cinthia, 80126, Napoli, Italy
\label{11}
\and
IPHC, Université de Strasbourg, CNRS/IN2P3, Strasbourg, France
\label{12}
\and
Institute of Nuclear Research of the Russian Academy of Science, Moscow, Russia. \\ National Research Nuclear University MEPhI,  115409 Moscow, Russia. \\ Moscow Institute of Physics and Technology (MIPT), 141701 Moscow.
\label{13}
\and
Phys. Dep. Università degli Studi di Bari, via Amendola 173, Bari, Italy
\label{14}
\and
INFN, Sezione di Roma 1, piazzale A. Moro 2, Rome, Italy
\label{15}
\and
LP2I Bordeaux, Universitè de Bordeaux, CNRS/IN2P3, 33175 Gradignan, FR
\label{16}
\and
INFN, Laboratori Nazionali di Frascati, via Fermi 40, Frascati (Rome), Italy
\label{17}
\and
Dip. di Fisica e Astronomia ``Augusto Righi''
Viale Berti-Pichat 6/2, Bologna, Italy
\label{18}
\and
Phys. Dep. Università degli Studi di Napoli Federico II, via Cinthia, 80126, Napoli, Italy
\label{19}
\and
%Center of Excellence for Advanced Materials and Sensing Devices, Ruder Boskovic Institute, HR-10000 Zagreb, KR
Center of Excellence for Advanced Materials and Sensing Devices, Ruđer Bo\v{s}kovi\'c Institute, 10000~Zagreb,~Croatia
\label{20}
\and
Aristotle University of Thessaloniki. Thessaloniki 541 24, Greece
\label{21}
\and
DII, Università degli studi di Brescia, via Branze 38, Brescia, Italy
\label{22}
\and
INFN, Sezione di Pavia, via Bassi 6, Pavia, Italy
\label{23}
\and
CEA, Centre de Saclay, Irfu/SPP, F-91191 Gif-sur-Yvette, France
\label{24}
}

\date{Received: date / Accepted: date}
% The correct dates will be entered by the editor

\maketitle

\begin{abstract}
The ENUBET project is aimed at designing and experimentally demonstrating the concept of monitored neutrino beams. The\-se novel beams are enhanced by an instrumented decay tunnel, whose detectors reconstruct large-an\-gle charged leptons produced in the tunnel and give a direct estimate of the neutrino flux at the source. These facilities are thus the ideal tool for high-precision neutrino cross-section measurements at the GeV scale because they offer superior control of beam systematics with respect to existing facilities.  In this paper, we present the first end-to-end design of a monitored neutrino beam capable of monitoring lepton production at the single particle level. This goal is achieved by a new focusing system without magnetic horns, a 20 m normal-conducting transfer line for charge and momentum selection, and a 40 m tunnel instrumented with cost-effective particle detectors. Employing such a design, we show that percent precision in cross-section measurements can be achieved at the CERN SPS complex with existing neutrino detectors.  
\end{abstract}

\section{Introduction}
\label{sec:intro}
Neutrino cross sections at the GeV scale are currently known with limited precision despite their prominent role in the present and future long-baseline experiments for the study of neutrino oscillations \cite{katori2018}. The T2K experiment is already dominated by systematic uncertainties for some oscillation parameters where cross-section uncertainties play a dominant role \cite{pandey_nufact2022}. Even more, the next generation DUNE \cite{Abi:2020wmh} and HyperKamiokande (HK) \cite{Abe:2018uyc} experiments will be limited by cross-section uncertainties and the lack of high-precision data hinders the development of reliable theoretical models to describe GeV neutrino interactions with nuclei \cite{Alvarez-Ruso:2017oui,Branca:2021vis}.
Neutrino cross sections from a few hundred MeV up to tens of GeV have been measured by many experiments in the last few decades and are an essential ingredient for studying Standard Model processes, nuclear physics, and physics beyond the Standard Model. These measurements \cite{Alvarez-Ruso:2017oui} are limited by uncertainties in the flux and energy of initial-state neutrinos and the precision that can be achieved is generally larger than 10\%. There is a broad consensus on the fact that "to
extract the most physics from DUNE and Hyper-Kamiokande a complementary programme of experimentation to determine neutrino
cross sections and fluxes is required" \cite{EUdeliberation2020}.
ENUBET (Enhanced Neutrino Beam with kaon Tagging) tackles this challenge by designing a new generation of short-baseline neutrino experiments with unprecedented control of flux, flavor, and energy of neutrinos \cite{ENUBET_proposal,ENUBET_spsc_2020,ENUBET_spsc_2021}. Unlike conventional beams, monitored neutrino beams \cite{Longhin:2014yta} employ a decay tunnel that is fully 
instrumented with detectors for diagnostics \cite{app11041644}. Charged leptons produced in association with neutrinos are recorded by the instrumentation of the tunnel and provide a direct measurement of the neutrino flux and flavor. The ERC ENUBET project \cite{ENUBET_ERC} focused on the identification of positrons from the three-body semileptonic decay of kaons ($K_{e3}$): $K^+ \rightarrow e^+ \nu_e \pi^0$. More recently the CERN NP06/ENUBET experiment \cite{ENUBET_proposal} extended the ENUBET reach to monitor muons from $K_{\mu2}$ ($K^+ \rightarrow \ \mu^+ \nu_\mu $) and pion decays. As a consequence, the ENUBET beamline can presently reconstruct both electron and muon neutrinos produced in the instrumented decay tunnel. In the past, monitoring charged leptons in the decay tunnel at the single particle level was considered far-fetched due to the number of meson decays needed for neutrino applications. Current neutrino beams based on a fast extraction of protons from accelerators and a focusing horn produce charged leptons at a rate that conventional detectors cannot cope with. In 2020-22, ENUBET solved this issue by employing a beamline without a pulsed magnetic focusing horn, 
%with a horn-less beam
where protons can be slowly extracted in few-second spills and the lepton rate at the tunnel does not exceed 100 kHz/cm$^2$. This beamline is described in Secs.\ref{sec:proton_extraction} (proton extraction), \ref{sec:target} (target station), and \ref{sec:transfer_line} (transfer line and static focusing system). The instrumentation for the decay tunnel was simulated with GEANT4 \cite{GEANT4:2002zbu,Allison:2006ve,Allison:2016lfl} and validated with prototypes at testbeams; it is described in Sec.~\ref{sec:instrumentation}.
The radiation hardness of components is instrumental to reliably building monitored beams and is discussed  in Sec.\ref{sec:doses}.     
This paper presents the full analysis chain to record candidate events in the instrumented tunnel (Sec. \ref{sec:eventbuilder}) and identify positrons and muons from kaon decays. Background arises from particle misidentification, pile-up, other kaon decay modes, and tertiary particles ($\pi^\pm$, $\gamma,p$, and halo muons) produced in the beamline. Secs. \ref{sec:positrons} and \ref{sec:muons} summarize the identification capability and background suppression performance of ENUBET. The physics performance of the beamline is assessed in Secs. \ref{sec:neutrino_fluxes} and \ref{sec:neutrino_energy} in terms of $\nu_\mu$ and $\nu_e$ fluxes at the neutrino detector and energy reconstruction capability. Neutrino energy reconstruction is based on the Narrow-Band Off-Axis technique described in Sec. \ref{sec:neutrino_energy}. It does not rely on the identification of final state particles in the neutrino detector but only on the radial position of the neutrino interaction vertex.

\section{Monitored neutrino beams and the ENUBET beamline}
\label{sec:monitored}

Beam diagnostics have been employed to estimate the neutrino flux and characterize beam properties since the inception of accelerator neutrino physics \cite{Danby:1962nd}.
Over the years, the power of neutrino beams has increased by orders of magnitude \cite{kopp2006}. As a consequence, the focus of beam diagnostics moved toward delivering monitoring devices capable of withstanding the charged particle rates and doses of modern beams, while improvement in precision has been quite limited \cite{app11041644}. At the time of writing, diagnostics determine the flux with a precision of $\sim$ 10\% but this value is no more acceptable to cross-section experiments that must deliver measurements at the percent level to match the needs of oscillation physics. The world-record measurement of flux in an accelerator neutrino beam was achieved by MINER$\nu$A in 2022 by combining beam diagnostics, state-of-the-art beam simulation, and neutrino interaction generators, and elastic $\nu$-e scattering in the neutrino detector \cite{MINERvA:2022vmb}. By employing such a sophisticated method,  MINER$\nu$A  achieved a precision of 3.3\% for $\nu_\mu$-enriched runs and 4.7 \% for $\bar{\nu}_\mu$-enriched runs.\footnote{$\nu$-e scattering cannot be employed to measure the $\nu_e$ flux due to the smallness of the $\nu$-e cross-section at the neutrino detector.} 
To achieve a substantial improvement in beam diagnostics and deliver cross sections with a $<1\%$ precision, we must rely on observables that are directly linked to the neutrino flux on a particle-by-particle basis. The most straightforward technique to address this challenge was envisaged in the 1970s \cite{Hand1969,Pontevcorvo1979} but could not be implemented due to technical limitations. In the decay tunnel of a neutrino beam, every neutrino created by the decay of a charged meson produces one charged lepton. Counting charged leptons in the decay tunnel thus provides a direct measurement of the flux because the $\nu_\mu$ ($\bar{\nu}_\mu$) yield is proportional to the $\mu^+$ ($\mu^-$) yield in the tunnel and the $\nu_e$ ($\bar{\nu}_e$) yield is proportional to the $e^+$ ($e^-$) yield in the tunnel. Several authors have proposed to measure these leptons in the tunnel together with the corresponding neutrino interactions in the neutrino detector and correlate in time the occurrence of these events, i.e. associate every neutrino interaction with its charged lepton \cite{Pontevcorvo1979,Bernstein,Ludovici:1996sx}. The measurement of the lepton provides the initial neutrino flavor at the source. The lepton energy combined with a high-precision measurement of the parent meson momentum uniquely determines the neutrino energy on an event-by-event basis. Such a bold facility is called a {\it tagged neutrino beam}. Tagged neutrino beams were developed in USSR in the 1980s but never achieved their physics goals because of the enormous charged particle rate in the decay tunnel and the limited 4D (i.e. time and space) precision of trackers \cite{ammosov}. To cope with the rate limitations of trackers, the USSR Tagged Neutrino Facility was descoped and served experiments for kaon physics. Tagged neutrino beams are still under consideration \cite{Longhin:2014yta,Terranova:2015nsa,Perrin-Terrin:2021jtl,NUTECH,Longhin:2022tkk} and we discuss this option in the framework of ENUBET in Sec. \ref{sec:time_tagged} but high-precision cross-section measurements can be performed with less demanding facilities: the {\it monitored neutrino beams} \cite{Longhin:2014yta}. 

%\begin{figure}[h!]
\begin{figure*}
    \centering
          \includegraphics[width=1.0\textwidth]{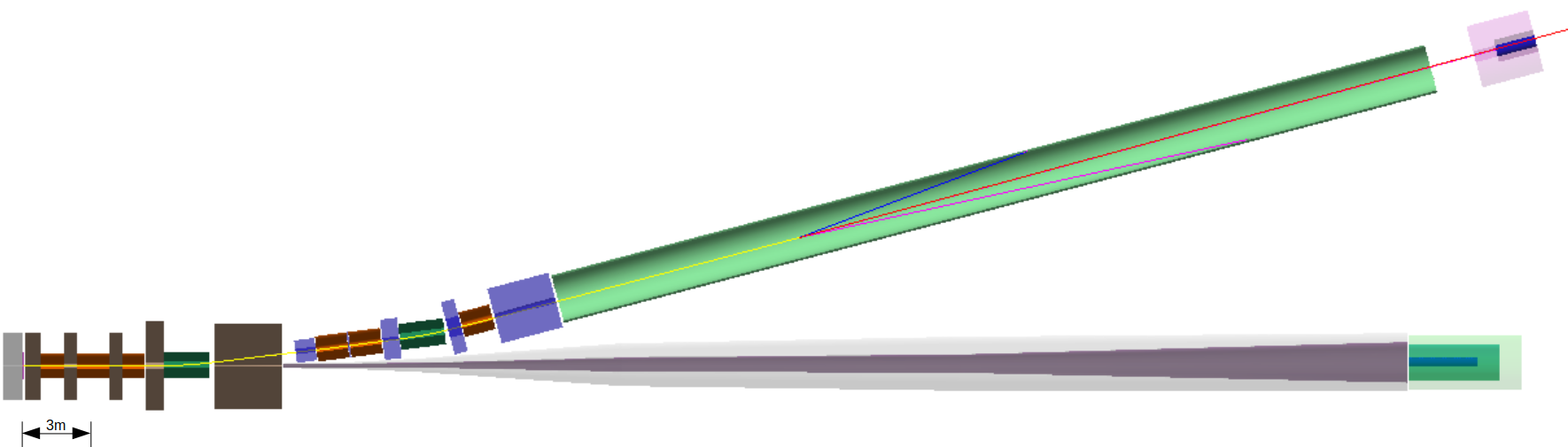}
    \caption{The ENUBET beamline as implemented in G4Beamline, GEANT4 and FLUKA. Primary protons impinge on the target  (not shown) located on the left of the figure; secondaries are sign and momentum selected by the transfer line and transported at the entrance of the instrumented decay tunnel (light green). From left to right: the quadrupole triplet (orange) and copper collimators (brown), double-bend momentum selector section composed of two dipoles (green), two quadrupoles, and one momentum copper collimator (brown), and the last quadrupole surrounded by two Inermet180 collimators (blue). The light-green cylinder represents the instrumented decay tunnel where a $K_{e3}$ decay is shown; the red track corresponds to the $\nu_{e}$ traveling towards the neutrino detector. The hadron dump that stops all particles but neutrinos is located at the tunnel exit. Non-interacting protons and forward particles travel inside the proton pipe (in gray) and are stopped by the proton dump (dark green). The neutrino detector considered in ENUBET (mass: 500 tons - not shown in the figure) is located 50 m far from the decay tunnel exit.}
    \label{fig:tlr6v5}
\end{figure*}

A monitored neutrino beam (Fig.~\ref{fig:tlr6v5}) is a conventional beam where the cylindrical walls of the decay tunnel are equipped with low-cost particle detectors that act as beam diagnostics.  
Protons are extracted by a proton driver and steered to a solid-state target through a transfer line. Proton interactions with the target produce short- and long-living hadrons. Short-li\-ved hadrons like $K_S$ or charmed mesons may produce neutrinos in the forward direction that propagate parallel to non-interacting primary protons and high-momentum se\-con\-da\-ries. These particles are stopped in the proton dump of Fig.~\ref{fig:tlr6v5} and their neutrinos are unlikely to reach the neutrino detector, which is located off-axis with respect to the proton dump. After the target, secondaries within the momentum bite of the beamline (8.5 $\pm 10\%$ GeV/c in ENUBET) are collected and sign-selected by the focusing system that transports these particles to the entrance of the decay tunnel. At the entrance of the tunnel, charged particles are collected inside the momentum acceptance together with background particles: tertiary particles ($e^\pm$, $\gamma$, muons, $\pi^\pm$), off momentum mesons that are produced along the transfer line, and halo muons. Halo muons are produced by pion decays in the transfer line, cross the collimators and enter the decay tunnel even if they are outside the momentum acceptance of the beamline. Background particles have a smaller momentum than the central momentum of the beamline (8.5 GeV/c). If the beamline selects positive (negative) hadrons as in a neutrino (anti-neutrino) run, particles within the momentum bite entering the decay tunnel are mainly protons, $K^+$, and $\pi^+$ ($\bar{p}$, $K^-$, and $\pi^-$ in an antineutrino run). As a consequence, neutrinos are mostly due to kaon and pion decays, plus contamination originating from early decays of off-momentum particles in the beamline and low-energy neutrinos from the proton and hadron dumps. The beamline presented in this paper was optimized to produce muon and electron neutrinos in the energy range of interest for the DUNE experiment (1-4 GeV).\footnote{A dedicated low-energy optimization of the same beamline that increases the neutrino flux in the region of interest for HyperKamiokande will be presented in a forthcoming publication.}  
The beamline optics, length, acceptance, momentum bite, and the neutrino detector distance were optimized to maximize the number of $\nu_\mu$ and $\nu_e$ with energy between 1 and 4 GeV that reach the neutrino detector and originate from pions and kaon decays. In particular, we maximized:
\begin{itemize} 
    \item the number of kaons and pions transported at the entrance of the tunnel
    that go through it without crossing the instrumented walls; 
    \item the number of kaon decays inside the tunnel before the hadron dump
\end{itemize}
and minimized
\begin{itemize}
    \item the number of muon decays in flight ($\mu^+ \rightarrow e^+ \nu_e \bar{\nu}_\mu$)
    \item the number of background and off-momentum particles crossing the walls of the decay tunnel.
\end{itemize}
The suppression of muon decays in flight is instrumental to having kaons as the only source of $\nu_e$ in the neutrino detector. Since the kaon mass is much larger than pions, charged leptons created by $K^+ \rightarrow \mu^+ \nu_\mu$ and $K^+ \rightarrow e^+ \nu_e \pi^0$ are produced at a larger angle than muons created by pion decays. As a consequence, if we monitor large-angle leptons we can uniquely determine the $\nu_e$ flux and the high-energy component of the $\nu_\mu$ flux ($\nu_\mu$ from $K_{\mu2}$). In addition, if we instrument the hadron dump with detectors that can withstand the muon flux, we can monitor the low energy component of the $\nu_\mu$ flux ($\nu_\mu$ from $\pi^+ \rightarrow \mu^+ \nu_\mu$). The high- and low-energy $\nu_\mu$ spectra are completely separated, as in any ``narrow-band beam'' \cite{kopp2006} (see also Fig. \ref{fig:numuCC} below),  The constraints above fix the optimal central momentum of ENUBET (8.5 GeV) because lower momenta reduce the number of produced kaons and increase the number of kaon decays before the entrance of the tunnel. Higher momenta bring the neutrino energy above the region of interest for DUNE. Since kaons must decay inside the tunnel but muons must reach the hadron dump, the tunnel length is shorter than a standard narrow-band beam at 8.5~GeV/$c$. The optimal value is around 40~m. This is fortunate because a short tunnel decreases the cost of the instrumentation. The optimal distance ("baseline") between the entrance of the tunnel and the neutrino detector is $\sim 100$ m. In ENUBET, the length of the tunnel is 40 m and the distance between the end of the tunnel and the neutrino detector is $50$~m. In this work, we considered a 500-ton neutrino detector with dimensions comparable to ProtoDUNE-SP and ProtoDUNE-DP at CERN~\cite{protodune:tdr}.   
In short, ENUBET is a 10\% momentum-bite narrow-band-beam that serves a short-baseline ($L=90$ m) neutrino experiment. In Secs. \ref{sec:neutrino_fluxes} and \ref{sec:neutrino_energy}, we will show  that these findings are key to measuring the neutrino flux (energy) with a precision of $< 1\%$  ($\sim 10\%$)  on an event-by-event basis. 

Secondaries (pions and kaons) that are transported at the entrance of the tunnel must have a small divergence because the beam envelope cannot be larger than the width of the tunnel. This is mandatory in a monitored neutrino beam where most of the instrumentation is located in the cylindrical walls of the tunnel.
The ENUBET instrumentation is made of iron-scintillator sampling calorimeters that are longitudinally segmented into modules (see Sec.~\ref{sec:instrumentation}).
This array of modules is complemented by a photon veto made of plastic scintillator tiles positioned in the innermost part of the calorimeter, as shown in Figs. \ref{fig:schematics_instrumented_tunnel} and \ref{fig:demonstrator}. The tunnel instrumentation can thus record large-angle charged leptons, perform particle identification, and measure the $e^\pm$ energy. The use of iron-scintillator calorimeter modules, whose light is read by SiPMs located in the outer rim of the tunnel (see Sec. \ref{sec:instrumentation}), keeps the cost of diagnostics well below ($\sim 10$\%) the overall cost of the neutrino beam even if the tunnel is instrumented along its whole length (40 m). 

The monitored neutrino beam concept was introduced in 2015 and investigated in 2016-2022 by the ERC ENUBET project. The project aimed at demonstrating that the rate of large-angle positrons from $K_{e3}$ (i.e. $K^+ \rightarrow e^+ \nu_e \pi^0$ decays)  produced in the decay tunnel provides a measurement of the $\nu_e$ flux $\phi(\nu_e)$ with a precision better than 1\%. In 2020-22, the ENUBET Collaboration achieved a breakthrough in the field by delivering a design that fulfills this requirement with adequate intensity without employing a focusing horn. The beamline described in Sec.~\ref{sec:transfer_line} produces pions and kaons that are momentum- and sign-selected by a static focusing system. The system comprises only normal-conducting elements: six qua\-dru\-poles and two bending dipoles.
Unlike horns, dipoles and quadrupoles are DC 
%https://www.overleaf.com/project/61683727c207eec8daf6a09b 
powered and there is no constrain on the duration of the proton extraction: protons can be slowly  %https://www.overleaf.com/project/61683727c207eec8daf6a09b
steered into the target (2 s in ENUBET) and produce a steady stream of mesons like in fixed-target experiments~\cite{slawg:2019}. 
Unlike tagged neutrino beams, leptons do not need to be associated in time with neutrinos interacting at the neutrino detector. This feature relieves the time resolution requirement of the calorimeter to a few ns instead of tens of ps. Timing is only needed to associate modules belonging to the same event (charged lepton) and remove pile-up. 
4D reconstruction in a monitored neutrino beam is thus performed in a neighborhood of the lepton impact point, while tagged neutrino beams must provide a {\it global} event reconstruction because a $\nu$ interaction at the neutrino detector must be uniquely linked in time with a charged lepton. This lepton must be picked up from $\sim 10^6$ potential candidates: a task that can be accomplished only by fast ($O$(100 ps)) trackers. Thanks to the static focusing system and the loose requirements on time resolution, ENUBET can monitor not only large-angle positrons but the vast majority of particles produced in the tunnel. This feature is exploited to monitor all neutrino flavors including $\nu_\mu$ (see Sec. \ref{sec:muons}) and provide a complete beam characterization for a new generation of high-precision cross-section experiments.

\section{Proton extraction}
\label{sec:proton_extraction}
The fast-extraction scheme of the primary protons, chosen by the majority of recent neutrino beam experiments and projects (e.g.~T2K, MINER$\nu$A, OPERA, MicroBOOnE, DU\-NE), cannot be employed for a successful operation of the ENUBET monitored neutrino beam. In a fast-extraction sche\-me, the full primary proton beam is extracted onto the target in tens of microseconds or less: this would generate a pile-up rate not sustainable in the instrumentation of ENUBET, making a direct neutrino flux estimation impossible, let alone the possibility of time-tagging. The ideal extraction method for ENUBET is the slow resonant ('multi-turn') extraction of the primary protons, where the full intensity is extracted continuously in a time interval of a few seconds \cite{app11041644}. Currently, the CERN SuperSynchrotron (SPS) already provides the fixed target experiments at the North Experimental Area (NA) with a $400$~GeV slowly extracted proton spill of $4.8$~s length, for a maximum intensity of about $4.5\times10^{13}$ protons per spill. This beam represents an excellent candidate for ENUBET not only for the high energy and intensity that guarantee a high kaon yield: the absence of a Radio Frequency structure and low spill ripples would ensure a stable operation, and it would also be straightforward to use a $\sim 2$~s spill, which has an appropriate length for ENUBET \cite{Kain:2019qxl}.

\begin{figure}[h!]
    \centering
    \includegraphics[width=\columnwidth]{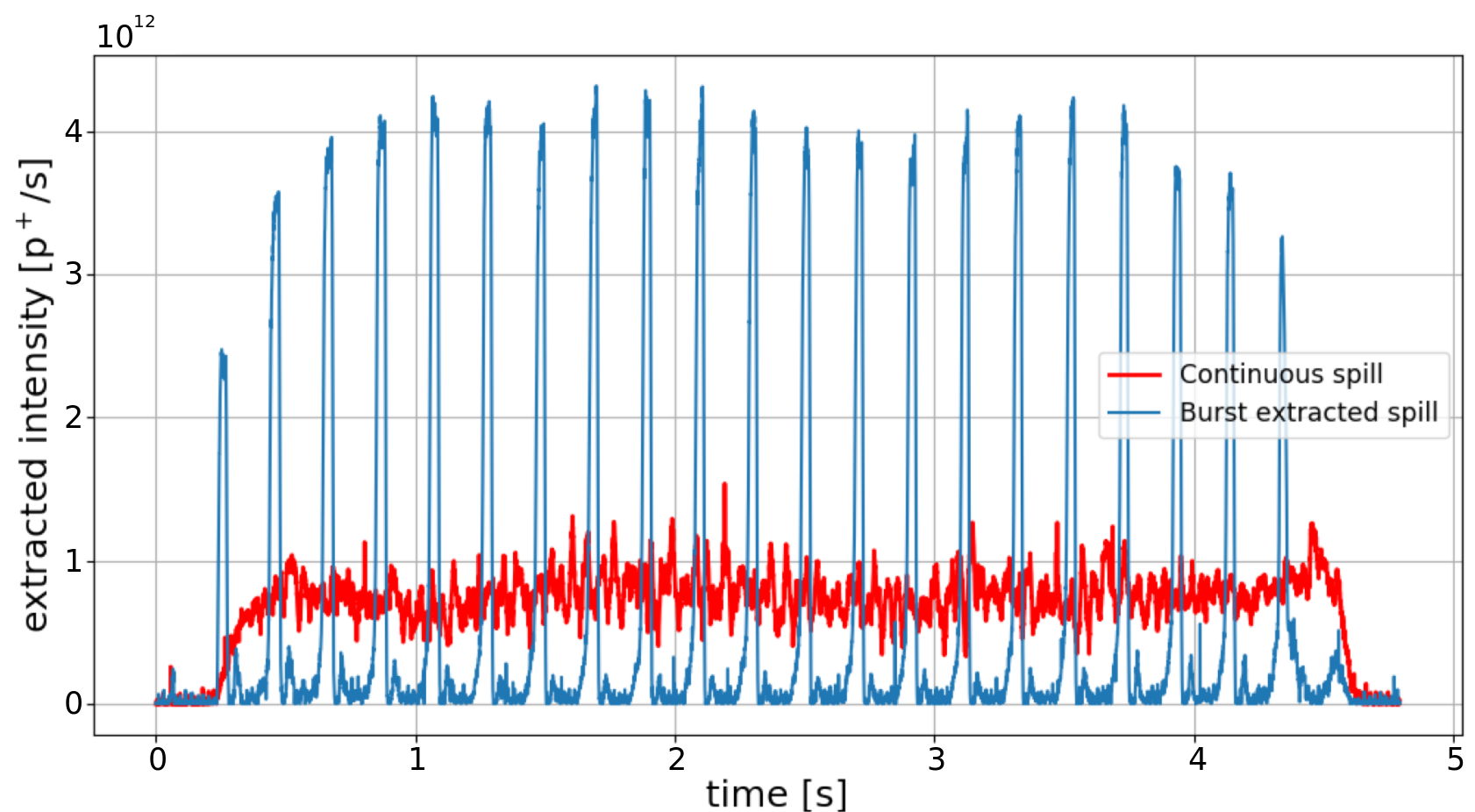}
    \caption{Example of nominal SPS slow extracted spill (red) and pulsed version (blue). The data has been measured with a secondary emission intensity monitor during dedicated machine developments.}
    \label{fig:burst}
\end{figure}

While this $2$~s SPS slow extraction is assumed as the default one for ENUBET, further in-depth studies on alternative slow extraction schemes and their possible improvements have also been performed~\cite{pari:tesi} in the framework of the ENUBET project. A pulsed version of the slow extraction (burst-mode slow extraction, shown in Fig.~\ref{fig:burst}) has been designed, developed, and successfully tested at the CERN-SPS with the goal of pairing it with stronger focusing devices as magnetic horns. 
This pulsed beamline does not fall into the scope of the paper but is being considered for other applications at the CERN SPS. 
Dedicated studies on the nominal SPS slow extraction have also shown how to further reduce its frequency noise, pointing to optimized extraction configurations not only for the case of ENUBET but for any fixed target experiment with tight rate requirements~\cite{pari:ripples}.

Simulation studies have proven that both for the nominal slow extraction and the burst-mode scheme (e.g.~with $10$~ms bursts repeated at $10$~Hz) the maximum particle rate in any calorimeter module of the ENUBET decay tunnel remains well below a few hundred of~kHz/cm$^2$, allowing ENUBET to perform lepton monitoring with moderate pile-up effects (see Secs. \ref{sec:positrons} and \ref{sec:muons}).

\section{Target}
\label{sec:target}
The target optimization for ENUBET was performed using the FLUKA~\cite{FLUKA1, FLUKA2} and G4beamline~\cite{g4beamline} Monte-Carlo simulation codes and was carried out assuming the CERN SPS as a proton driver (proton momentum: 400 GeV/c). Even if the ENUBET beamline can be implemented in any proton driver with an energy of around 100 GeV, the choice of the CERN SPS is particularly attractive when the mean secondary momentum corresponds to 8.5 GeV/c because the number of pions produced per proton hitting the target scales roughly with the energy of the primary protons~\cite{Feynman1969} and the secondary yield at 8.5 GeV/c is close to maximum \cite{parozzi_thesis}.
 
\begin{figure}[!hb]
\centering
\includegraphics[width=\columnwidth]{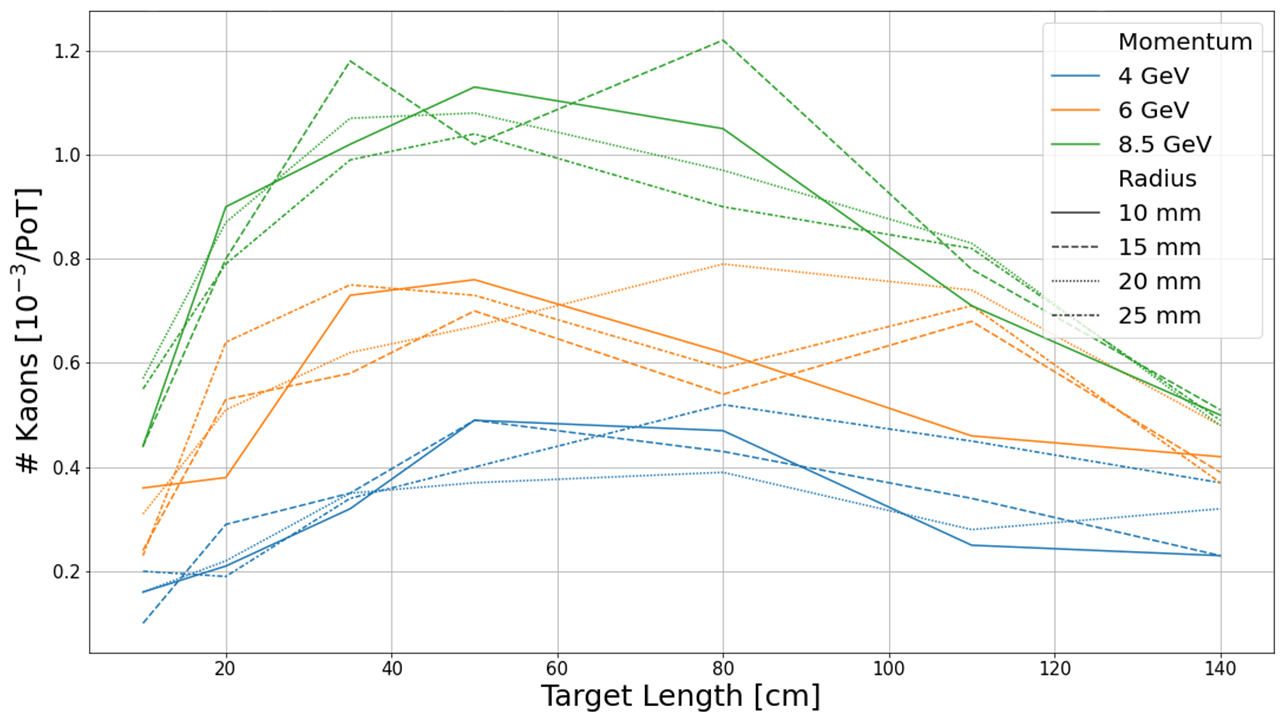}
    \caption{Kaon yields as a function of the graphite target length for a $400$~GeV/c proton beam. The figure of merit for the optimization is the number of kaons (10\% momentum bite) entering an ideal beamline with $\pm 20$~mrad angular acceptance in both planes placed $30$~cm after the target.  Colors refer to different kaon's momenta while the line style identifies the target radius. The error bars (not plotted to ease the reading) are dominated by the FLUKA interaction modeling systematics and amounts to $\sim 20$\%.
  } 
    \label{fig:K_prod}
\end{figure}

The most promising materials studied for ENUBET were graphite, 
%(2.23 g/cm$^{3}$ density) 
beryllium, %(1.85 g/cm$^{3}$)
and Inconel-718, %(8.19 g/cm$^{3}$). 
These materials are employed in moderate-power neutrino beams (like ENUBET) and Superbeams, that is the beams currently serving the T2K and NO$\nu$A experiments and their upgrades~\cite{T2K:2011qtm,T2K:2019eao,Adamson:2015dkw}.  
Each target prototype was modeled as a cylinder with various lengths (1-140 cm) and radii (10-30 mm). %Materials considered were gold, graphite, beryllium, and tungsten. 
Material comparison has shown a clear advantage of graphite over other materials even in kaon-enriched beams, whi\-ch confirms expectations since graphite is the material of choice for high-power pion en\-ri\-ched beams. Inconel shows slight\-ly less kaon  yields than graphite and, given the limited operational experience with this material for high-intensity neutrino beams, has not been investigated any further.   
Fig.~\ref{fig:K_prod} summarizes the kaon yields for different graphite target configurations. The figure of merit (FOM) for the optimization is the number of kaons (10\% momentum bite) that enters an ideal beamline with $\pm 20$~mrad angular acceptance in both planes placed $30$~cm after the target.
The leading parameter is the target length since 8.5~GeV particles are produced in the forward direction and the target radius plays a minor role in the re-interaction probability. FLUKA results indicate an optimal length of $\sim 70$~cm. We thus chose a 70~cm length, 3~cm radius as the reference target for the ENUBET beamline because smaller radii increase the mechanical complexity of the target station without improving the kaon yield. 

%\begin{figure}[!htb]

\section{Transfer line and static focusing system}
\label{sec:transfer_line}

\begin{figure}
    \centering
    \includegraphics[width=\columnwidth]{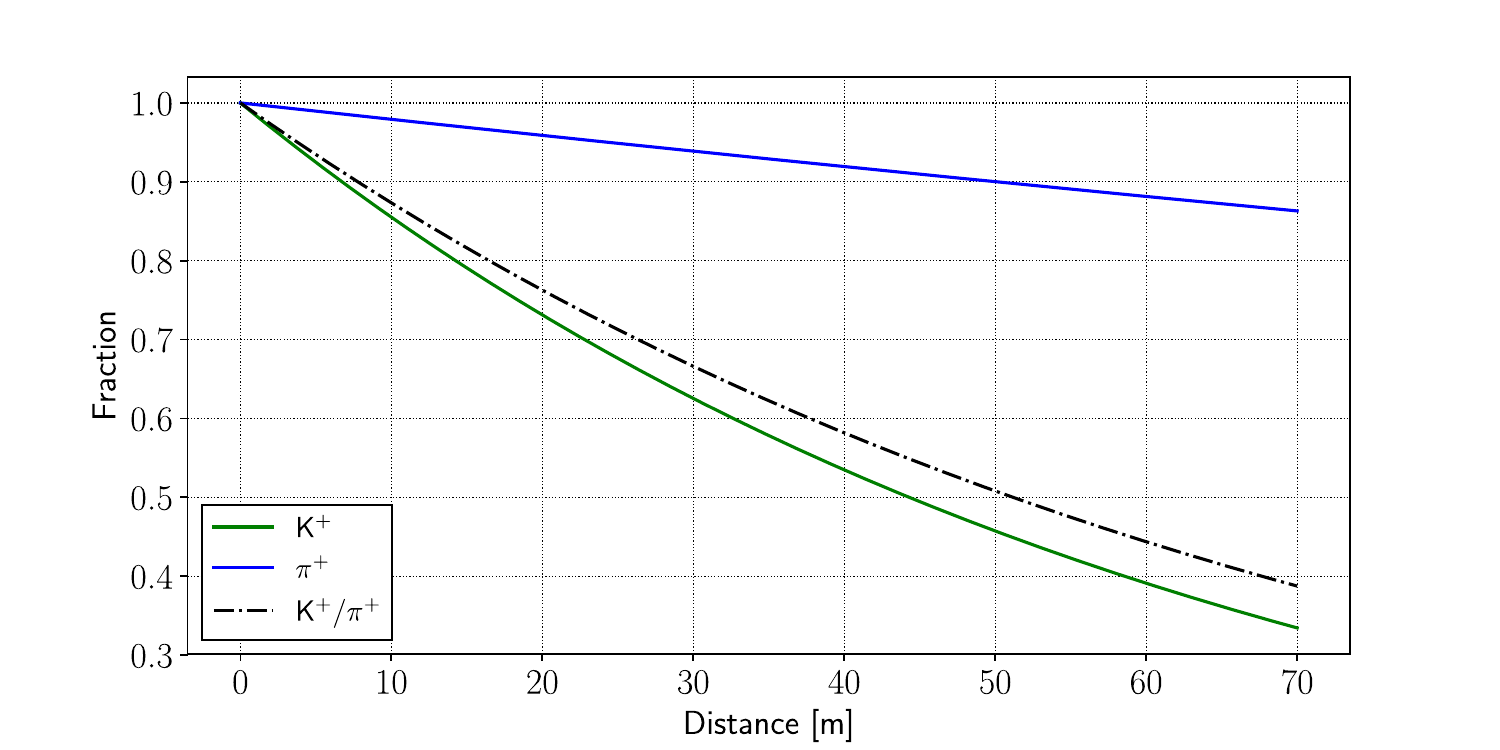}
    \caption{Decay rate and relative abundance of kaons and pions as a function of their travelled distance along the beamline at $8.5$~GeV.}
    \label{fig:kpiratios}
\end{figure}

\begin{figure}[h]
    \centering
    \includegraphics[width=0.9\columnwidth]{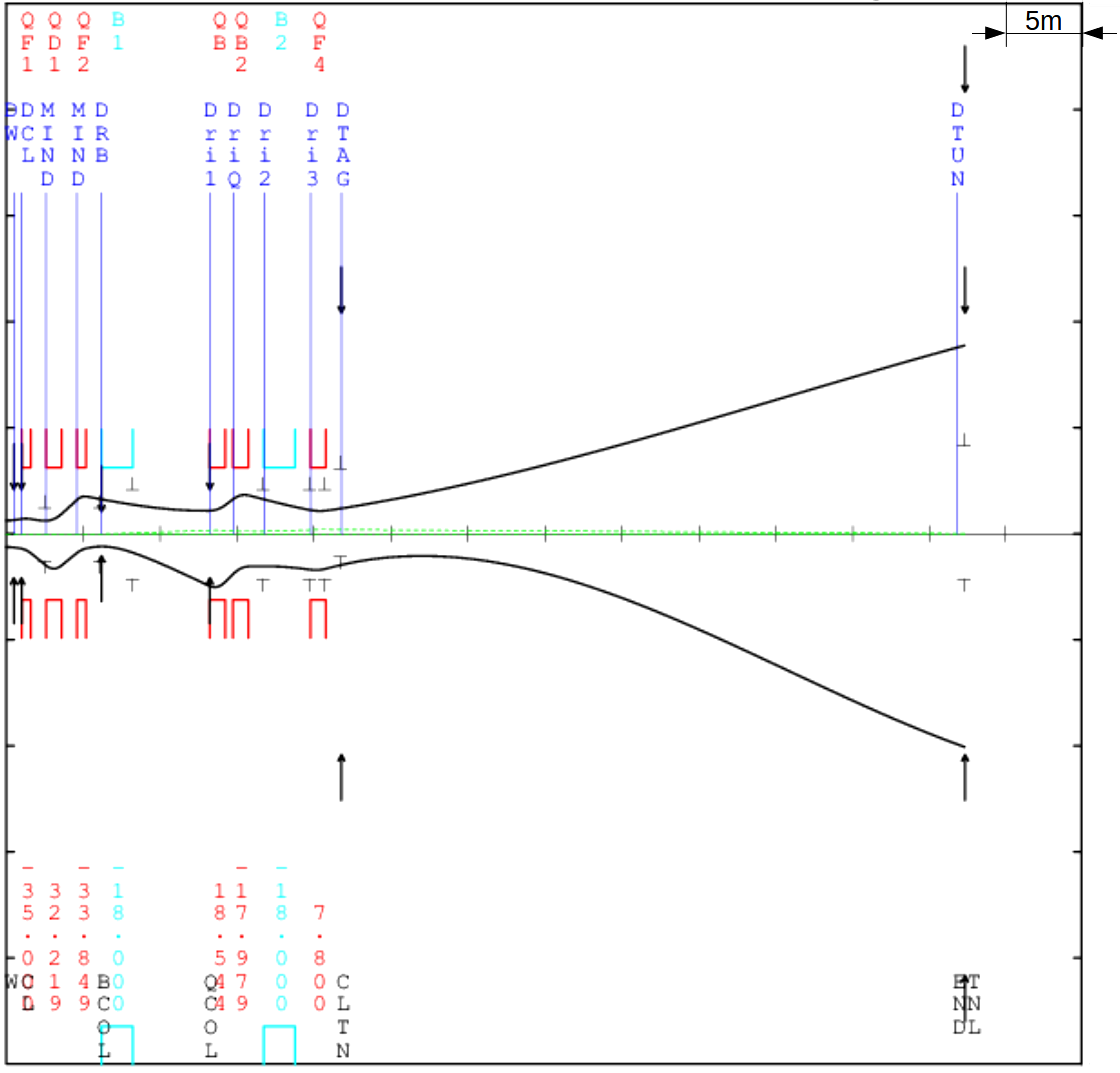}
    \caption{Beam optics and envelope (black lines) for the $X$ (up) and $Y$ (down) planes designed with TRANSPORT. The red rectangles indicate quadrupoles and their $X$ and $Y$ apertures, while the light-blue ones indicate dipoles. The black ticks mark the requirements to the optimizer for the beam size along the beamline. The green line represents the beam centroid, the black arrows at the end of the beam line correspond respectively to the 1 m radius of the tunnel (external arrows) and to 50 cm (internal arrows).}
    \label{fig:transport}
\end{figure}

As noted in Sec.~\ref{sec:monitored}, the ENUBET transfer line was designed to maximize the monitoring performance. 
This is a complex optimization process that involves several variables and constraints.
First of all, the detector technology at the instrumented decay tunnel can only withstand a particle rate $\lesssim 100$~kHz/cm$^2$ for successful lepton monitoring. This limit comes from the detector granularity ($3 \times 3$ cm$^2$), intrinsic time resolution ($\simeq 400$ ps), the front-end electronics and digitizer sampling rate ($\simeq 1$ Gs/s), as discussed in Secs.~\ref{sec:instrumentation} and~\ref{sec:eventbuilder}. Such a low particle rate can be reached with the aforementioned slow extraction of the primary protons.
%, where the accelerated protons are de-bunched and continuously extracted onto the ENUBET target in a few seconds \cite{Kain:2019qxl,app11041644}. 
The proton extraction mode sets an important design constraint for the ENUBET neutrino beamline, which relies on static focusing elements, i.e.~dipoles and quadrupoles, and does not employ a magnetic horn. Magnetic horns are difficult to be operated with a proton extraction longer than a few ms due to Joule heating of the conductors~\cite{app11041644}. Since ENUBET is designed as a narrow-band neutrino beam, this solution is quite natural: precise focusing and selection of secondary particles with a narrow momentum bite would require the use of quadrupole and dipole magnets even if a horn + fast proton extraction were employed.
The use of a magnetic horn in a monitored neutrino beam is possible but non-trivial, both for the hardware-side R\&D required to pulse a conventional horn for several ms and for the coupling with the static focusing elements~\cite{kopp2006}.  
The design of an effective, purely static beamline as the one presented in this paper has been a breakthrough in the ENUBET R\&D because it allows for very low pile-up levels, removes the operational complexity of a magnetic horn, and turned out to be highly cost-effective.
The fact that the ENUBET signal comes from kaon decays also puts a tight constraint on the beamline length, which has to be short enough to minimize kaon decays before the entrance of the decay tunnel. 
Hence, the length of the transfer line plays a pivotal role in the optimization process. Even with a transfer line as short as $20$~m, about $30\%$ of kaons are lost, and the K/$\pi$ abundance ratio drops by $\sim 25\%$ (see Fig.~\ref{fig:kpiratios}).
These constraints are imposed already in the first stage of the beamline design defining the layout and optics of the transfer line from the target to the decay tunnel. Beamline optics were simulated using the TRANSPORT code~\cite{Brown:1973jce} and validated with MAD-X~\cite{MAD-X}. This design process was driven by the following goals: 
\begin{itemize} 
    \item  maximize the number of $K^{+}$ and $\pi^{+}$ in the momentum range of interest for ENUBET at the entrance of the decay tunnel;
    \item focus the beam so that non-decaying particles exit the decay tunnel without hitting the tunnel walls, and constrain the beam envelope to be fully contained inside the aperture of all beamline elements (quadrupole, dipoles, collimators);
    \item use field and aperture values of the magnets compatible with existing technology. In ENUBET, we only employed normal-conducting, conventional devices to reduce the cost and complexity of the neutrino beamline.
\end{itemize}
The optics were optimized requiring TRANSPORT to 
minimize the total length of the transfer line and constrain the beam size in both planes at the tunnel exit (see Fig.~\ref{fig:transport}). 

\begin{figure}[h!]
    \centering
    \includegraphics[width=0.9\columnwidth]{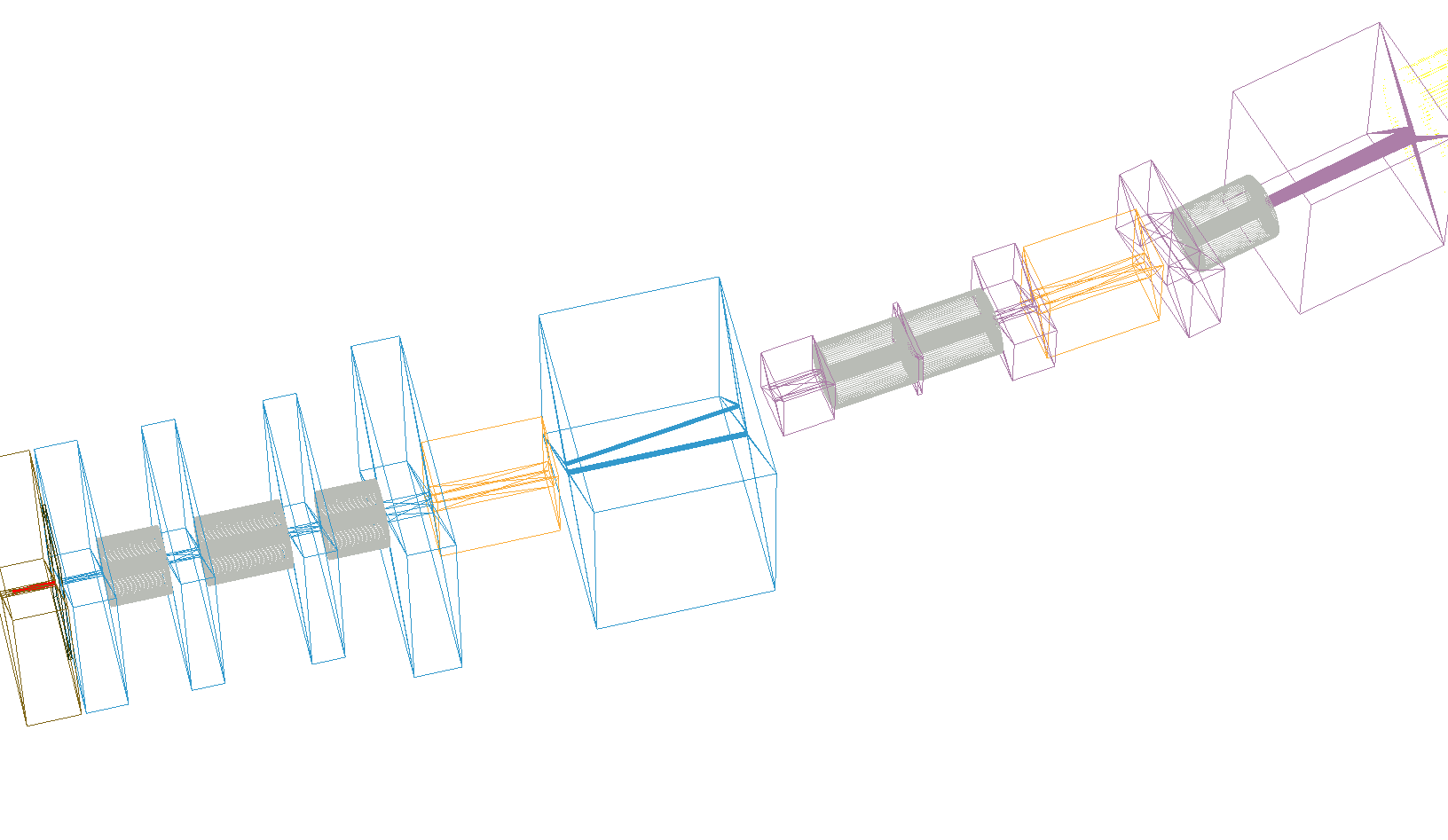}
    \caption{Details of the beamline: in blue the copper collimators, in purple the Inermet180 collimators. Quadrupoles and dipoles are shown respectively in gray and orange. The target (red cylinder) is surrounded by concrete shielding.}
    \label{fig:bl_details}
\end{figure}

The reference TRANSPORT beam has a momentum of $8.5$ GeV/c with a momentum bite of $5$\%. The momentum bite considered in the optics optimization is smaller than the final momentum bite used by ENUBET (10\%) due to tertiary interactions, i.e.~the spread that is expected moving from the optics simulation to the full tracking and interaction simulation. There are several possible configurations in terms of magnetic elements that can be used to transport kaons and pions to the instrumented decay tunnel. The best one achieved in this optics optimization phase consists of a quadrupole triplet followed by a bending dipole, a pair of quadrupoles, another bending dipole identical to the first one, and a final quadrupole (see Fig.~\ref{fig:bl_details}). The two dipoles are based on existing CERN magnets that can be operated up to $1.8$~T with a field length of $2.038$~m and an aperture of $300$~mm. Each dipole provides a bending angle of 7.4$^\circ$, for a total bending of the beam with respect to the primary proton line of $14.8^\circ$. Fields at poles in the quadrupoles are kept below $11$~kG, for an aperture radius of $15$~cm. The result of the optics design is shown in Fig.~\ref{fig:transport}.

\begin{figure}[h]
    \centering
    \includegraphics[width=0.8\columnwidth]{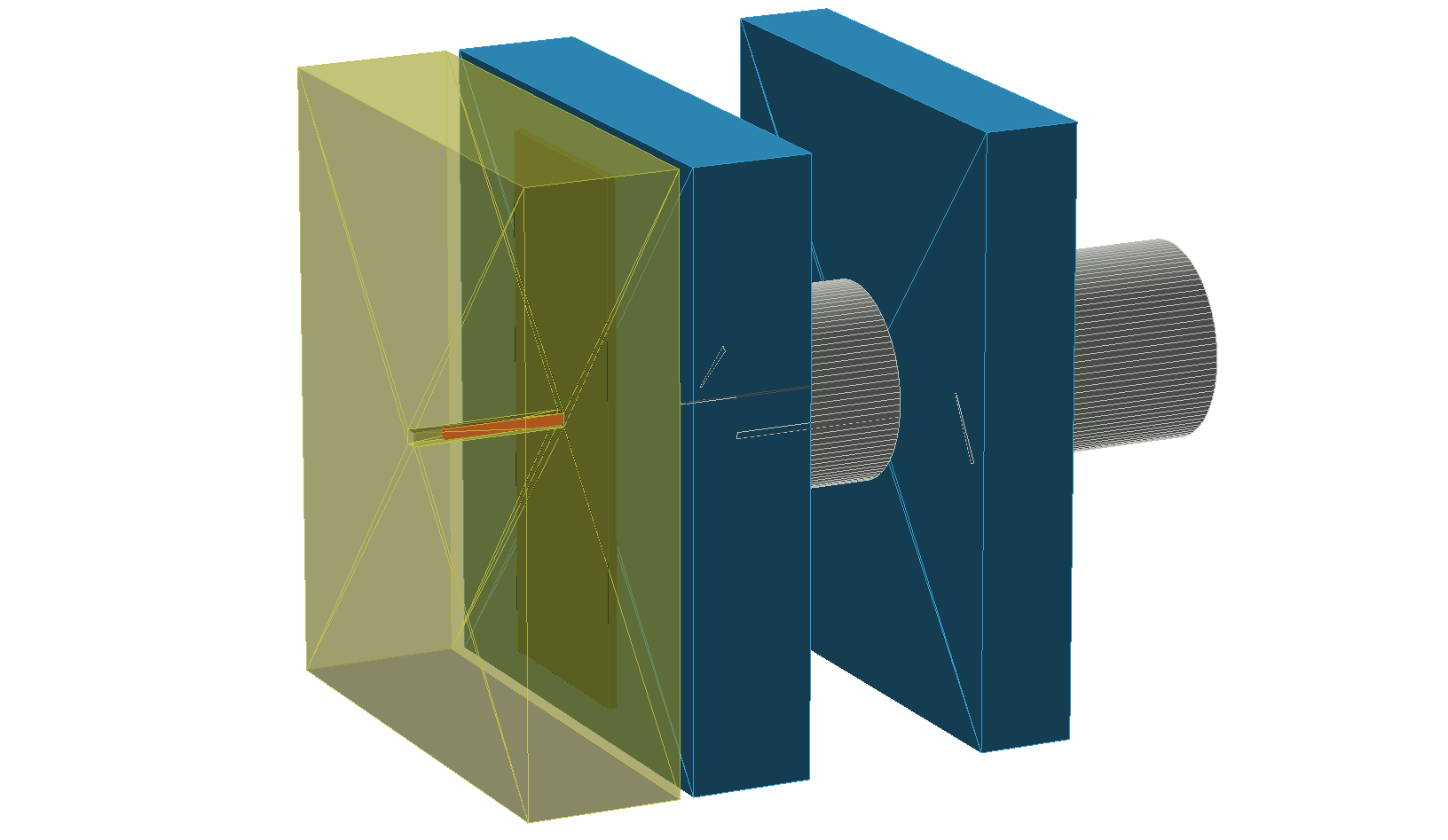}
%    \includegraphics[width=0.45\columnwidth]{Figures/tlr5Wbox.png}
%    \caption{Collimation details of the ENUBET beamline. Left: target (red) and concrete shielding (yellow), first quadrupoles copper shielding (blue) and Tungsten foil for absorpion of target positrons (black). Right: detail of last two Inermet180 collimators before the tunnel entrance. Quadrupoles are shown in gray.}
    \caption{Target (red) and concrete shielding (yellow), first quadrupoles copper shielding (blue) and Tungsten foil for absorption of target positrons (black layer downstream the yellow shield).}
    \label{fig:wfoil}
\end{figure}

\begin{figure}[h]
    \centering
    \includegraphics[width=\columnwidth]{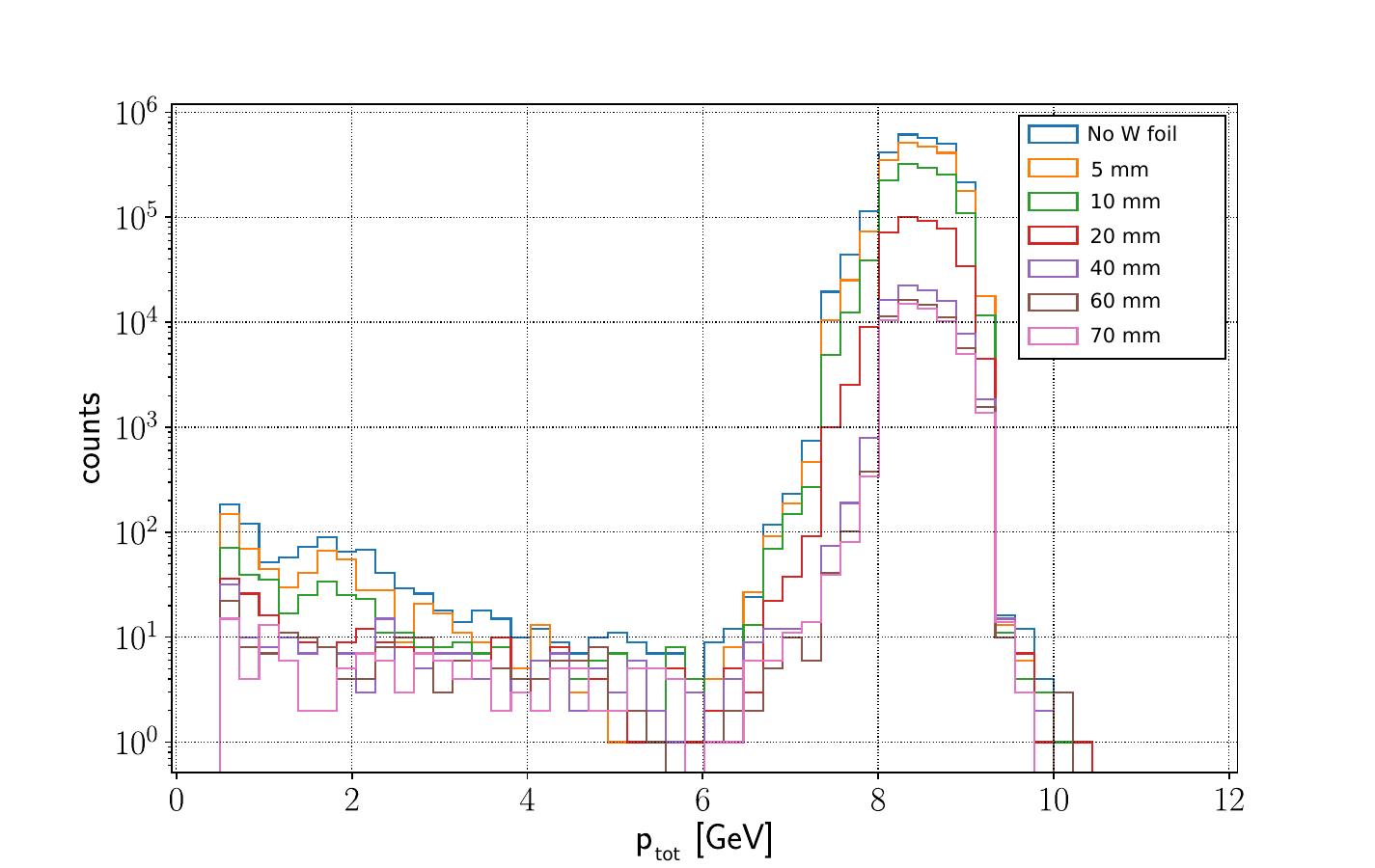}
    \caption{Momentum distributions of the positrons entering the decay tunnel for different values of the Tungsten filter thickness.}
    \label{fig:wfscan_pt}
\end{figure}

\begin{figure}[h]
    \centering
    \includegraphics[width=\columnwidth]{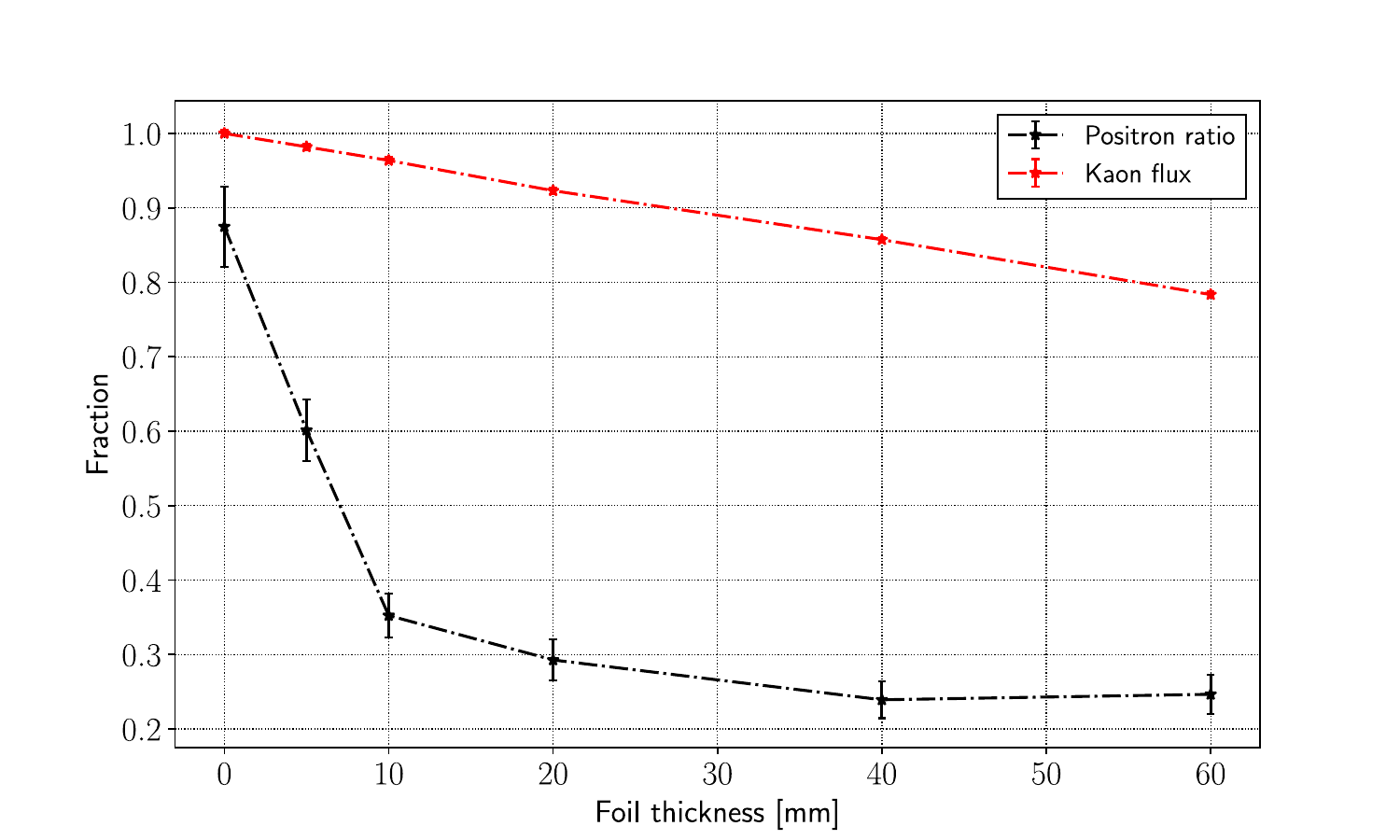}
    \caption{Results of the tungsten foil thickness scan. The foil is placed just after the graphite target. Black line: ratio of background positrons to signal positrons (i.e. $K_{e3}$ positrons) hitting the walls of the instrumented decay tunnel. Red line: surviving fraction of the kaon flux at the entrance of the decay tunnel.}
    \label{fig:wfscan}
\end{figure}

Once the beam optics is designed, the full beamline was implemented and simulated into particle tracking and interaction codes: FLUKA, GEANT4, and G4beamline~\cite{Roberts:2007nte}, with the addition of absorbers and collimators between elements, a Tungsten foil after the target to screen positrons that would otherwise reach the tagger and contribute to the background, a low power hadron dump at the end of the decay tunnel and the proton dump. This is a fundamental step of a secondary beamline design since the optics design programs are limited to the decay- and interaction-less propagation of the particles within the momentum bite. In a real secondary beamline as the ENUBET transfer line, many undesired particles move from the target down to the decay tunnel, together with the products of interactions with collimators, shielding, and materials around the beamline. 
The beamline was implemented both in G4Beamline and GEANT4.
G4Beamline is a standard tool used for secondary beamline simulation and development: it is built on top of GEANT4 and relies on it for all physics processes, providing a higher-level interface and several preset configurations to speed up code development and simulation times (e.g.~precise and automatic implementation of standard beamline magnets and their fields). Using this program was thus ideal for the rapid development and first optimization of the ENUBET beamline. On the other hand, using directly GEANT4 allows for superior control and customization of the whole simulation process: the GEANT4 implementation of the ENUBET beamline allowed to retain and analyze the full particle history and better assess the systematics of the neutrino beam (see Sec.~\ref{sec:neutrino_fluxes}). This model was also used to perform the final optimization phase of the beamline.
The FLUKA code was instrumental in the simulation of the meson yields after the target (Sec.~\ref{sec:target}) and to evaluate the doses at the beamline elements and the instrumentation (Sec.~\ref{sec:doses}). 
The interaction of the primary proton beam in the target is thus simulated with FLUKA and the exiting particles are used as input for the complete transfer line simulation. The graphite target is shielded with concrete on the upstream side, while it faces the first copper collimator of the quadrupole triplet on the downstream side. A $5$~cm thick tungsten foil ($\sim14.3$~X$_0$, $0.5~\lambda_0$) is placed just after the target in order to filter out excess positrons produced by the $400$~GeV protons interaction with graphite. This solution is commonly employed in fixed target experiments and, in particular, in kaon physics experiments when positrons and electrons originating from the target constitute a potential background~\cite{NA62:2017rwk}. 
The target block with the tungsten foil is shown in Fig.~\ref{fig:wfoil}.
Differently from the usual case, where the positron filter is placed on a beam waist, ENUBET employs it almost as a second target. The overall positron energy reduction from a standard-placed filter along the beamline would have increased the number of low-energy background positrons that hit the tunnel walls or significantly impact the hadron beam parameters if too thick. Conversely, using the filter just after the graphite target does not cause a shifting of the positron spectrum, but leads to an overall decrease of the positron background in the target (see  Fig.~\ref{fig:wfscan_pt}).
The transfer line is designed to transport on-momentum positrons (peak centered at $8.5$~GeV/c in Fig.~\ref{fig:wfscan_pt}) across the decay tunnel without hitting the walls, so the main contribution to the background is given by the low-momentum tail (below 6 GeV/c), despite its lower number of particles. The final thickness of the ENUBET tungsten filter was chosen with a dedicated study based on GEANT4, in which the signal-to-noise ratios and the overall flux reductions were analyzed as a function of the filter thickness, as shown in Fig.~\ref{fig:wfscan}.

\begin{figure}[h!]
    \centering
    \includegraphics[width=\columnwidth]{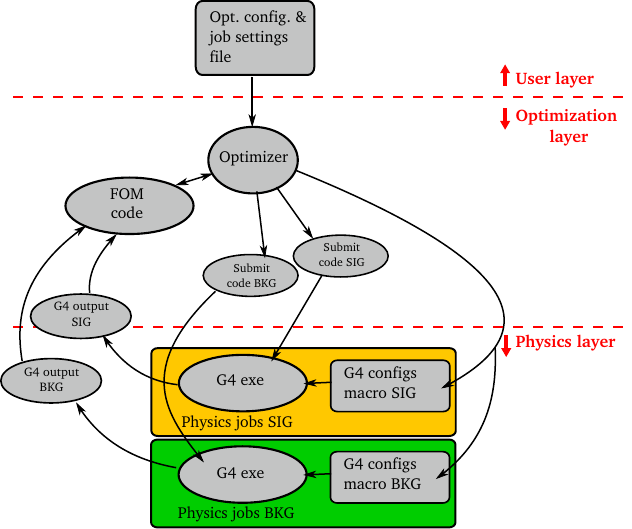}
    \caption{Scheme of the upgraded ENUBET optimization framework: the user can fully perform optimization by using a custom high-level instruction language (user layer). Multiple physics simulations can be developed and prepared for optimization (physics layer). The optimizer code will automatically handle the physics parameters, the computing cluster operations, and the FOM evaluation and combination for the optimization algorithm.}
    \label{fig:optimizer}
\end{figure}

\begin{figure*}[h!]
    \centering
    \includegraphics[width=0.33\textwidth]{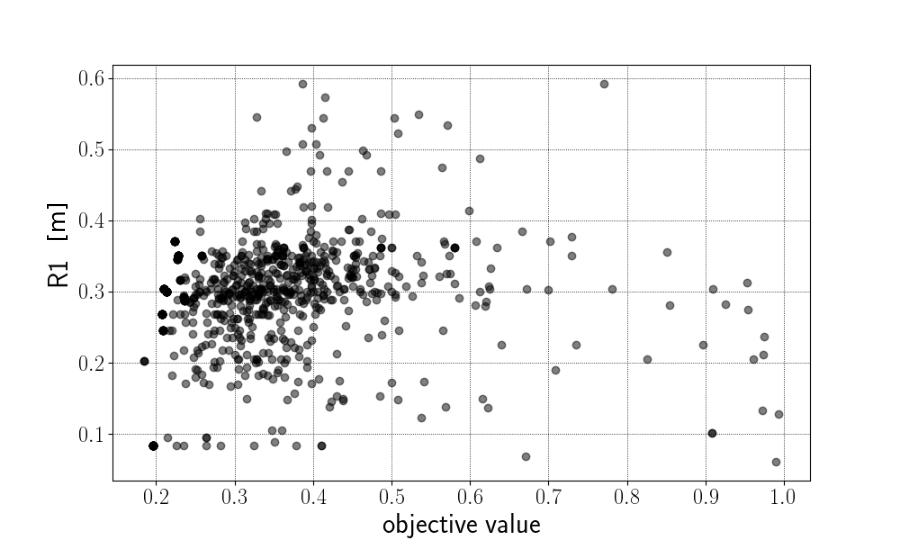}
    \includegraphics[width=0.33\textwidth]{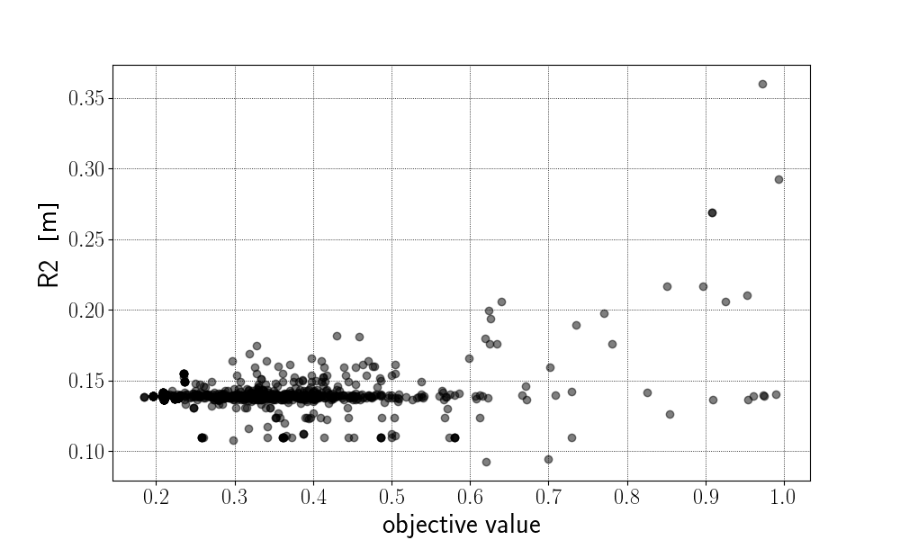}
    \includegraphics[width=0.33\textwidth]{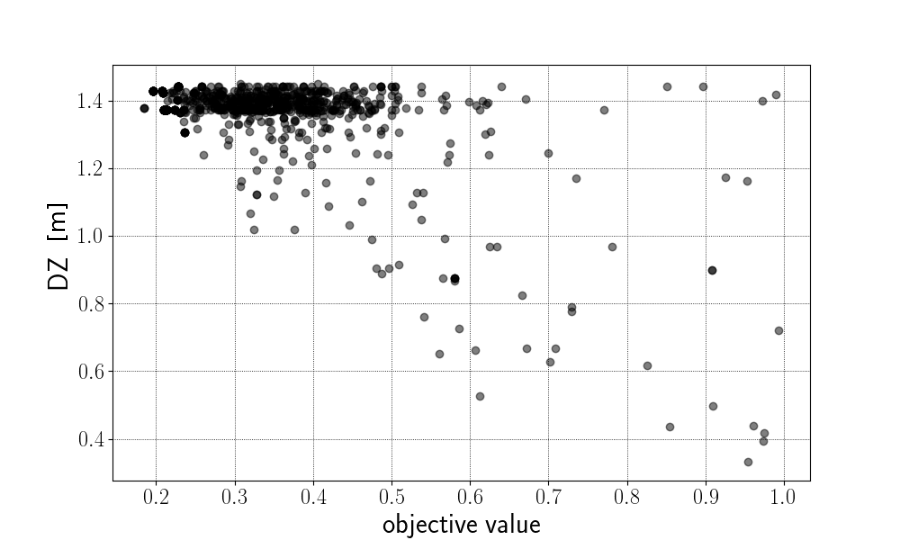}
    \caption{Figure of merit (objective value) as a function of the optimization parameters of the last collimator, which proved to be the most critical one for the final result. 
    R1 and R2 are the upstream and downstream aperture radii of the last collimator, and DZ its half-length. 
    }
    \label{fig:scatter}
\end{figure*}

\begin{figure}[h!]
    \centering
    \includegraphics[width=0.85\columnwidth]{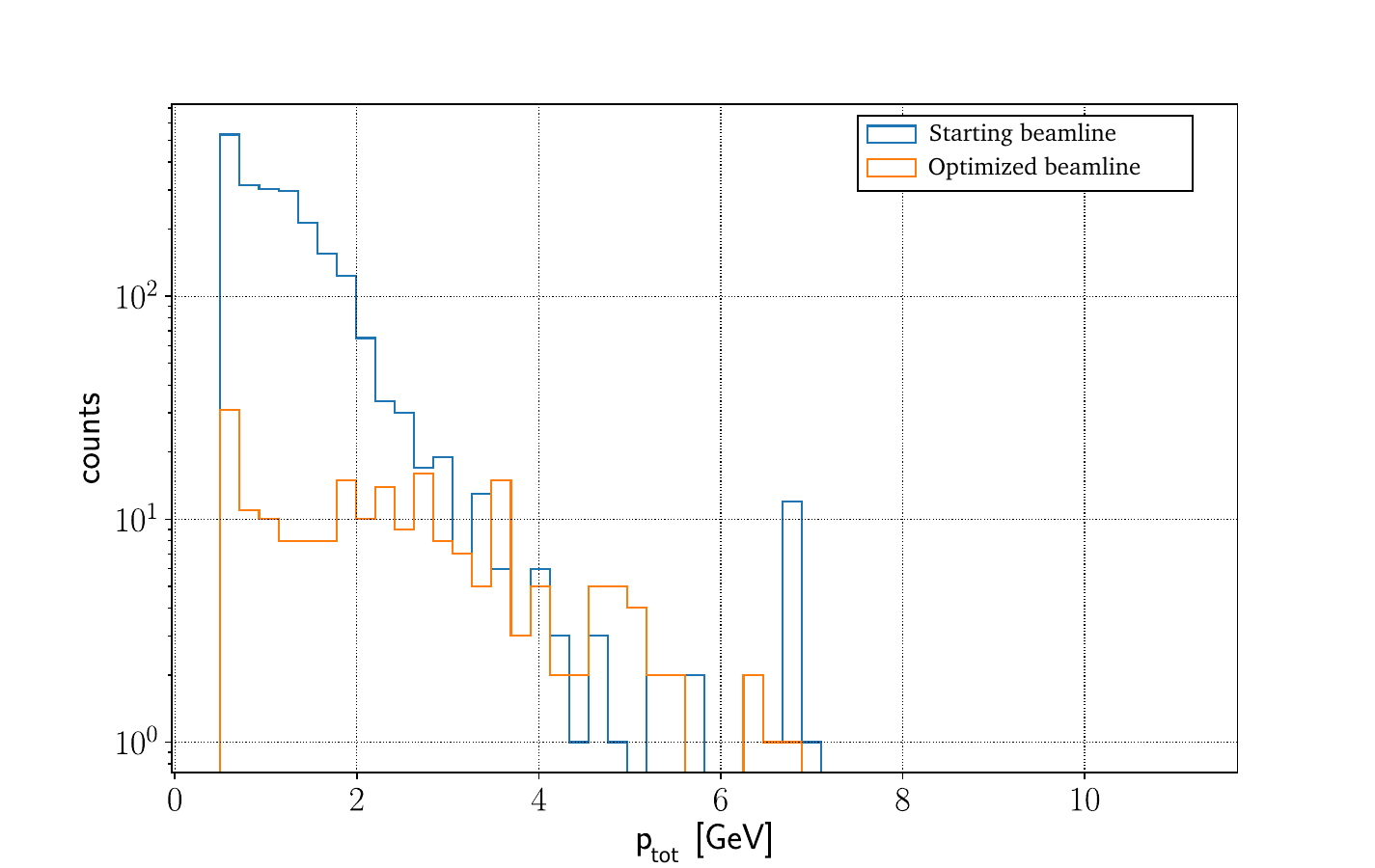}
    \includegraphics[width=0.85\columnwidth]{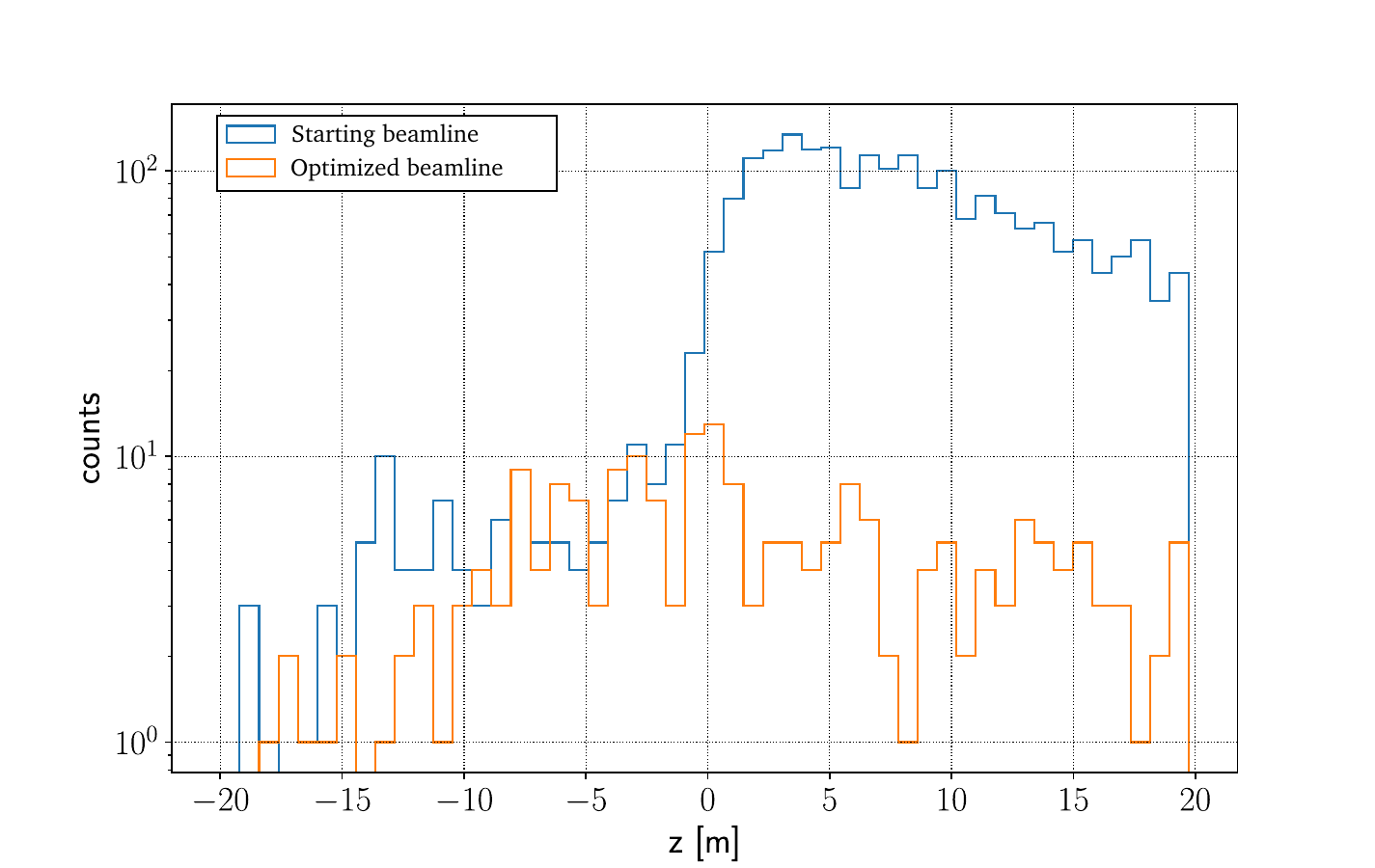}
    \caption{Background positrons hitting the tunnel walls. Top: momentum distribution. Bottom: impact point along the tunnel.}
    \label{fig:bkgposopt}
\end{figure}

\begin{figure}[h!]
    \centering
    \includegraphics[width=0.85\columnwidth]{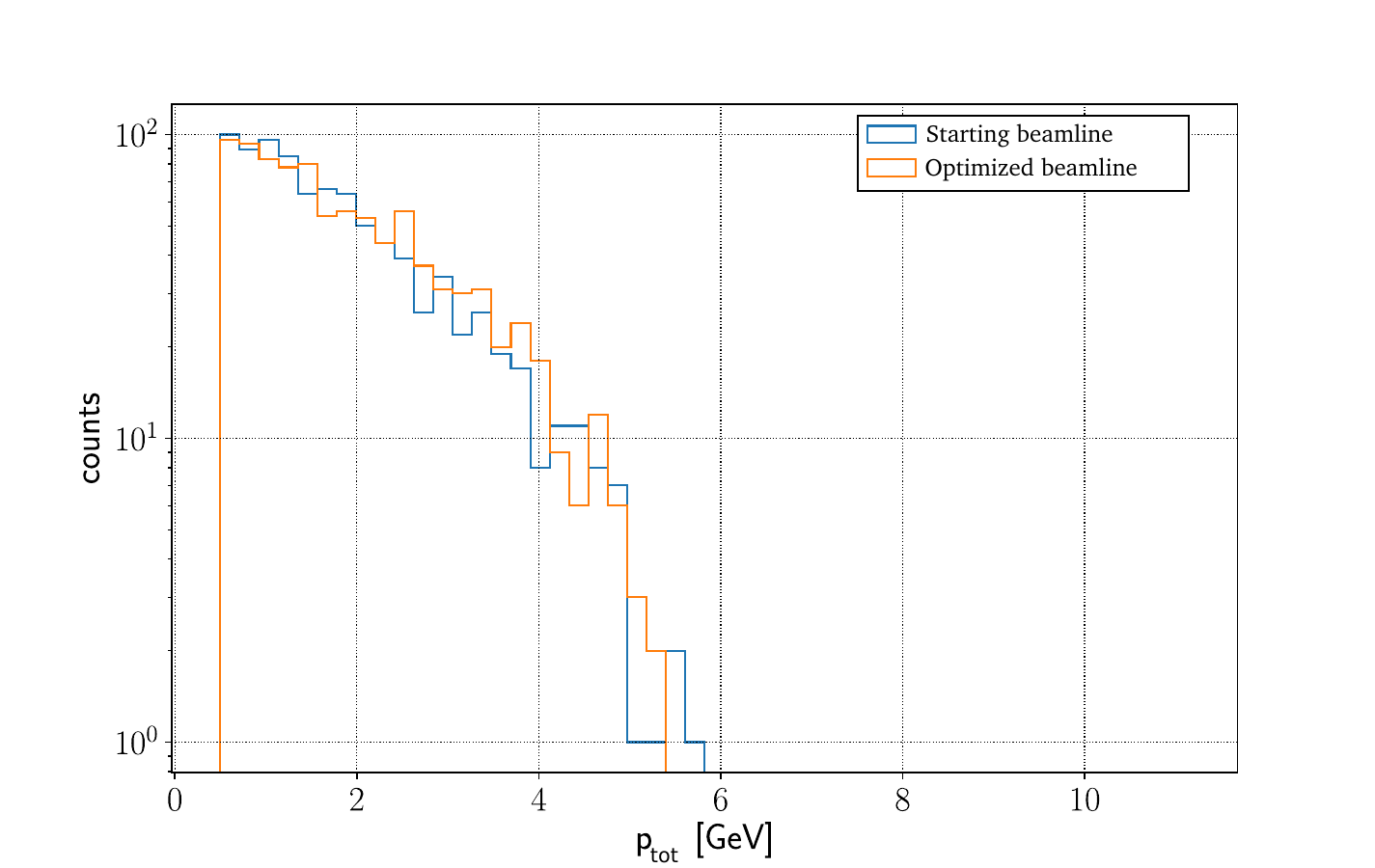}
    \includegraphics[width=0.85\columnwidth]{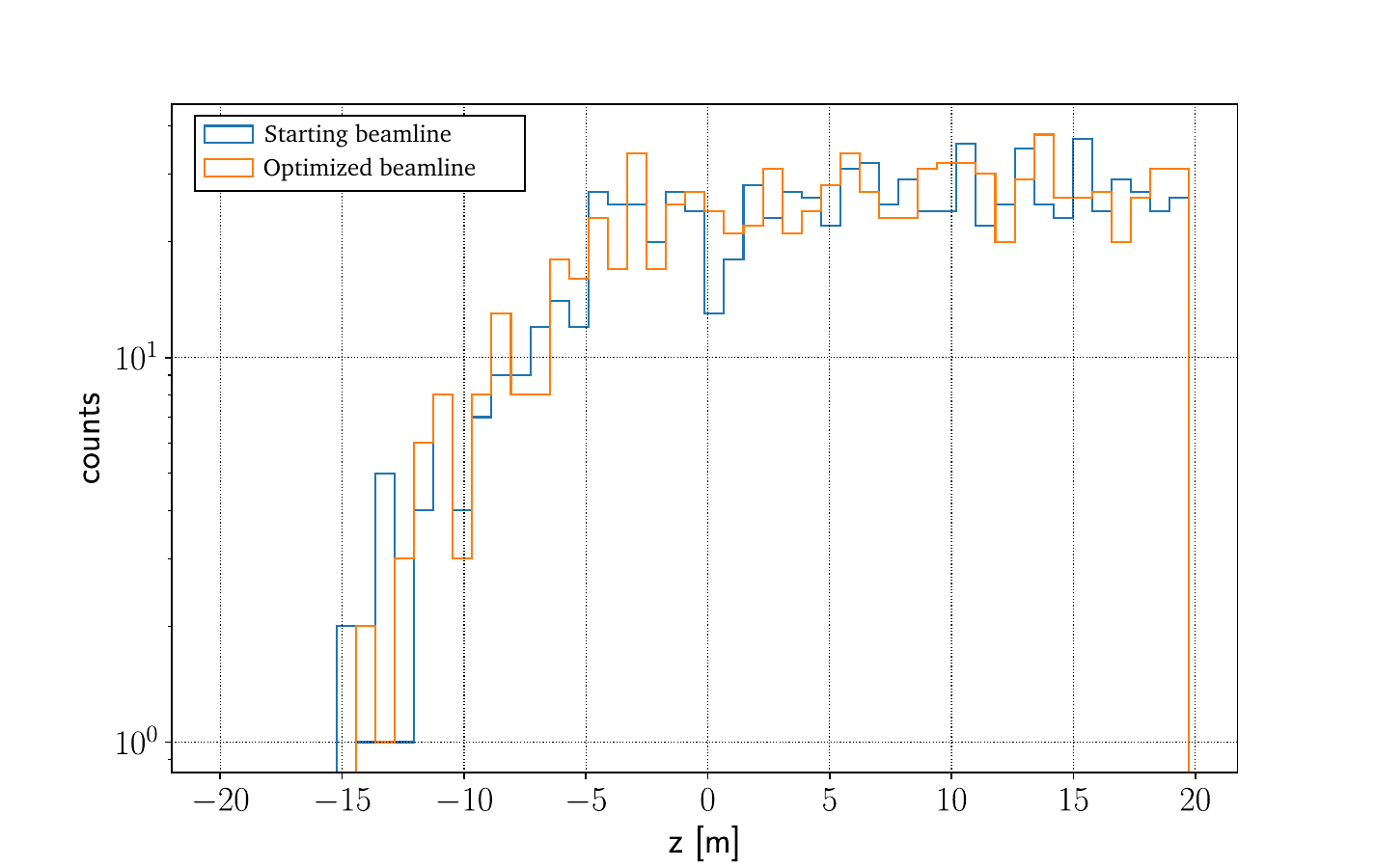}
    \caption{Signal positrons hitting the tunnel walls. Top: momentum distribution. Bottom: impact point along the tunnel.}
    \label{fig:sigposopt}
\end{figure}

The beamline collimation has been designed using copper blocks for the first quadrupole triplet and the momentum collimator, while in the second half of the beamline, the collimators are made of Inermet180, a heavier Tungsten-based alloy employed for the LHC collimators~\cite{inermet}. 
As shown in Fig.~\ref{fig:bl_details}, all collimators but two are implemented with rectangular apertures. The momentum collimator, located after the first dipole, contains two pipes, one for the bent $8.5$~GeV beam and the other for the straight $400$~GeV protons. The last Inermet collimator features a conical aperture, larger on the downstream side.
Finding the best collimation configuration of a secondary beamline like the one of ENUBET is a challenging task: a significant number of unwanted and off-momentum particles travel to the decay tunnel, interacting with the beamline elements and generating even more background particles. It is a multi-parameter problem that can only be solved with intensive numerical simulations. For this reason, we employed a numerical optimization approach relying on a meta-heuristic population-based method like the genetic algorithm also used for the nuSTORM horn optimization \cite{kochenderfer,Liu:2015ylc,pari:tesi}. The optimization framework developed in~\cite{pari:tesi} for the horn optimization in ENUBET was significantly upgraded for this task since the beamline simulation is much more CPU intensive. In the computing cluster that we used for the optimization (CC IN2P3~\cite{in2p3}) the simulation of a single job of $10^5$~protons-on-target (pot) takes from $8$ to $14$ hours of CPU time and must be parallelized up to $3000$ jobs.
The optimizer workflow that carries out this optimization task is depicted in Fig~\ref{fig:optimizer}. It is based on a high-level instruction language (user layer) that defines the parameters and the figure of merit (FOM). The optimizer generates a set of parallel jobs by means of an optimization layer to achieve the configuration with the optimal FOM. 
For the case of the ENUBET beamline, the collimators most influencing the signal-to-noise ratio were the last two Inermet ones before the decay tunnel: the optimization was thus set as a $5$-parameters problem consisting in the radii of the conical aperture of the last collimator, its length and the half apertures of the next to last collimator. The figure of merit is the ratio between the background and signal positrons hitting the tunnel walls. 
To further speed up the optimization process:
\begin{itemize}
    \item for the signal positrons, the K$_{e3}$ branching ratio was set to $100\%$ once kaons enter the decay tunnel; this leads to a statistics increase of about a factor $20$, with no modifications in any positron distribution;
    \item 
    background positrons were selected after a cut in the momentum of particles coming from the target (from $7$ to $100$~GeV/c) allowing electromagnetic processes only.
    This leads to a speed-up of about a factor $10$  with a statistics reduction of a factor $\lesssim 2$.
\end{itemize}
These two configurations - corresponding to signal and background respectively - were run as two separate simulations (see Fig~\ref{fig:optimizer}). $20$ beamlines were run in parallel and set the population size: $20$ Mpot for the signal case and $25$ Mpot for the background case. Convergence was reached in about $50$ iterations.
The impact of different parameters on the figure of merit is shown in Fig~\ref{fig:scatter}. The optimizer favors a longer last collimator and a tighter downstream radius before the decay tunnel. The second to last collimator, on the other hand, better performs when we set a high aperture.
The signal and background positron distributions at the end of the optimization process (Figs.~\ref{fig:bkgposopt} and  Fig~\ref{fig:sigposopt})  show the improvement with respect to the non-optimized version of the beamline. 
The final length of the last collimator is $\sim 2.8$~m, for a total length of the transfer line up to the tagger entrance of $26.7$~m.
%\vspace{1cm}

\begin{figure}[h]
    \centering
    \includegraphics[width=1.0\columnwidth]{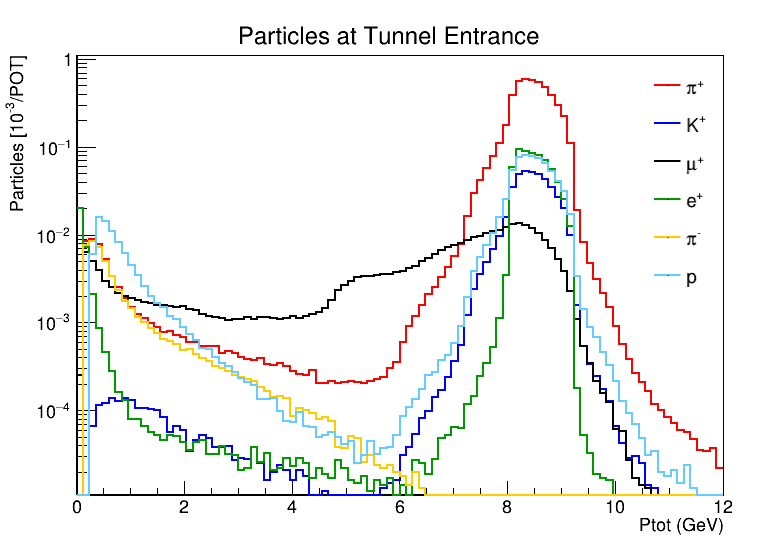}
    %{Figures/fluxesAtTagger.png}
    \caption{Momentum distribution of particles at the ENUBET tagger entrance as obtained from the GEANT4 simulation after all optimizations.}
    \label{fig:fluxesAtTagger}
\end{figure}

The flux at the tunnel entrance for 8.5 GeV/c particles in a  10\% momentum bite is $0.4 \times 10^{-3}$/pot for $K^{+}$ and $4.6 \times 10^{-3}$/pot for $\pi^{+}$. The momentum spectra 
at the tagger entrance with the GEANT4 simulation in standard mode (i.e. with the nominal $K_{e3}$ BR) of the final optimized beam\-line are shown in Fig.~\ref{fig:fluxesAtTagger}.

\begin{figure}[h]
    \centering
   \includegraphics[width=1.0\columnwidth]{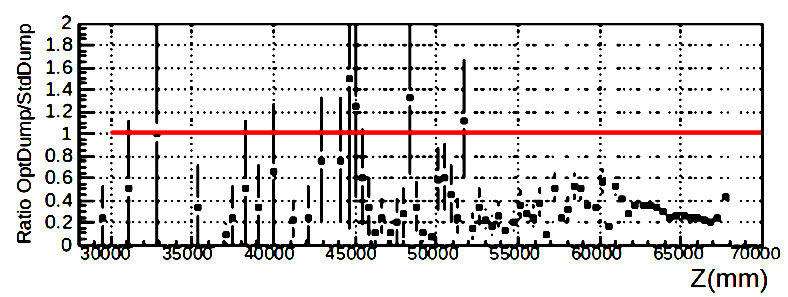}
    \caption{Ratio of neutron fluence resulting from the backscattering for the optimized hadron dump and an iron dump as a function of the longitudinal position along the tagger. In the last few meters - where the neutron background is dominated by backscattering - the ratio is $\sim$0.2.}
    \label{fig:neutrons}
\end{figure}

The final design of the hadron dump at the tagger exit is the result of a dedicated study. It is placed $2$~m after the tagger exit and was designed to reduce the background due to backscattering particles reaching the decay tunnel instrumentation. It is composed of a graphite core ($50$~cm diameter) placed inside a layer of iron ($1$~m diameter) covered by borated concrete ($4$~m diameter). Additional borated concrete ($1$~m thick) is placed in front of the hadron dump leaving an opening for the beam (see Fig.~\ref{fig:tlr6v5}). This dump configuration is commonly employed in secondary beamlines to reduce neutron flux. In the last few meters of the tunnel where the neutron fluence is more critical, the ratio between neutrons from the hadron dump hitting the tagger with respect to those one would get with a simpler block of iron is $\sim 0.2$, as shown in Fig.~\ref{fig:neutrons}.

The proton dump has a similar design with three cylindrical layers: a $3$~m long graphite core, surrounded by aluminum, which in turn is covered by iron. It is placed at the end of a pipe for the primary $400$~GeV proton beam that is filled with air and surrounded by concrete walls (see Fig.~\ref{fig:tlr6v5}).

\section{Decay tunnel instrumentation}
\label{sec:instrumentation}   

\begin{figure}[htb!]
\centering
  \includegraphics[width=\linewidth]{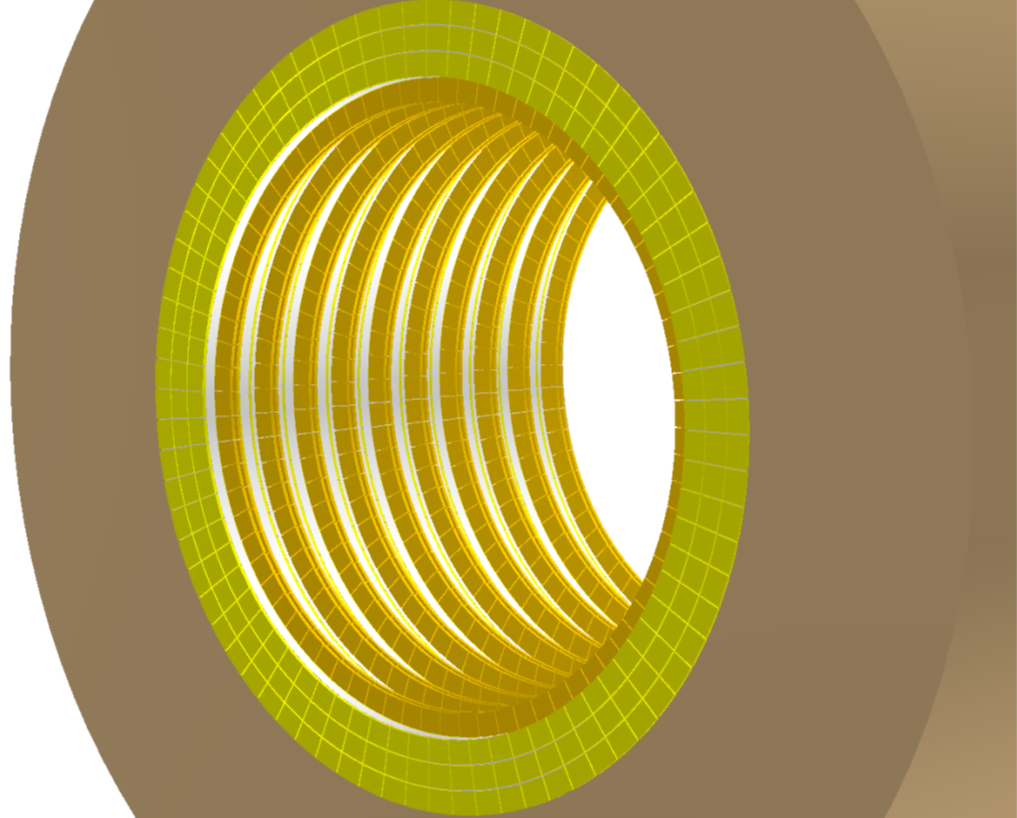}
  \caption[]{Schematics of the ENUBET instrumented decay tunnel. The three layers of modules of
the calorimeter (light green) constitute the inner wall of the tunnel. The rings of the scintillator tiles (doublets) of the photon veto (yellow) are located just below the modules. The
optical fibers (not shown) bring the light to the outer part of the tunnel in the radial direction. They cross the neutron shielding (light brown) where the SiPMs (not shown) are positioned.}
  \label{fig:schematics_instrumented_tunnel}
\end{figure}

\begin{figure}[htb!]
\centering
  \includegraphics[width=\linewidth]{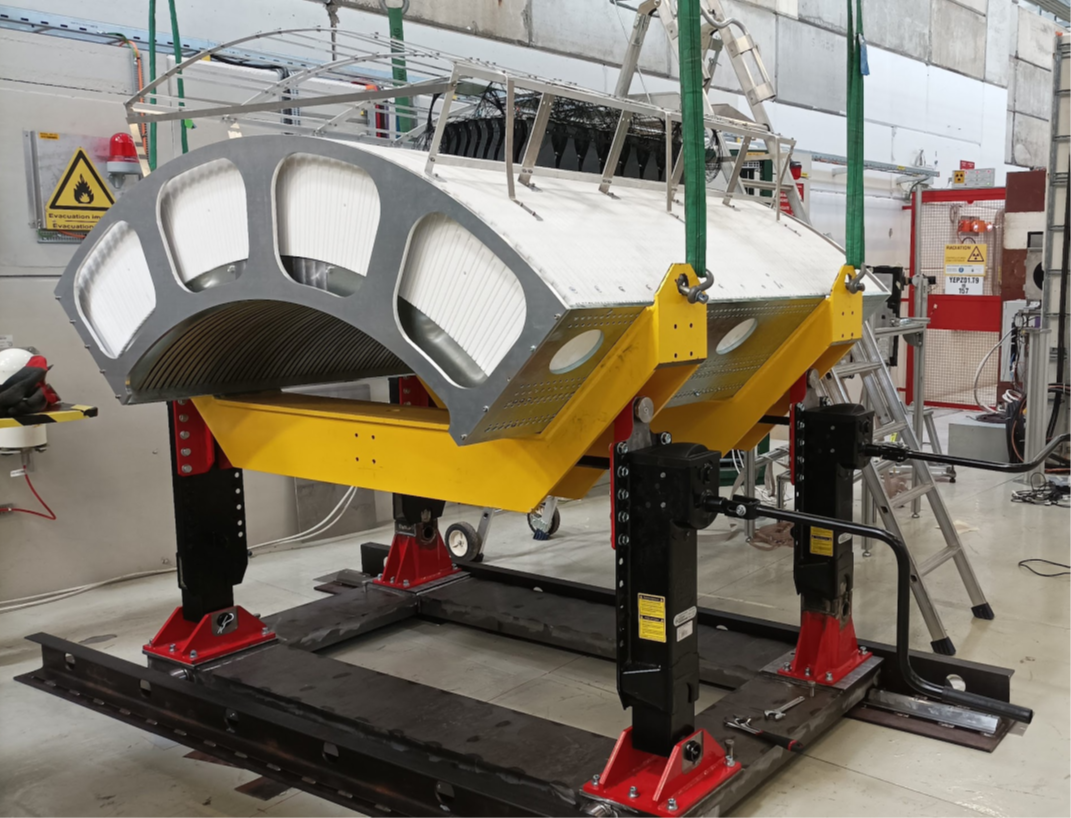}
  \caption[]{The ENUBET demonstrator in the CERN T9 beamline. The white layer between the inner part of the detector and the SiPM electronics is the borated polyethylene shielding. During the test, only 20\% of the detector (length: 80 cm, $18^\circ$ coverage in the azimuthal angle) was instrumented grouping the fibers and steering them to the SiPM boards located on top of the black plastic guides above the shielding. A test with an increase of the readout fraction by a factor 3 has been performed in August 2023.}
  \label{fig:demonstrator}
\end{figure}

The detector technology for the instrumentation of the decay tunnel was investigated by the ENUBET collaboration from 2016 to 2022 \cite{Berra:2016thx,Berra:2017rsi,Ballerini:2018hus,Acerbi:2019wti,Acerbi:2020itd,Acerbi:2020nwd} and culminated with the construction of a Demonstrator: a 1.7 m long section of the instrumented tunnel validated with a charged particle beam at the CERN East Experimental Area \cite{Torti:2023hsc,demonstrator}. The instrumentation comprises two detectors: a modular sampling calorimeter and a photon veto (``$t_0$ layer''). Each calorimeter module (Lateral-readout Compact Module - LCM) is made of five slabs of iron interleaved with tiles of plastic scintillators (Eljen EJ-200). The slab and tile cross section is $\sim 3\times3$ cm$^2$ and the thickness is 1.5 and 0.7 cm, respectively. Each LCM thus samples electromagnetic and hadronic showers every 4.3 $X_0$. Three radial layers of LCMs cover the walls of the tunnel at R=100, 103, and 106 cm as in Fig. \ref{fig:schematics_instrumented_tunnel}.  
Each LCM covers an azimuthal angle of 31 mrad and we have 200 LCM per layer at fixed $z$, where $z$ is the position along the tunnel axis. The dimension of the LCM is a compromise between the need for high-granularity modules for pile-up reduction and particle identification and the total cost of the tunnel instrumentation ($<10$\% of the cost of the facility). In the ENUBET Demonstrator, the light produced by charged particles in the tiles is trapped inside the tiles by a diffusive coating (Eljen EJ-510) deposited on the tile surfaces. The only areas that are not covered by the diffuser are the grooves positioned in the back of the tile. Two WLS optical fibers (Y11, Kuraray) are glued to the grooves using an Eljen optical glue (EJ-500) with a refraction index similar to the plastic scintillator. As a consequence, part of the light impinges on the WLS fibers and is re-emitted at $\lambda \sim 440$ inside the fiber, transported outside the calorimeter, and recorded by a $4\times 4$ mm$^2$ Silicon Photomultiplier (SiPM). 
All (ten) fibers belonging to the same LCM are grouped and optically connected to the same SiPM after crossing a
30 cm borated polyethylene shielding (see Figs \ref{fig:schematics_instrumented_tunnel} and \ref{fig:demonstrator}) that protects the photosensors from neutron irradiation in the tunnel (see Sec. \ref{sec:doses}).  Each calorimeter module corresponds to one electronic channel. The SiPM output signal is connected to a digitizer that samples the waveform and sends them to the DAQ. The photon veto is made of a doublet of tiles identical to the tiles of the calorimeter. Each tile is read out by two WLS fibers and one $1\times 1$ mm$^2$  SiPM. Since each tile of the doublet is read out separately, the $t_0$ layer provides a detector to veto photons (no signal in any tile of the doublet) and converted photons (two mip-like signals in one or two tiles of the doublet) against charged particles (one mip-like signal per tile). The light produced at the tiles of the photon veto is transported by Y11 Kuraray WLS fibers running through the first two LCM tiles of each layer by means of additional grooves. Since each doublet is positioned just below the first tile of each LCM of the innermost layer (See. Fig. \ref{fig:schematics_instrumented_tunnel}), the photon veto also provides the absolute time when a charged particle (a candidate positron or muon) impinges on the tunnel wall. Testbeam data collected at CERN in 2018 and 2022 show a time resolution of 400~ps when the waveform is sampled by a 1~GS/s digitizer \cite{Acerbi:2020nwd}. 

The simulation of the detectors has been performed using GEANT4. The simulation software includes the material description (scintillator, passive materials, WLS bars, grooves, and optical coatings), and it records the energy deposition from the particles impinging in the calorimeter. Particles are tracked from the tunnel's entrance using the output of the GEANT4 simulation of the ENUBET beamline described above. Tracking includes decay and interactions inside the tunnel and the creation of electromagnetic and hadronic showers originating from positrons, electrons, photons, and mesons. The energy deposition (MeV/LCM and MeV/tile in the photon veto) is converted into the number of photoelectrons (p.e.) at the SiPM using the test beam data collected at CERN and employed to account for p.e. statistical fluctuations (mean value: $\sim 15$ p.e./MeV in an LCM \cite{Acerbi:2020nwd}). The simulation of the front-end electronics (SiPM response, waveform generation, and peak reconstruction) is based on the GossIP software package \cite{Eckert:2012yr} and is described in Sec. \ref{sec:eventbuilder}. The performance of the Demonstrator was measured in the T9 beamline at the CERN-PS East Experimental area and is detailed in \cite{demonstrator}. These results confirm the GEANT4 detector simulation employed in this work. Thanks to the cost-effectiveness of the detector and the small size of the ENUBET tunnel, we consider here a beamline where the decay tunnel is fully instrumented by $2.2 \times 10^{5}$ LCMs. Full coverage is not mandatory, indeed, because monitoring can be performed by sampling a fraction of the tunnel length and the minimum number of LCMs needed to reach the goal systematic budget is under evaluation \cite{systematics}. 

\section{Doses at the instrumentation and at the quadrupoles}
\label{sec:doses}

The evaluation of ionizing and non-ionizing doses in the beamline is a prominent task for the design of ENUBET. Ionizing doses are key to evaluating the damage to the beam components and, in particular, the first quadrupole located in front of the target. Non-ionizing doses in the decay tunnel must be estimated to assess the potential damage to the photosensors over the entire duration of the ENUBET run.  

Doses  in ENUBET were evaluated using FLUKA and the beamline and detector geometry of Secs. \ref{sec:transfer_line} and \ref{sec:instrumentation} were ported to to this simulation framework, too. Since the FLUKA geometry description is semi-automatically generated within GEANT4 in the ENUBET software package, the FLUKA model faithfully reproduces the GEANT4 and G4Beamline ones. Based on the secondary meson yields described in Sec. \ref{sec:target}, we traced all particles down to the hadron dump and computed doses in any critical component, accounting for shielding in the hadron and proton beamline, and detector shielding.

\begin{figure*}[htb!]
    \centering
    \includegraphics[width=15cm]{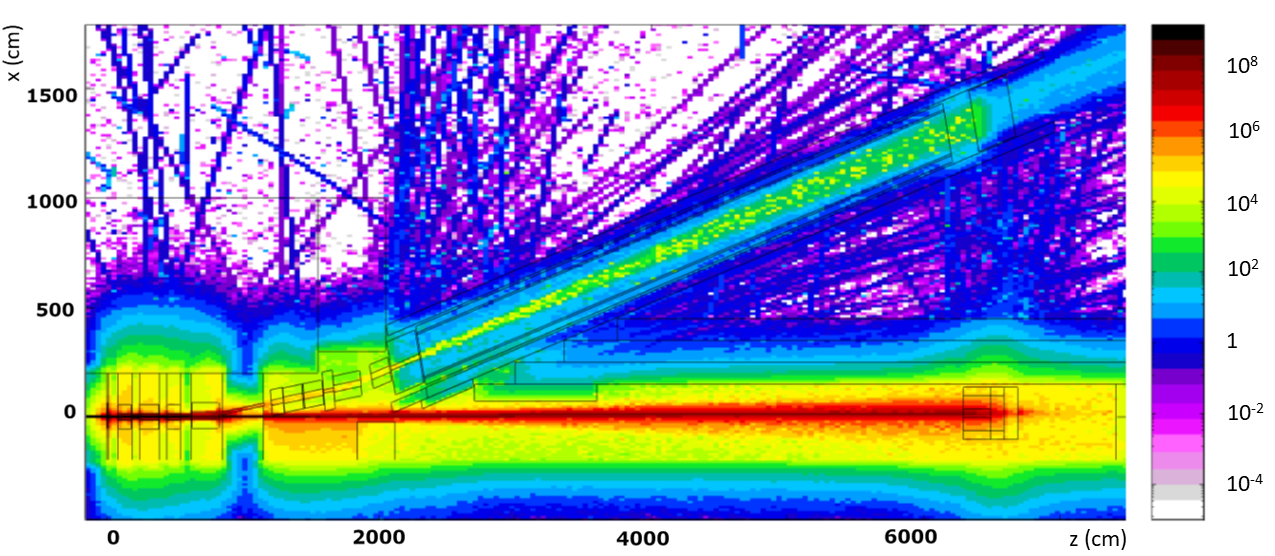}
 \includegraphics[width=15cm]{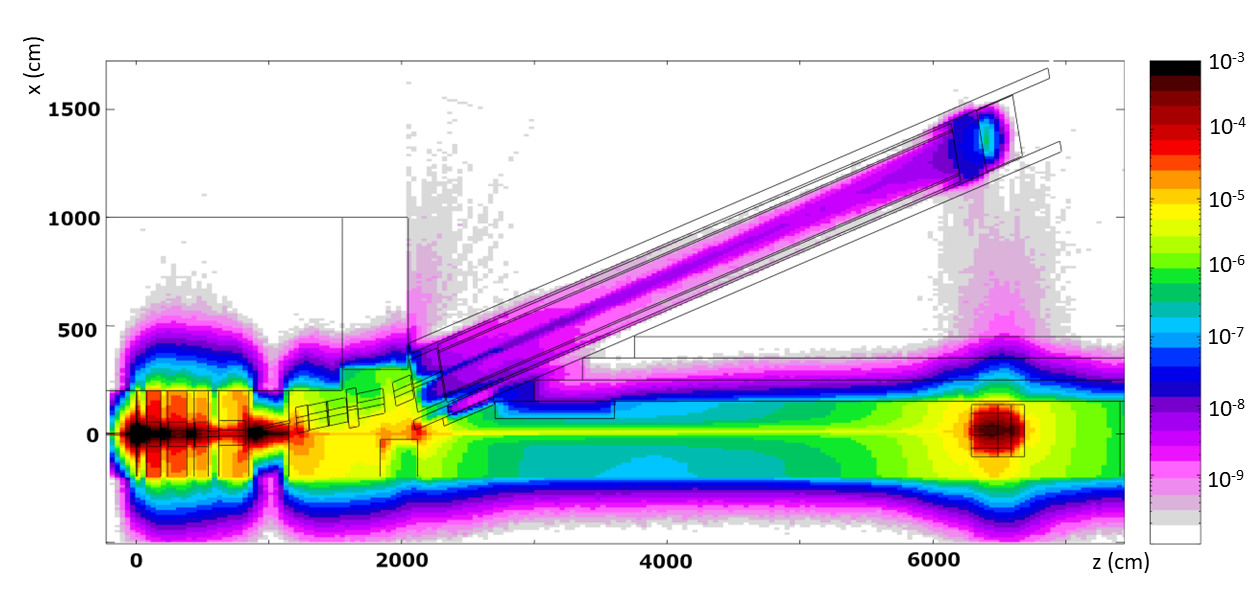}
    \caption{Top: dose map in Gy for 10$^{20}$~pot. The first quadrupole in the map is located between $z\simeq 200$ and 500~cm. Bottom: 1-MeV-eq neutron fluences.}
 \label{fig:mapdose}
\end{figure*}

The map of the accumulated ionizing dose in Gy obtained with FLUKA for 10$^{20}$~pot is shown in Fig.~\ref{fig:mapdose} (top plot). The dose at the hottest point of the quadrupole closest to the target is 
$<300$ kGy for 10$^{20}$~pot. The maps in the proximity of the target prove that conventional magnets can be operated without risk in a monitored neutrino beam like ENUBET for the entire duration of the data taking.
Neutron fluences (non-ionizing doses) are shown in Fig.~\ref{fig:mapdose} (bottom plot) in units of neutrons/cm$^2$/primary proton.

\begin{figure*}
%\begin{figure}[htb!]
\centering
  \includegraphics[width=0.5\linewidth]{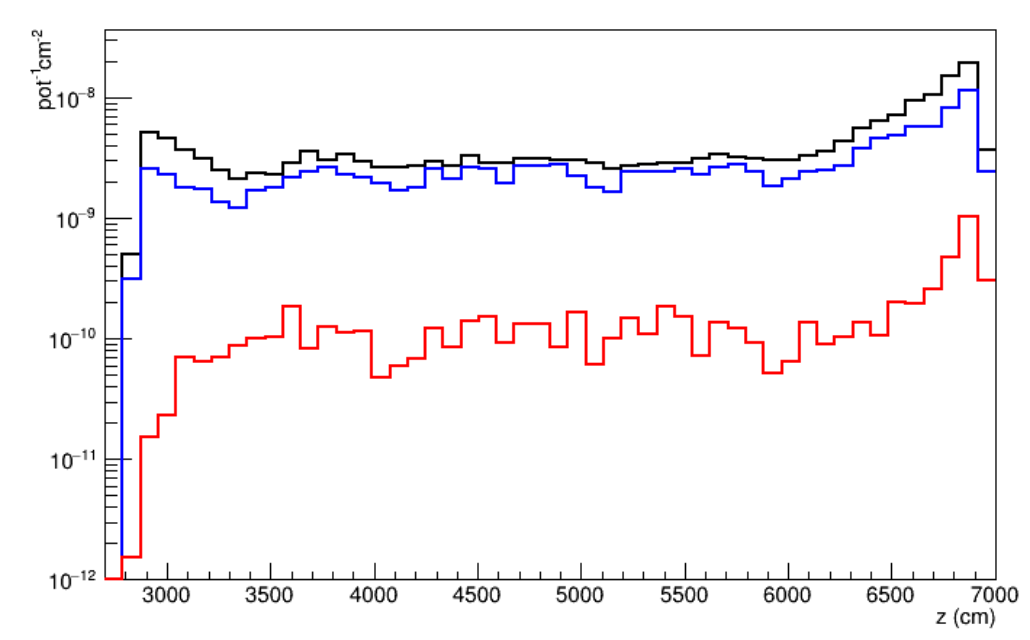}%
 \includegraphics[width=0.5\linewidth]{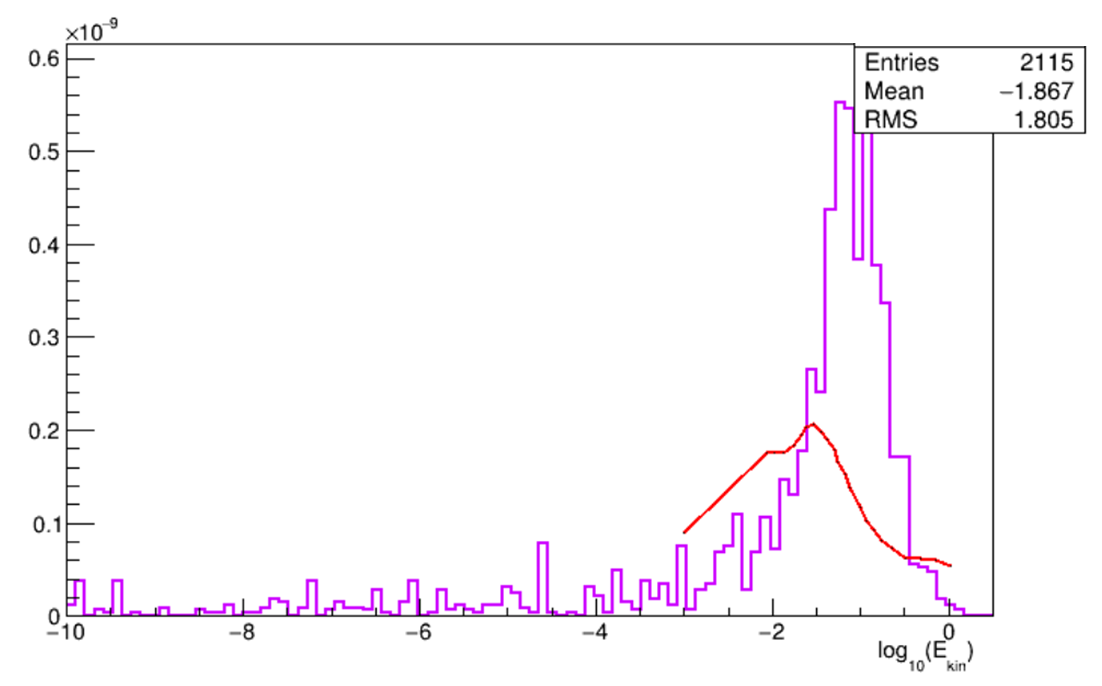}
  \caption[]{Left: FLUKA estimate of neutrons/pot/cm$^2$ as a function of the longitudinal coordinate along the tagger. The black line represents the fluence at the inner surface of the calorimeter, the blue one is computed at the surface on top of the three LCM layers, the red one is at the outer surface of the  30~cm of BPE i.e. the region where the SiPMs operate. Right: the kinetic energy spectrum of neutrons (GeV) reaching the SiPMS (magenta) and the neutron damage function for Silicon  (red) \cite{Leroy:2011goz}.}
  \label{fig:tlr5_fluka_neutronsT}
\end{figure*}

Even if non-ionizing doses are not critical for the ENUBET beamline, special attention must be paid to neutrons reaching the outer part of the tunnel instrumentation where the SiPMs are located.
As mentioned in Sec. \ref{sec:instrumentation}, the SiPMs are protected by a shielding layer  
of Borated polyethylene (BPE, 5\% Boron concentration) with a thickness of 30~cm.
%as shown in Fig.~\ref{fig:neutronS}, left.
In Fig.~\ref{fig:tlr5_fluka_neutronsT} we show the distribution of the neutron fluence (neutrons/pot/cm$^2$) as a function of the longitudinal coordinate along the tunnel ($z$), at the inner surface of the tagger (black), at the surface between the iron and the BPE (blue) and at the outer surface of
the BPE (red). The neutron reduction induced by adding the BPE layer amounts to 
a factor of $\sim$~18, averaging over the expected energy spectrum and it settles at about $7\times 10^{-11}$~n/pot/cm$^2$ in the middle region of the tagger
($7\times $~10$^9$~n/cm$^2$ for 10$^{20}$~pot). The bulk of neutrons reaching the SiPMs have kinetic energies of $\mathcal{O}$(10-100)~MeV/c$^2$ (Fig.~\ref{fig:tlr5_fluka_neutronsT}, right). Modern SiPMs developed for collider physics can stand $>10^{12}$ neutrons/cm$^2$ \cite{Musienko:2017znn,Garutti:2018hfu} and these sensors can be employed without risk in the ENUBET instrumented decay tunnel. Commercial SiPMs with a radiation hardness similar to the photosensors employed in the ENUBET Demonstrator were irradiated up to $10^{11}$ neutrons/cm$^2$ in a dedicated campaign performed in 2018 \cite{Acerbi:2019wti}. In ENUBET, these SiPMs retain sensitivity to minimum-ionizing particles for neutron fluences that are $>3$ times larger than the expected fluence in ENUBET.

\section{Event simulation and reconstruction in the tagger}
\label{sec:eventbuilder}
The ENUBET tagger uses G4TAG, a GEANT4 package that simulates the detectors in the decay tunnel (Sec.~\ref{sec:instrumentation}).  
The simulation manages the propagation and decay up to the hadron dump of particles provided by the GEANT4 simulation of the transfer line with the static focusing system (Sec.~\ref{sec:transfer_line}) and assumes a 2~s slow extraction with 4.5$\times$10$^{13}$~pot per spill. 
The detector response is simulated at hit level, with corrections from test beam data \cite{Acerbi:2020nwd}.  
A software framework based on GosSiP \cite{Eckert:2012yr} simulates the sensor response and the pile-up effects by generating waveforms from hits on each channel.
A full waveform reconstruction is being developed by the ENUBET Collaboration to simulate and reconstruct raw data. The results in the following use a simpler algorithm that handles pile-up effects 
by adding the energy and averaging the time of neighboring hits with a time difference below 1~ns.

The first step for the identification of leptons in the decay tunnel is the event building; we correlate in space and time hits in different LCMs and t$_0$-layer tiles to select those belonging to the same particle. Cuts at the event-builder level are optimized to suppress beam halo particles and other me\-sons’ decay modes. Reconstructed events are then processed with a multivariate analysis based on the Multilayer perceptron Neural Network (NN) provided by the TMVA toolkit \cite{Hocker:2007ht} in order to disentangle the signal from the background.
The signal sample for the NN training consists of leptons from kaon decays, while the background sample includes all particles produced by the full GEANT4 simulation of the ENUBET beamline except for signal events. Both samples are produced by G4TAG from a large subsample of particles at the tunnel entrance reconstructed with the same event-building algorithm, resulting in $\sim$$10^4$ events for each sample.
Two different reconstruction and analysis chains, with specific cuts, NN training samples, and discriminating variables, are used for positrons and muons from kaon decays, respectively. They are explained in the next two sections.

\subsection{Particle selection and background: positrons}
\label{sec:positrons}

\begin{figure*}[h!]
    \centering  
    \includegraphics[width=2\columnwidth]{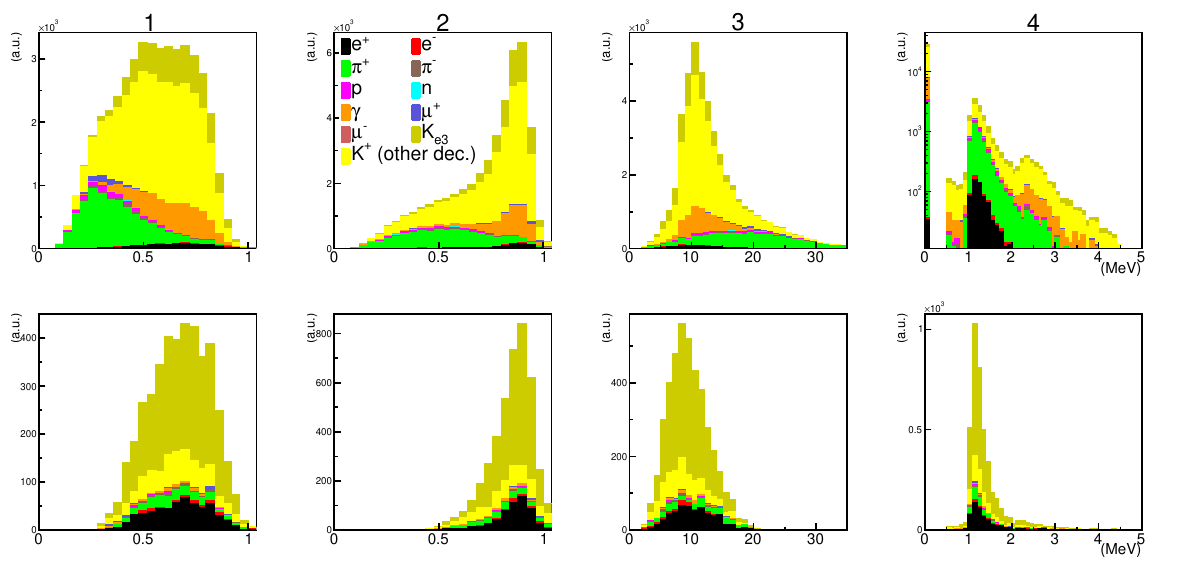}
    \caption{The four most representative variables fed to the input layer of the Neural Network for positron identification before (top row) and after (bottom row) the cut on NN classifier. See the text for the definition of variables. Variable 4 (energy deposited in the most upstream t$_{0}$-layer) is reported in log-scale on the top row to highlight the main structures of the distribution, including the two mip-like signal at $\sim$2.5~MeV due to photon conversions in the scintillator.}
    \label{fig:NNbest}
\end{figure*}

The event building algorithm for $K_{e3}$ is aimed at preselecting positron candidates through the identification of a visible energy deposit in the LCMs of the innermost layer exceeding 28 MeV\footnote{The Landau fit of the energy released by a mip in an LCM has a most probable value of $\sim$6.5~MeV.} as a ``seed'' for the event reconstruction. LCM and t$_0$-layer signals correlated to the seed are then clustered taking into account their position and timing.
All the LCMs in $\pm$5 azimuthal sectors with respect to the seed one  and in longitudinal planes in the interval [-3,10] around the
seed plane are taken into account for the event building. The nine upstream t$_0$-layer tiles in the same $\phi$ sector as the seed are also sought for compatible hits. After correcting the time of each hit for its distance from the seed, the ones in a time window of $\pm$1~ns with respect to the seed are considered. A reconstructed event is selected for further analysis if at least 10 hits are clustered. This procedure is iterated over all
recorded signals, discarding already clustered ones. On average $\sim$95\% of the clustered hits belongs to the same event as the seed.

The request of at least one calorimeter module (the event seed) with an energy deposit largely exceeding the typical mip signal determines a relevant reduction of the halo muons and non-interacting hadrons background that can be further suppressed by exploiting their single-track topology. The photon background originated from interactions of stray particles with the elements of the beamline and from $\pi^0$ produced in kaon decays, can be effectively suppressed by means of the t$_0$-layer. The rejection of the hadronic background, due to off-momentum particles and to pions generated in most of the kaon decay modes, benefits from the longitudinal, transverse, and radial segmentation of the calorimeter since electromagnetic showers develop inside a few LCMs, while the energy deposit pattern of a hadronic shower encompasses tens of LCMs.
We defined a set of 19 discriminating variables in order to exploit differences in the energy deposition pattern in the tagger. These quantities feed the input layer of the NN. This layer also includes the event position in the transverse and longitudinal direction since positrons from kaon decays impact mostly the downstream part of the tunnel with a uniform distribution in $\phi$, while off-momentum particles lay mostly on the bending plane intercepting the initial part of the tunnel. The NN implemented in this way allows monitoring $K_{e3}$ positrons with an efficiency of 20\% implying a signal-to-noise (S/N) ratio exceeding 2. 

\begin{figure}[h]
    \centering
    \includegraphics[width=1\columnwidth]{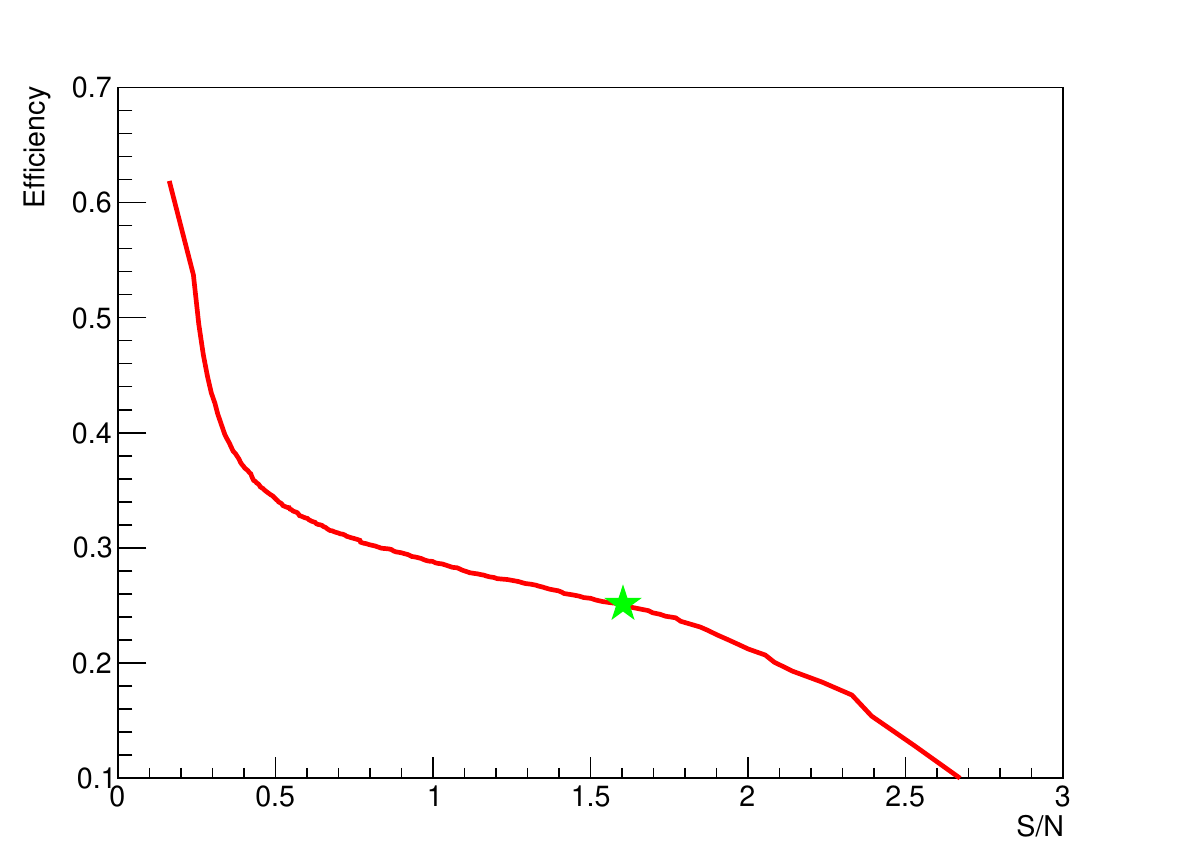}
    \caption{
%    Performance (efficiency versus signal-to-noise ratio) of the event classifier for $K_{e3}$ positrons selection.
    Signal efficiency versus signal-to-noise ratio for 
    the $K_{e3}$ event selection. The green marker corresponds 
    to the working point for signal selection, that is the point in the curve maximizing the product between signal efficiency and purity.
   }
    \label{fig:effSNnoerr}
\end{figure}

Fig.~\ref{fig:NNbest} shows four of the NN variables: from left to right, the fraction of energy in the LCM of the first calorimeter layer with the largest energy deposition, and in the downstream one (1), the fraction of energy of the LCMs in the same $\phi$ sector as the seed (2), the total number of LCM hits clustered in the event (3), the energy deposition in the most upstream t$_0$-layer tile (4).
The top row shows variables at the event building level, while on the bottom row the same variables are depicted after applying a cut on the NN classifier that maximizes the product of the efficiency in the selection of $K_{e3}$ events and its purity. The working point chosen this way is reported with the green marker over the ROC curve of the NN shown in Fig.~\ref{fig:effSNnoerr}. 
The NN at the working point gives a positron selection efficiency, including the geometrical one ($\sim$53\%), of 25.1\%, and a S/N=1.6.

\begin{figure}[h]
    \centering
    \includegraphics[width=1\columnwidth]{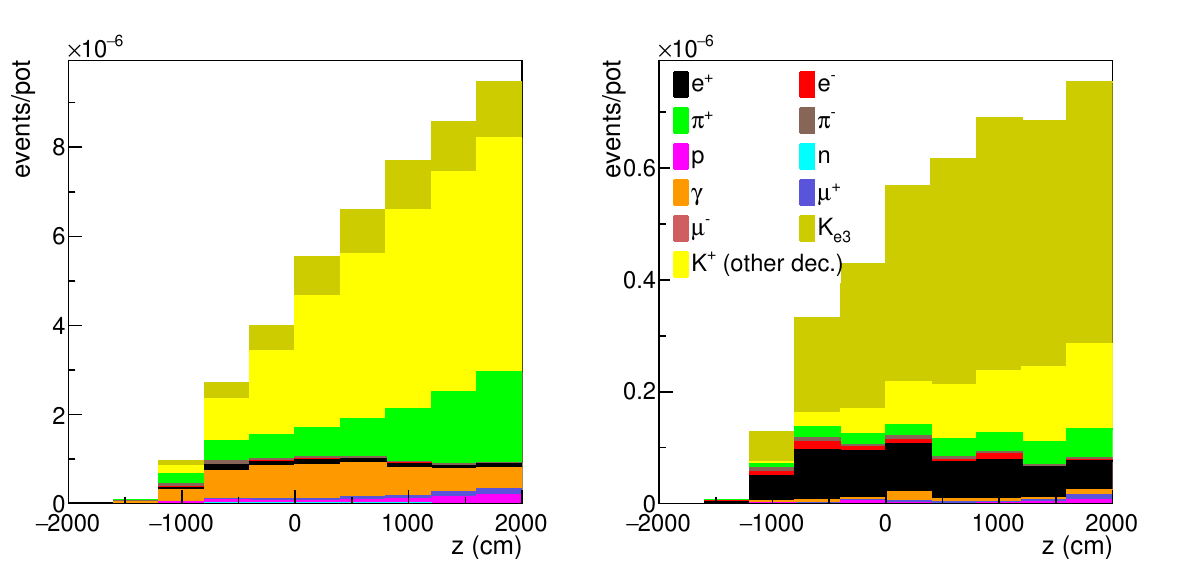}
    \caption{Longitudinal position of the reconstructed events, before (left) and after (right) the cut on the NN classifier.}
    \label{fig:zpos}
\end{figure}

\begin{figure}[h]
    \centering
    \includegraphics[width=1\columnwidth]{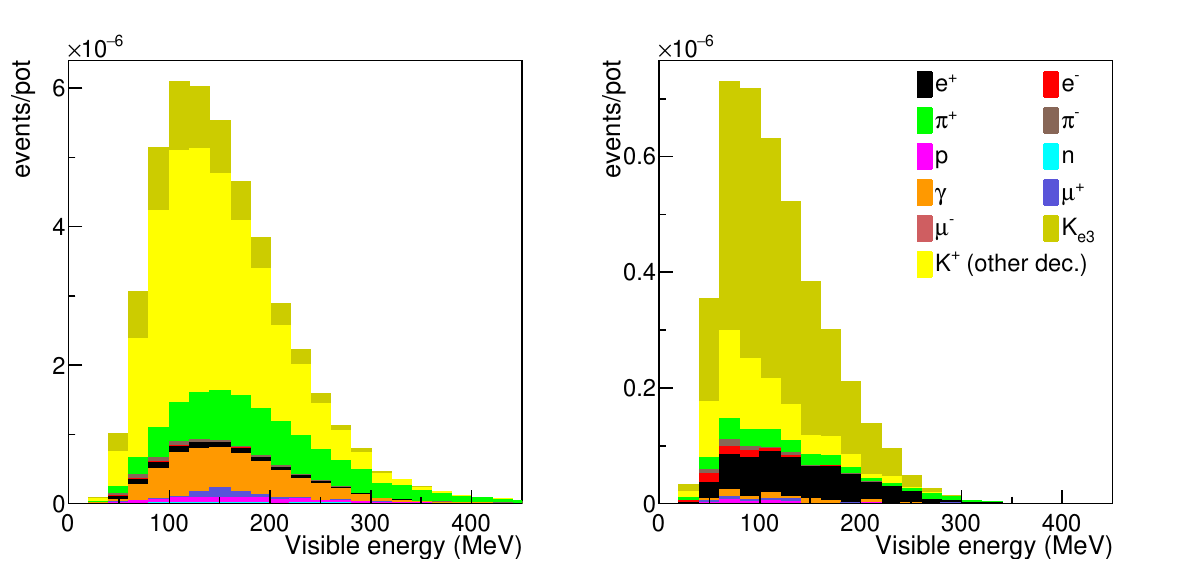}
    \caption{Visible energy of the reconstructed events, before (left) and after (right) the cut on the NN classifier.}
    \label{fig:visE}
\end{figure}

Fig.~\ref{fig:zpos} shows the longitudinal position $z$ along the decay tunnel of the particles before and after the NN cut. The total visible energy of the events is shown in Fig.~\ref{fig:visE}. These two observables play a special role. They are used as priors to reduce the systematic uncertainty in the $\nu_e$ flux. The fitting procedure described in Sec.~\ref{sec:neutrino_fluxes} combines standard information available in any neutrino beam (hadroproduction data, estimated flux from MC simulation, etc.) with the unique features of a monitored neutrino beam: the space and energy distribution of the positrons produced in the decay tunnel and their absolute rate. As discussed in Sec.~\ref{sec:neutrino_fluxes}, monitoring positrons with the afore-mentioned efficiency and S/N is enough to reduce the leading systematics uncertainties on the flux below 1\%. Further improvements are obtained by combining the muon sample described below.
As can be seen from the figures, the dominant background at the event building level, represented by hadronic decays of kaons (in yellow), non-collimated pions coming directly from the target or from early decays in the transfer line (in green), and photons from the beamline (in orange) are efficiently suppressed by the NN classifier, while halo positrons produced in the beamline and transported to the walls of the tunnel (black) are left as the main component of the background.

\subsection{Particle selection and background: muons}
\label{sec:muons}
As anticipated in Sec.~\ref{sec:intro}, the goal of ENUBET has been extended to monitor the $\nu_{\mu}$ component of the neutrino flux at the detector. The $\nu_{\mu}$ flux is given by two well-separated populations. A low-energy component originating from the two-body decays of $\pi^{+}$ and a high-energy component from $K^{+}$. The two components can be easily separated at the neutrino detector by a visible energy cut ($\sim 4$ GeV - see Fig.~\ref{fig:numuCC}) for charged-current (CC) events. 

As for the case of the $\nu_{e}$ flux, the goal of monitoring the $\nu_{\mu}$ flux is achieved by identifying the muons produced together with the neutrinos in the two-body decays of $K^{+}$ and $\pi^{+}$. Thanks to the larger kaon mass with respect to pions, the former are less boosted, and their decay is such that the produced muons are inside the geometrical acceptance of the instrumented decay tunnel (the average emission angle for muons from kaon is 60~mrad). Conversely, the highly boosted pions emit muons in the very forward region, outside the calorimeter geometrical acceptance. Monitoring of the low energy $\nu_{\mu}$ component can then be achieved by instrumenting the hadron dump, placed right after the decay tunnel, with muon stations to measure the range-out of muons from pions. An R\&D study for the development of such a system is being pursued in the framework of the ENUBET and PIMENT (Picosecond Micromegas for ENUBET) \cite{PIMENT} projects and will be the topic of a forthcoming publication.

\begin{figure*}[h]
    \centering
    \includegraphics[width=2\columnwidth]{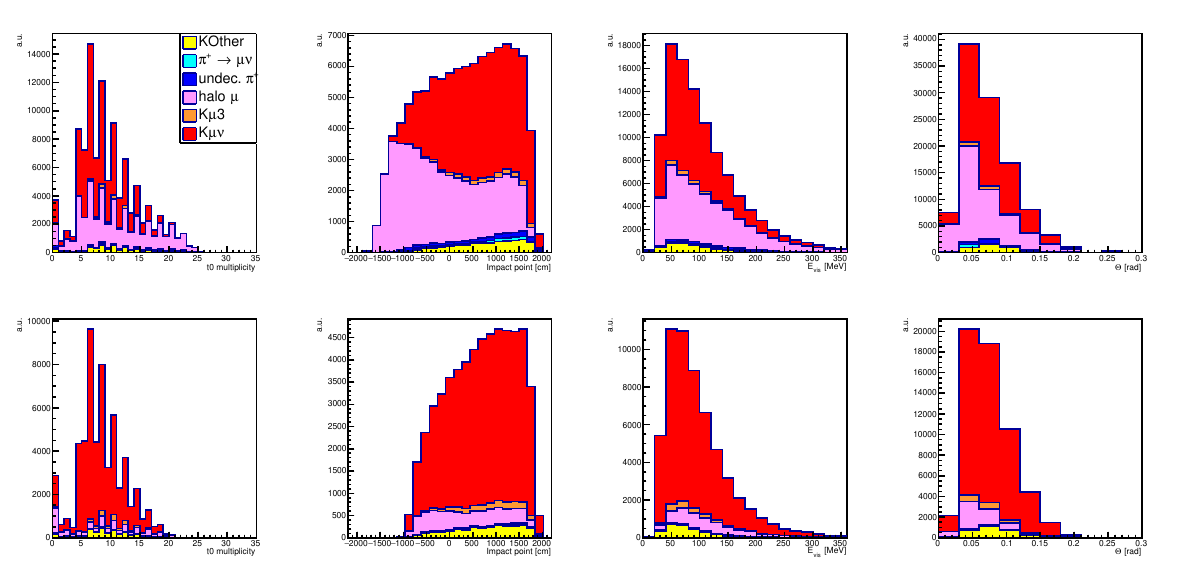}
    \caption{Example of 4 out of the 13 observable distributions identified for the training of the NN for muon PID analysis. The distributions are shown before (top row) and after (bottom row) performing the cut on the NN classifier output. The observables are computed from the clustered tracks of muon candidates, and from left to right are: the multiplicity of energy deposition in the $t_0$ layer; the impact point along the beam direction in the first layer of the calorimeter; the total visible energy deposition; the angle of the track with respect to the beam direction.}
    \label{fig:NNmu_vars}
\end{figure*}

Muons generated by kaon decays have an energy range of $\sim 0.3-9$~GeV, and interact as mips when crossing the calorimeter, releasing about 7 MeV in an LCM independently of the muon energy until they escape the calorimeter or range out. Their peculiar energy deposition pattern is key to muon identification in ENUBET and the NN classifier achieves even better efficiency and purity. 
%For this purpose, a pattern recognition algorithm has been developed. 
ENUBET is equipped with a dedicated event-building algorithm for muons.
The expected pattern for a muon crossing the calorimeter is a set of adjacent modules, each with an energy deposit in the scintillators of $\sim 6-8$~MeV, forming a straight line that starts from the muon impact point in the inner calorimeter wall and proceeds outward the three radial layers in the forward direction. A muon reconstruction starts as soon as a visible energy deposit compatible with a mip (energy between 5 and 15~MeV) is found in an LCM in the innermost layer of the calorimeter. This energy deposit is the seed for the hit clustering algorithm, where all other LCMs and photon-veto doublets whose energy deposits have a position in space compatible with that of a muon track are grouped together in a muon event candidate. The absolute time of the LCM energy deposits (precision: 1 ns) is exploited in the clusterization, and the final muon candidate is a collection of only those energy deposits compatible with the propagation of a muon both in space and time. 
Thanks to the unique topology of muon energy deposits compared with those produced by electromagnetic and hadronic showers (way more confined in space and involving different close-by LCMs), a good rejection of backgrounds is already obtained at the event building level. A muon candidate is required to have at least 7 LCMs clustered to the track along the longitudinal direction. This selection allows rejecting most of the electromagnetic and hadronic showers, reaching a signal efficiency of $\epsilon$=40.6$\%$ and a S/N$\sim$1.1 already at this stage.  
An important contribution to the background that cannot be reduced by the event builder is due to muons produced along the beamline (purple in Fig.~\ref{fig:NNmu_vars}), which do not originate from kaon decays in the tunnel. The primary sources of this background are halo muons from mesons decays along the transfer line, plus a subdominant contribution from decays of off-momentum pions in the tunnel, whose muons may cross the tunnel walls. Other minor background contributions are punch-through pions, i.e. non interacting pions produced by hadronic showers in the calorimeter ($\pi \rightarrow \mu$ misidentification -- cyan in Fig.~\ref{fig:NNmu_vars}) and muons from other  kaon decay modes than $K_{\mu2} \equiv K^+ \rightarrow \mu^+ \nu_\mu$ (e.g. muons from the decay-in-flight of pions in $K^+$ hadronic decays -- yellow in Fig.~\ref{fig:NNmu_vars}).

As for positrons, an enhancement of the muon PID performance is achieved by training a Neural Network for the selection of muons from $K^{+}$ decays. Therefore, a crucial task is the identification of observables showing a difference in the shape of their distribution between signal muons and backgrounds. 
Further suppression of $\pi \rightarrow \mu$ misidentified events is achieved by exploiting variables that account for the energy deposition pattern in the calorimeter. On average muons deposit less energy in the calorimeter with respect to pions, thus energy related variables have a good discrimination power for these events. Furthermore, the NN exploits isolation variables since the energy around the clustered track is low for candidate muons.  
Concerning the halo muons, the differences in their topology with respect to signal muons help to select a high-purity signal sample, while keeping a high selection efficiency. Given the kinematics of $K^{+}$ decays inside the tagger, signal muons impact the calorimeter mostly in the forward region and uniformly in the azimuthal coordinate $\phi$. Halo muons are particles that are not focused by the transfer line and lay outside the momentum bite. They mostly impact in the first half of the calorimeter region or have$\phi$=0,$\pi$. These $\phi$ values correspond to the bending plane, where halo muons are clustered.

\begin{figure}[h]
    \centering
    \includegraphics[width=1\columnwidth]{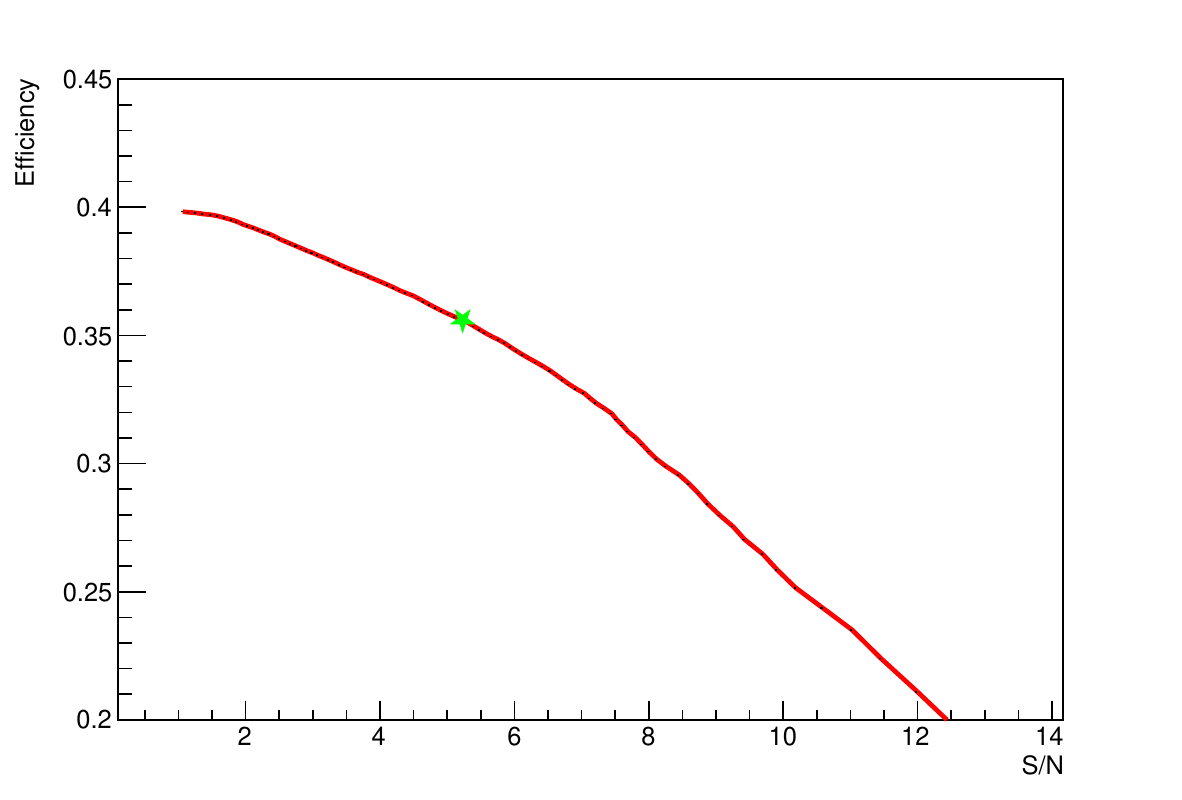}
    \caption{Signal efficiency versus signal-to-noise ratio for the 
    K$_{\mu2}$ and K$_{\mu3}$ event selection. The green marker corresponds 
    to the working point for signal selection, that is the point in the curve maximizing the product between signal efficiency and purity.
    }
    \label{fig:ROC_mu}
\end{figure}

A subset of the identified variables with good discrimination power is shown in 
Fig.~\ref{fig:NNmu_vars}-top. These variables are fed to the input layer of the NN and used to train the NN on clustered muon tracks from a pure signal sample, made out of muons from $K^{+}$ two body decay and $K_{\mu3}$, and a pure background sample, comprising all other particles impacting on the calorimeter. The trained network is then applied to the sample of candidate clustered muons, to select a high-purity signal sample by applying a cut on the network classifier. The working point cut is optimized to maximize both signal selection efficiency and purity. The ROC curve of the trained NN, reporting the full PID performance of the muon analysis, is shown in Fig.~\ref{fig:ROC_mu}. At the working point, the NN shows a signal selection efficiency of $\epsilon$=35.6$\%$, including the geometrical acceptance, and an S/N=5.2.  In the bottom row of Fig.~\ref{fig:NNmu_vars} we show the input variable distributions after performing the NN selection. After the NN classifier cut, the remaining backgrounds are halo muons with the same topology as $K_{\mu2}$ muons and other kaon decay modes. 

\section{Neutrino fluxes}
\label{sec:neutrino_fluxes}
\begin{figure*}[h]
    \centering
    \includegraphics[width=1.5\columnwidth]{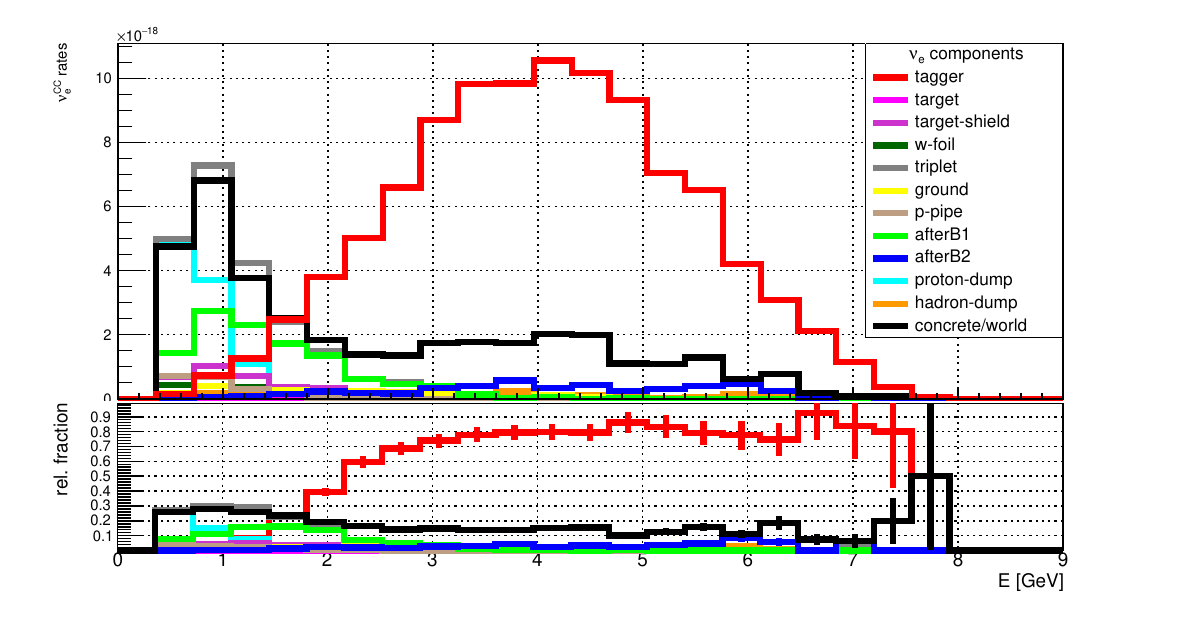}
    \caption{Top: energy spectra of the $\nu_{e}$ interacting in the detector via CC, normalized to 1 pot. We consider a 500 t detector with 6$\times$6~m$^2$ front face, orthogonal to the neutrino beam, placed 50~m from the decay tunnel end. Each spectrum corresponds to the $\nu_{e}^{CC}$ interactions for which the origin point of the neutrinos is located in a specific region of the ENUBET facility. The red spectrum represents $\nu_{e}^{CC}$ interactions where the neutrinos are produced within the decay tunnel volume (see text for further information). Bottom: fraction of each spectrum relative to the total $\nu_{e}^{CC}$ interactions.}
    \label{fig:nueCC}
\end{figure*}

\begin{figure*}[h]
    \centering
    \includegraphics[width=1.5\columnwidth]{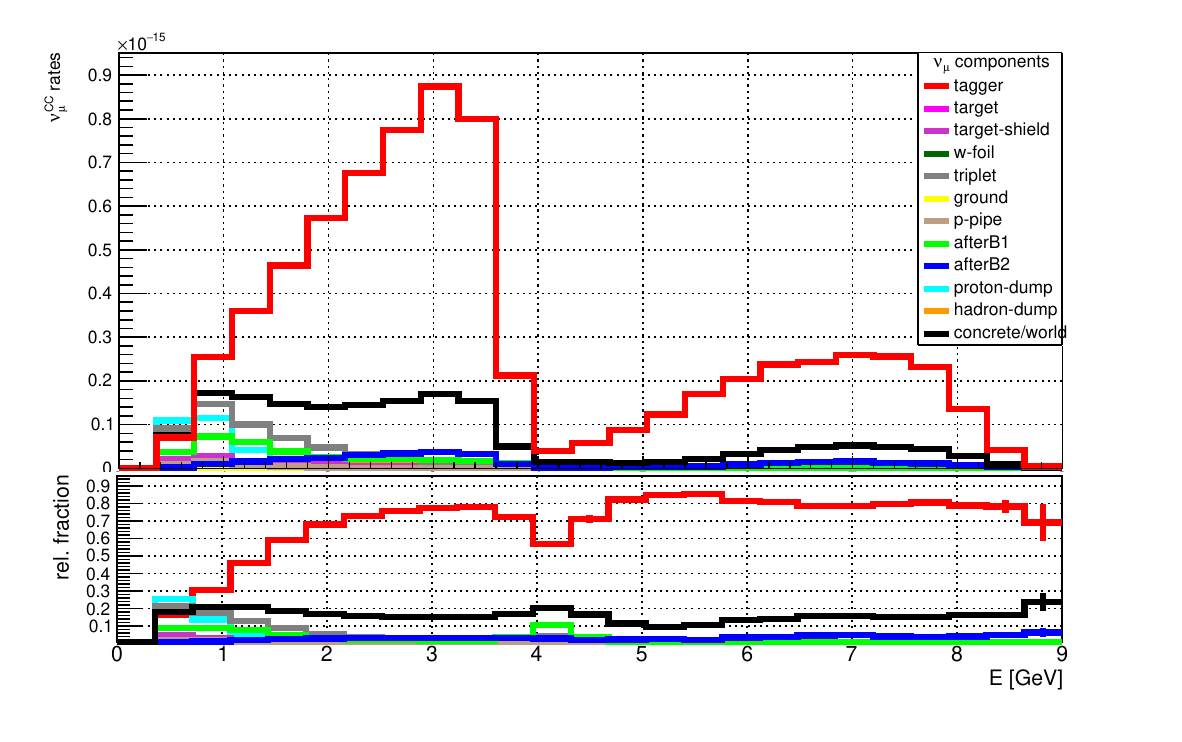}
    \caption{Top: energy spectra of the $\nu_{\mu}$ interacting in the detector via CC, normalized to 1 pot. The two populations are related to the neutrinos produced in the two body decay of pions, lower energy spectra (E$_{\nu_{\mu}}$ < 4~GeV), and to the two body decay of kaons, higher energy spectra (E$_{\nu_{\mu}}$ > 4~GeV). We consider a 500-ton detector with 6$\times$6~m$^2$ front face, orthogonal to the neutrino beam, placed 50~m from the decay tunnel end. Each spectrum corresponds to the $\nu_{\mu}^{CC}$ interactions for which the origin point of the neutrinos is located in a specific region of the ENUBET facility. The red spectrum represents $\nu_{\mu}^{CC}$ interactions where the neutrinos are produced within the decay tunnel volume (see text for further information). Bottom: fraction of each spectrum relative to the total $\nu_{\mu}^{CC}$ interactions.}
    \label{fig:numuCC}
\end{figure*}

The beamline developed for monitored neutrino beams and presented in this paper has been optimized to maximize the number of neutrinos that are produced within the decay tunnel. At the same time, the beamline design is such as to keep as low as possible the neutrinos with energy in the region of interest produced from decays outside the tunnel and reaching the detector. Considering a neutrino detector with a mass of 500~t, a front face of 6$\times$6~m$^2$ orthogonal to the neutrino beam, and placed 50~m from the tunnel end, the energy spectra of $\nu_{e}$ and $\nu_{\mu}$ interacting through charged-current (CC) within the detector are shown in Figs.~\ref{fig:nueCC} and \ref{fig:numuCC}, respectively. The total spectra are split into components defined by the point of origin of neutrinos within the facility. Focusing on the $\nu_{e}^{CC}$ in Fig.~\ref{fig:nueCC}, the neutrinos produced within the tagger volume (shown in red) can be directly monitored by measuring the corresponding positrons produced in the decays if they are in the acceptance of the tunnel instrumentation, and are well separated in energy from the lower energy neutrinos. The latter, are originated mainly from early decays happening in the first part of the beamline, before the first bending (section comprising the quadrupole triplet) and decays in the concrete shielding. Other contributions come from the proton-dump, where secondaries are produced by the dumping of the primary protons, and the region of the beamline between first and second bending. A simple energy cut at E$_{\nu}$ > 1.5~GeV, allows discarding the non-taggable low-energy neutrino component. After the cut, the selected $\nu_{e}^{CC}$ sample consists of 67.8\% neutrinos from kaon decays in the tunnel volume. This is the component that can be directly monitored. A subsample of 3.8\% neutrinos is still produced within the tunnel region but from muon decays in flight. The remaining 28.4\% neutrino sample cannot be monitored, given that they are produced outside the instrumented tunnel. Nevertheless, their contribution can be accounted for by the simulation. The dominant contribution to this sample comes by decays in the concrete shielding and by focused particles that keep decaying in the space between the tunnel end and the hadron-dump, 15\%; by decays in the straight part of the beamline after the second bending dipole, facing the tunnel entrance, and in the hadron-dump, 4.3\%; and by decays in the region of the beamline between the two bending dipoles, 3.2\%.  
Assuming 4.5$\times$10$^{19}$~pot/year at the SPS, a total of 10$^4$ $\nu_{e}^{CC}$ interactions in the detector, due to neutrinos produced by kaon decays within the tunnel volume, can be reached in 2.3 years.

Thanks to the narrow band beam, the $\nu_{\mu}^{CC}$ spectra reported in Fig.~\ref{fig:numuCC} are characterized by two populations well separated in energy, due to the two body decays of pions, at lower energies (E$_{\nu}$ < 4~GeV), and kaons, at higher energies (E$_{\nu}$ > 4~GeV). Unlike the case of $\nu_{e}^{CC}$, among the two components of $\nu_{\mu}^{CC}$ from neutrinos produced within the tagger volume (shown in red), only the one produced by kaon decays can be directly monitored by measuring the corresponding muons crossing the calorimeter, as discussed in Sec.~\ref{sec:muons}. 
The sample of $\nu_{\mu}^{CC}$ with E$_{\nu}$ > 4~GeV consists of 79.7\% neutrinos produced by kaon decays in the tunnel volume. The remaining 20.3\% is produced outside the decay tunnel and thus cannot be monitored, but can be accounted for by the simulation. Similarly to the $\nu_{e}^{CC}$, the larger contribution to this sample comes by decays in the concrete shielding and by focused particles that keep decaying in the space between the tunnel end and the hadron-dump, 14.5\%, and by decays in the straight part of the beamline after the second bending dipole, facing the tunnel entrance, and in the hadron-dump, 4.9\%. Assuming again the SPS, a total statistic of 10$^5$ $\nu_{\mu}^{CC}$ interactions in the detector, due to neutrinos produced by kaon decays within the tunnel volume, can be reached in 1.1 years.

As pointed out in Sec.~\ref{sec:monitored}, the neutrino flux at the detector would be limited in precision to 5-10\% in a conventional neutrino beam. Uncertainties on the hadroproduction, that is the production of mesons due to the interactions of primary protons in the target material, dominate this systematic value on the neutrino flux. By counting the leptons coming from mesons decays, the ENUBET monitoring technique allows overcoming the hadroproduction uncertainties, thus considerably enhancing the neutrino flux precision. The strategy that we developed to assess the impact of the monitoring technique on the neutrino flux precision is based on the following workflow. A realistic hadroproduction model \cite{Bonesini2001} is fitted to data from NA56/SPY \cite{Ambrosini:1999id} and NA20 \cite{Atherton:1980vj} experiments that used 450 and 400 GeV/$c$ protons on target, respectively. Thanks to the GEANT4 simulation of the ENUBET facility, mesons outgoing the target, leptons measured in the instrumented tagger, and neutrinos crossing the detector are directly linked, through their evolution histories in the simulation. Therefore, the effect of the hadroproduction uncertainties can be propagated to the lepton observables measured in the calorimeter and to the neutrino flux. The uncertainty propagation is performed by reweighting the Monte Carlo events using the hadroproduction model, where the model parameters are varied taking into account the covariance matrix from the fit to data. For each extraction of the hadroproduction parameters, a new set of MC events is obtained, corresponding to a possible realization of the hadroproduction data within their errors. This method is known as multi-universe \cite{multiuniverse,MINERvA:2021ddh}. From the sets of MC events, the covariance matrices of the lepton observables are computed. Moreover, nominal distributions for lepton observables are built by reweighting the MC events using the hadroproduction model with nominal parameters. A signal plus background model probability density function (PDF) is then assembled by combining the nominal lepton observables and their variations computed from the observables covariance matrices. Pseudo-data are generated out of one of the MC sets, and fitted by building an extended maximum likelihood (EML) from the model PDF. The EML fit is validated through toy-MC experiments generated with the MC sets obtained by applying the multi-universe method. The RooFit package \cite{Verkerke:2003ir} from ROOT is exploited to build the model PDF, perform EML fits, and generate pseudo-data. Given the high correlation between the lepton observables and the produced neutrinos, the EML fit result allows us to set a strong constraint on the flux. The model PDF parametrizes the variation of the lepton observables induced by the hadroproduction and directly constrains the hadroproduction yields through the EML fit. The propagation of the residual uncertainties from the constrained hadroproduction model through the multi-universe method allows for computing the neutrino flux covariance matrix after the lepton monitoring is introduced (post-fit result). The result from the described workflow shows that the residual systematic on the neutrino flux due to hadroproduction is of $O(1\%)$, for both $\nu_e$ and $\nu_{\mu}$. The same workflow can be extended to determine the impact of the detector effect and beamline subdominant systematics on the neutrino flux. A detailed account of the described procedure, including the results of the assessment of all neutrino flux systematics, will be the topic of a forthcoming publication \cite{systematics}.

\section{Energy measurement with the off-axis narrow band beam
technique}
\label{sec:neutrino_energy}
\begin{figure}[h]
    \centering
    \includegraphics[width=1.0\columnwidth]{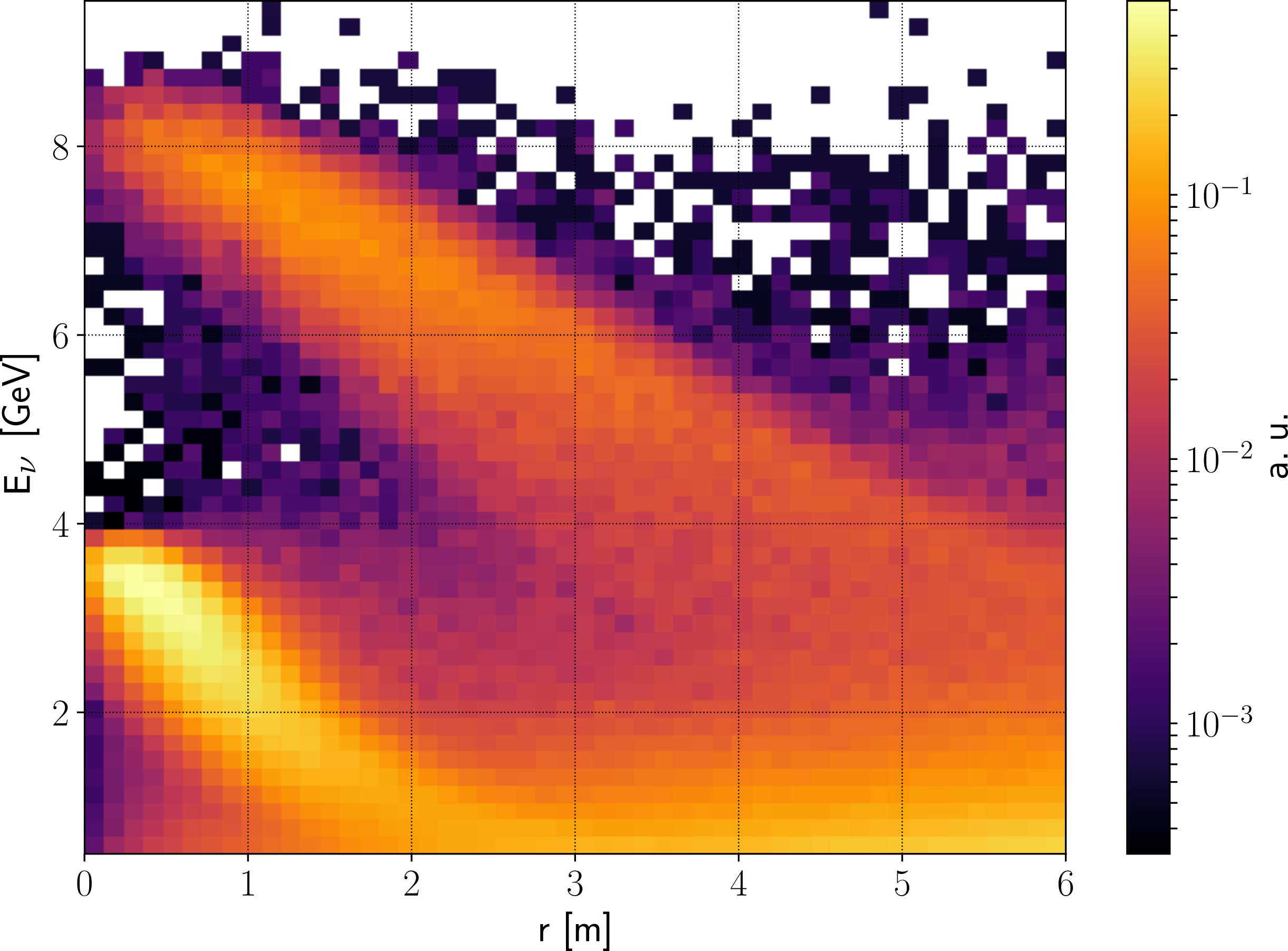}
    \caption{Distribution of muon neutrinos events at the neutrino detector in the energy versus impact radius plane. The kaon and pion neutrinos from the decay tunnel have a clear radius-energy correlation and can be well separated. The background neutrinos coming from the rest of the beamline and target station mostly impact the lower energies and higher radii.}
    \label{fig:nu2D}
\end{figure}

\begin{figure}[h]
    \centering
    \includegraphics[width=1.0\columnwidth]{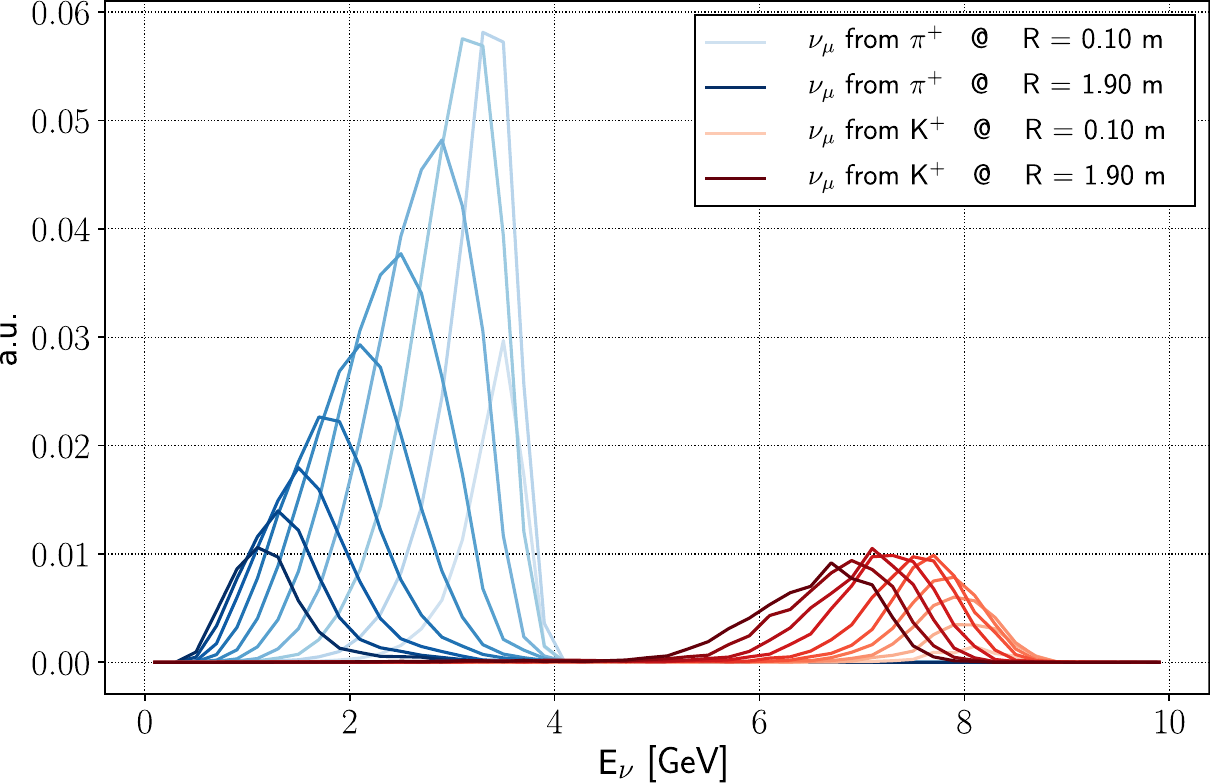}
    \caption{Energy distributions of $\nu_\mu$ events from the decay tunnel at different detector radii. Each distribution represents a $\pm 10$~cm radius interval around its central value.}
    \label{fig:nuHists}
\end{figure}

The narrow momentum width of the beam ($\mathcal{O}$(5-10\%)) can be exploited to provide the neutrino energy on an event-by-event basis, thanks to its correlation with the radial position of the interaction vertex in the neutrino detector. This would be valid for both muon neutrinos from pions and kaons produced by the ENUBET beamline, which can be fully separated in terms of the energy to impact radius dependency, as shown in Fig.~\ref{fig:nu2D}. We refer to this method as ``narrow band off-axis'' (NBOA).
The determination of the neutrino energy at the source with enough precision mitigates the uncertainties related to its measurement through the reconstruction of the final state, thus allowing to naturally produce differential cross sections as a function of neutrino energy without large model-dependent unfolding uncertainties. The kaon and pion neutrino components coming from the decay tunnel for different detector radii are shown in Fig.~\ref{fig:nuHists}, where it is possible to observe their separation in energy coming from the narrow band design of the beamline.
%The final radius-energy performance is summarized in Fig.~\ref{fig:nuPiKReal}.

\begin{figure}[h]
    \centering
    \includegraphics[width=1.0\columnwidth]{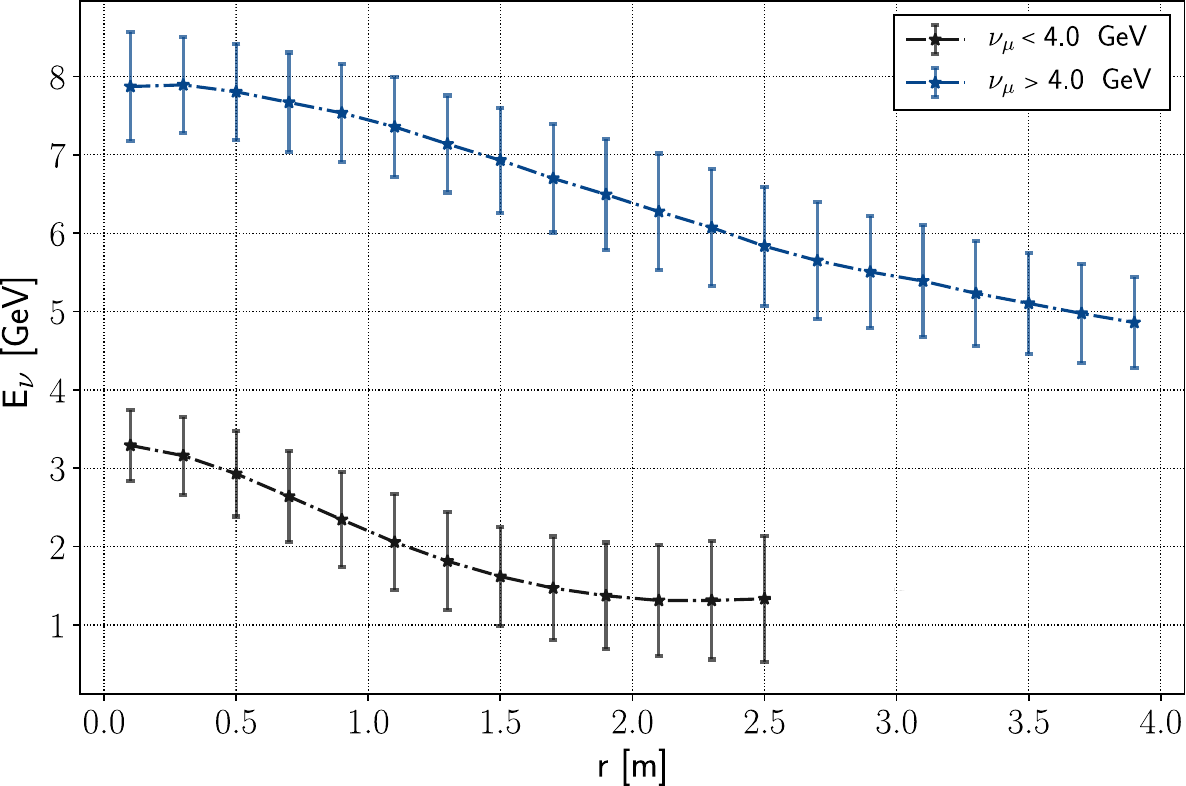}
    \caption{Energy to impact radius dependency for $\nu_\mu$ CC events for pion (black points laying at $E_\nu <4$ GeV) and kaon neutrinos (blue points laying at $E_\nu >4$ GeV). The events are integrated every $20$~cm along the detector impact radius dimension: the energy point represents the mean of the distribution, while the error bars its $\pm\sigma$ interval. For the pion neutrino case (low energy) the energy-radius dependency holds up to $2$~m radius, with fractional $1\sigma$ intervals that go from $\gtrsim 10\%$, to $25\%$ at $2$~GeV, and then worsening up to $45\%$ at higher radii and low energy. For the kaon case, it stays within $8$ and $15\%$ up to $4$~m.}
    \label{fig:nuPiKReal}
\end{figure}

The high energy kaon component of the spectra can be separated from the pion component and the low energy background with a cut on the reconstructed energy. The width of the pion peaks at different radii can then be used as an estimator of the precision on the incoming neutrino energy.
As shown in Fig.~\ref{fig:nuPiKReal}, the precision ranges from 10\% to 25\% in the DUNE energy domain (up to 2~GeV neutrino energy), for which the ENUBET beam is optimized. The worsening of the performance at higher radii and lower energy is due to the neutrino background coming from the beamline, target station, proton dump, and the divergence of the kaon beam. This contribution is heavily suppressed for the higher energy kaon neutrinos, where the performance is significantly better. For this sample, the energy resolution never exceeds $15\%$.

\section{Time tagged neutrinos}
\label{sec:time_tagged}

The combination of a slow proton extraction scheme together with a static transfer line not only provides a solid ground for the development of monitored neutrino beams but represents an opportunity to build for the first time a tagged neutrino beam (see Sec.~\ref{sec:monitored}). In tagged beams, each interacting $\nu_e$ at the detector is associated with the $K_{e3}$ decay that generated it through time coincidence between the interacting neutrino and the positron reconstructed at the tagger. The same method can be employed for $\nu_\mu$ events originating from pion or kaon decays.

\begin{figure}[h!]
    \includegraphics[width=1\columnwidth]{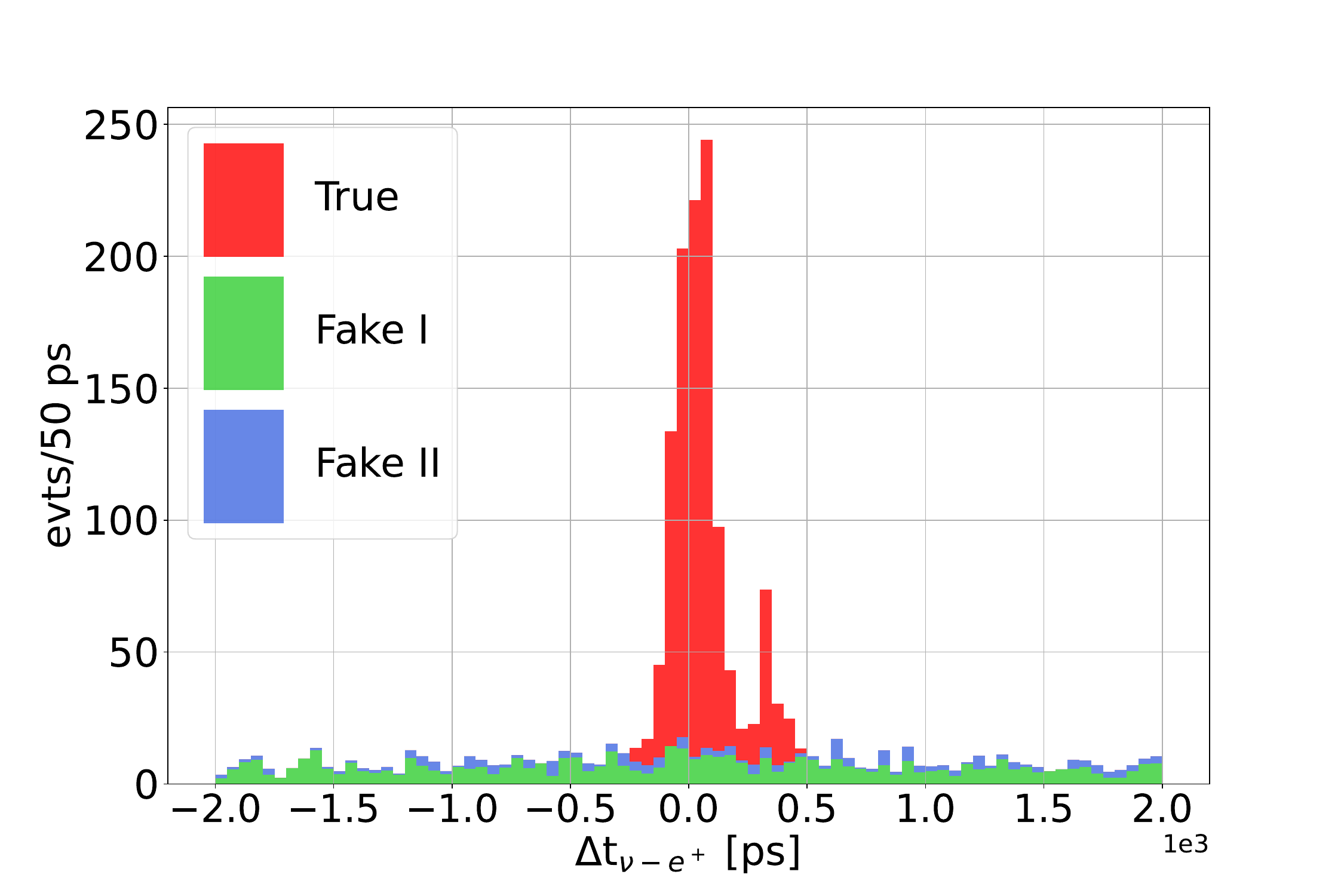}
    \caption{Distribution of time differences between all time tagged pairs of $\nu_e^{CC}$ and positron events within a $\pm$2~ns time window. True matches (red) are tagged pairs from the same $K_{e3}$ decays, whereas fake matches are tagged pairs between unrelated candidate positrons and $\nu_e^{CC}$ where the neutrino is produced inside (green) or outside (blue) the tagger volume. The spread of the peak from true matches is due only to the intrinsic resolution from kinematics. The subdominant peak around 300~ps is due to true matches from $K_{e3}$ happening outside the tagger volume (see text).}
    \label{fig:time_tagged_res1}
\end{figure}

\begin{figure}[h!]
    \includegraphics[width=1\columnwidth]{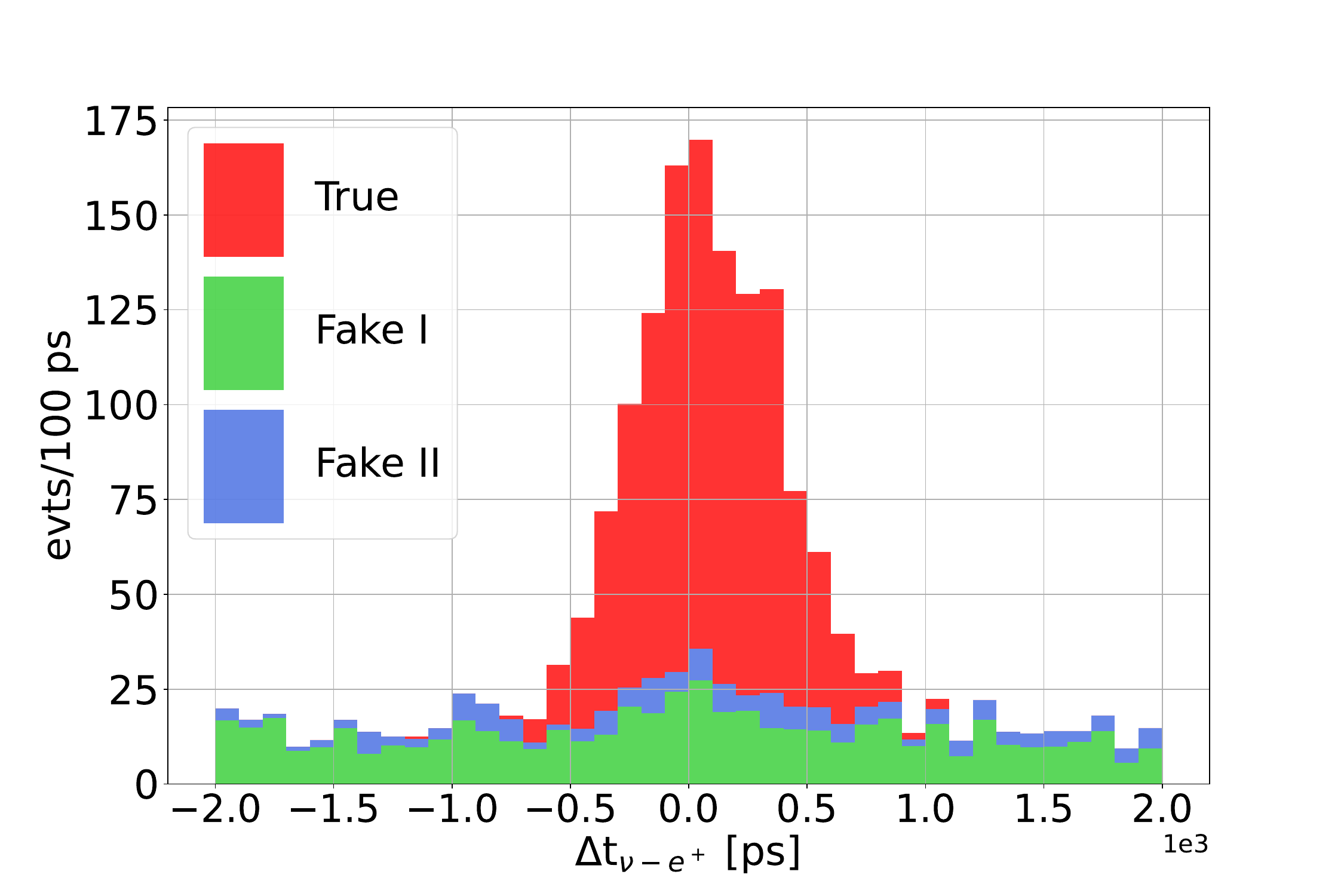}
    \caption{Distribution of time differences between all time tagged pairs of $\nu_e^{CC}$ and positron events within a $\pm$2~ns time window. True matches (red) are tagged pairs from the same $K_{e3}$ decays, whereas fake matches are tagged pairs between unrelated candidate positrons and $\nu_e^{CC}$ where the neutrino is produced inside (green) or outside (blue) the tagger volume. The spread of the peak from true matches is due to the intrinsic resolution from kinematics and the resolution of the detectors (see text).}
    \label{fig:time_tagged_res2}
\end{figure}

Tagged neutrino beams offer unprecedented opportunities to the physics of neutrinos because they uniquely identify the neutrino flavor at the source by the observation of the charged lepton. Further, the kinematic reconstruction of the particles in the decay tunnel constrains the neutrino energy well beyond the precision achieved in Sec.~\ref{sec:neutrino_energy}.  
For the ENUBET facility, Figs.~\ref{fig:time_tagged_res1} and \ref{fig:time_tagged_res2} show the distribution of the time difference between a $\nu_e^{CC}$ in the neutrino detector and the corresponding positron observed in the decay tunnel, for an extraction spill of 4.8~s.  Events are normalized to 1 year of data-taking (4.5$\times$10$^{19}$~pot). We assume perfect time resolution for the distribution of Fig.~\ref{fig:time_tagged_res1}, whereas in Fig.~\ref{fig:time_tagged_res2} we assume a time resolution of $\delta t$ = 200~ps for both the neutrino detector and the tunnel instrumentation. We consider a spatial resolution of 1~cm for the neutrino interaction vertex position, comparable with the resolution of ProtoDUNE-SP. Starting from the positron sample selected in Sec.~\ref{sec:positrons}, the time-tagging algorithm matches each positron to the closest in-time neutrino interaction. A cut on the neutrino energy of $E_{\nu}$>1.5 GeV is applied to discard the low-energy component from neutrinos produced outside the tagger, which would contribute to the fake match sample (the blue component in Figs.~\ref{fig:time_tagged_res1} and \ref{fig:time_tagged_res2}). Since the position of the neutrino production vertex is unknown, the time difference is computed from the time of the positron interaction in the tunnel wall (identified by the seed time) and the time of the neutrino interaction at the detector. The neutrino time is corrected by the time of flight, $\Delta/c$, between the positron impact point in the calorimeter (identified by the LCM corresponding to the seed deposit) and the neutrino interaction vertex.   

\begin{figure}[h!]
    \includegraphics[width=1\columnwidth]{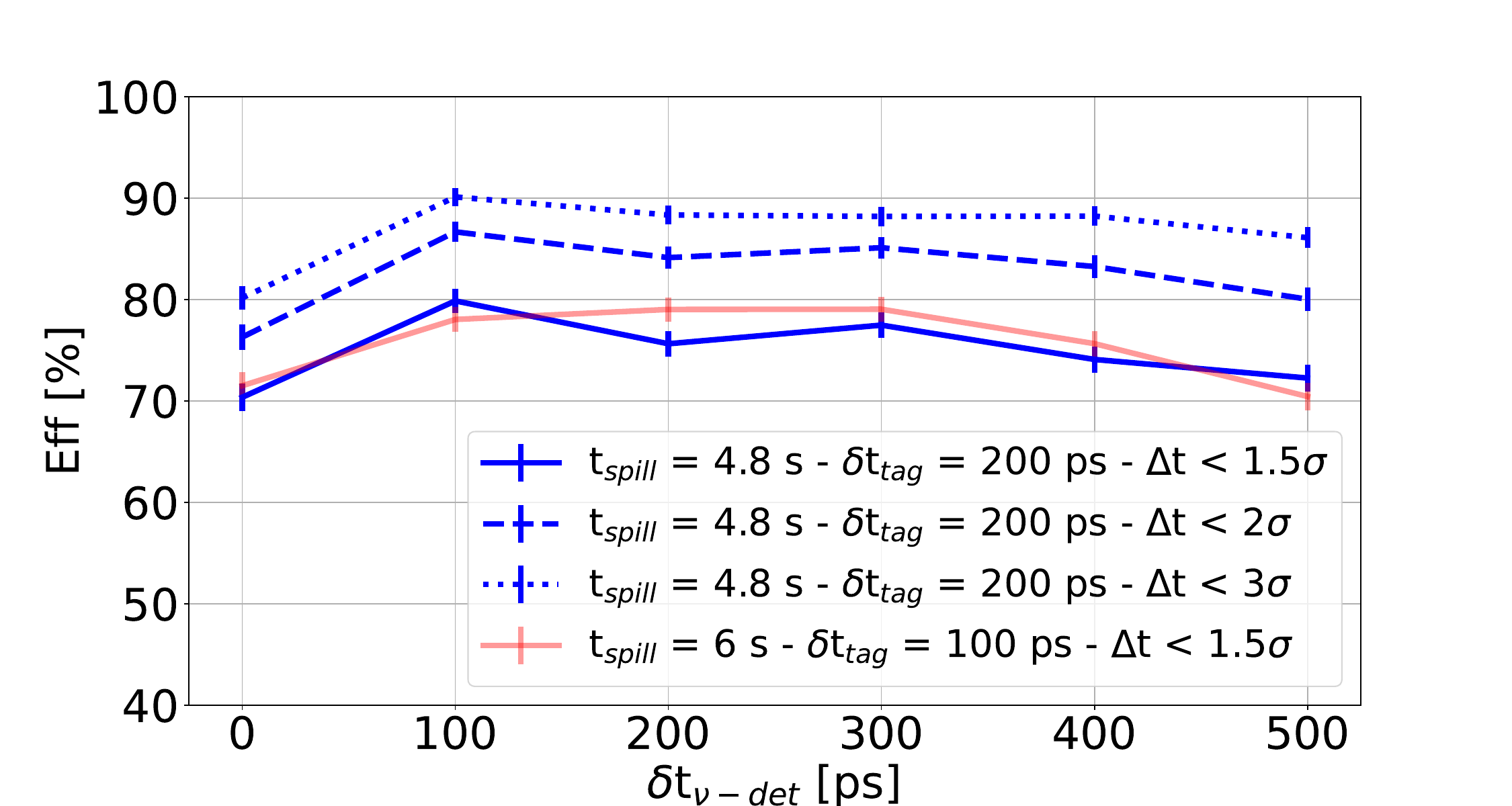}
    \caption{Efficiency of tagging a $\nu_e$-e$^+$ pair from the same $K_{e3}$ as a function of the neutrino detector resolution. An extraction spill of 4.8 s and a $t_0$ layer time resolution of $\delta t_{tag}$=200 ps (blue lines) are considered. The effect of a 1.5$\sigma$ (solid-line), 2$\sigma$ (dashed-line) and 3$\sigma$ (dotted-line) cut around the peak is also reported. As a comparison, the efficiency corresponding to an extraction spill of 6 s, a $\delta t_{tag}$=100 ps and a 1.5$\sigma$ cut around the peak is also shown (red line).}
    \label{fig:time_tagged_eff}
\end{figure}

\begin{figure}[h!]
    \includegraphics[width=1\columnwidth]{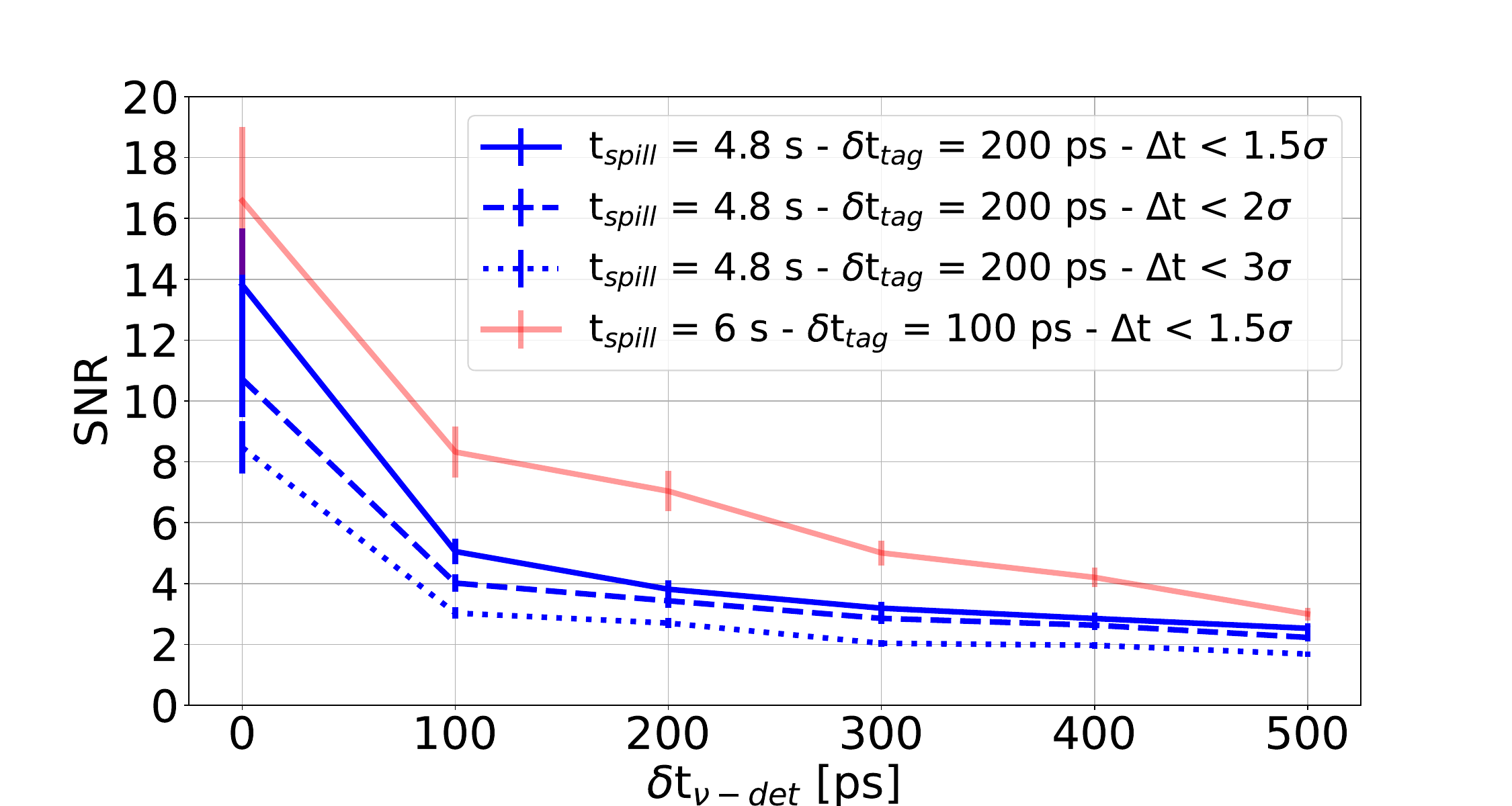}
    \caption{Signal to noise ratio of tagged $\nu_e$-e$^+$ pairs as a function of the neutrino detector resolution. An extraction spill of 4.8 s and a $t_0$ layer time resolution of $\delta t_{tag}$=200 ps (blue lines) are considered. The effect of a 1.5$\sigma$ (solid-line), 2$\sigma$ (dashed-line), and 3$\sigma$ (dotted-line) cut around the peak is also reported. As a comparison, the efficiency corresponding to an extraction spill of 6 s, a $\delta t_{tag}$=100 ps and a 1.5$\sigma$ cut around the peak is also shown (red line).}
    \label{fig:time_tagged_snr}
\end{figure}

A clear peak is visible in the distributions of Figs.~\ref{fig:time_tagged_res1} and \ref{fig:time_tagged_res2}, and is due to true matches between $\nu_e$ and positron pairs coming from the same $K_{e3}$ decay within the tunnel volume. Fig.~\ref{fig:time_tagged_res1} features a subdominant peak around 300 ps, due to true matches from $K_{e3}$ decays happening in the transfer line after the second dipole, before entering the tagger. The peak offset is related to the geometrical acceptance of this kind of events, which selects only $K_{e3}$ decays with peculiar kinematics, for which $\Delta$/c is underestimated. On the other hand, the flat distribution sitting below the peak is due to fake matches of $\nu_e$ and positron pairs, which are not produced in the same decay event. The main contributions to the fake matches come from (1) real $K_{e3}$ decays within the tunnel, where the neutrino and the positron are produced by two different decays, (2) other kaon decay channels, where the pions are misidentified as positrons, and (3) background positrons entering the tunnel and impinging in the tagger walls. 

The time resolution of time-tagged neutrinos, from a fit with a gaussian plus a flat background around the distribution peak of Fig.~\ref{fig:time_tagged_res1}, is $\sigma_{\Delta t}$=74~ps. This value is dominated by the intrinsic resolution due to the emission angle of the positron and the neutrino in $K_{e3}$ decays and by the length of the LCM, which introduces uncertainty on the actual positron impact point in the calorimeter and thus on $\Delta/c$. 
As a consequence, a detector resolution $<100$~ps does not compromise the timing performance, which is limited by the intrinsic time spread. 
Figs.~\ref{fig:time_tagged_eff} and \ref{fig:time_tagged_snr} show the signal efficiency and signal-to-noise ratio for 4.8~s pulse extraction length and different time windows around the peak as a function of the neutrino detector resolution. The performance are expressed for a time resolution of the $t_0$ layer equal to 200~ps. For the sake of comparison, we also show the performance corresponding to a longer extraction spill of 6~s and a time resolution of the $t_0$ layer equal to 100~ps.
A high-purity sample of time-tagged neutrinos can be obtained by performing a cut around the peak position within a time window of 1.5$\sigma_{\Delta t}$. With this cut, a time resolution of 200 ps for both the $t_0$ layer and the neutrino detector and a 1 cm vertex position resolution, the true match efficiency is $\epsilon$=75.6$\%$~\footnote{The true match efficiency is computed from the true time-tagged neutrinos and the total number of $\nu_e^{CC}$ at the detector from $K_{e3}$ with a positron reconstructed in the calorimeter.}, with a S/N = 3.8. The accidental probability, computed as the ratio of the total number of fake matches and the total number of neutrino interactions at the detector, is $A = 3.2\%$. These values become $\epsilon$ = 72.3$\%$, S/N = 2.5 and $A = 4.6\%$ for a neutrino detector resolution of 500 ps.

The results above support the possibility of upgrading an ENUBET-like monitored neutrino beam to a full-fledged tagged neutrino beam. Such an upgrade requires the improvement of the time resolution of the neutrino detector and the tunnel instrumentation down to 200 ps, i.e. a factor of 2 smaller than what has been achieved by the ENUBET demonstrator.

\section{Conclusions}
This paper reports the first end-to-end simulation of a monitored neutrino beam based on the ENUBET concept. The beamline results from the optimization of the target, optics, and collimators of the transfer line to maximize the number of neutrinos produced by kaon decays, whose leptons can be observed by the ENUBET instrumentation of the decay tunnel. The tunnel detector is able to recognize positrons and muons with a signal-to-noise greater than 2 and 5, respectively, while retaining a large ($>20$\%)  monitoring efficiency. Such an important achievement is attained thanks to the exploitation of a purely static focusing system combined with a long (2~s) proton extraction, which reduces the particle rate at the tunnel down to a level that is sustainable by conventional, low-cost, sampling calorimeters. By exploiting the positron and muon distributions to constrain hadroproduction uncertainties, this monitored neutrino beam offers a direct measurement of the $\nu_\mu$ and $\nu_e$ flux from kaons with a precision $<1$\%. The beam designed in this paper is optimized to provide a facility that can measure the $\nu_e$ and $\nu_\mu$ cross sections in the region of interest for DUNE with a precision of $\sim 1$\%. Its realization at CERN leverages the existing SPS accelerator and ProtoDUNE detectors. It can achieve its physics goals by accumulating $1.3 \times 10^{20}$ pot and running with ProtoDUNE-SP as the neutrino detector or, equivalently, with $6.5 \times 10^{19}$ pot when both ProtoDUNEs are in operation. More generally, this ENUBET-like facility represents an ideal framework for a new generation of cross-section experiments that will be running in parallel with DUNE and HK in the years to come.   

\begin{acknowledgements}
This project has received funding from the European Union's Horizon 2020 Research and Innovation programme under Grant Agreement no. 681647 and the Italian Ministry for Education and Research (MIUR, bando FARE, progetto NUTECH). It is also supported by the Agence Nationale de la Recherche (ANR, France) through the PIMENT project (ANR-21-CE31-0027) and by the Ministry of Science and Education of Republic of Croatia grant No. KK.01.1.1.01.0001.
%If you'd like to thank anyone, place your comments here
%and remove the percent signs.
\end{acknowledgements}

%\appendix
%
%\section{Appendix section}\label{app}
%
%Appendix text

% BibTeX users please use one of
%\bibliographystyle{spbasic}      % basic style, author-year citations
%\bibliographystyle{spmpsci}      % mathematics and physical sciences
\bibliographystyle{spphys}       % APS-like style for physics
\bibliography{bibliography}   % name your BibTeX data base

\begin{thebibliography}{10}
\providecommand{\url}[1]{{#1}}
\providecommand{\urlprefix}{URL }
\expandafter\ifx\csname urlstyle\endcsname\relax
  \providecommand{\doi}[1]{DOI \discretionary{}{}{}#1}\else
  \providecommand{\doi}{DOI \discretionary{}{}{}\begingroup
  \urlstyle{rm}\Url}\fi

\bibitem{katori2018}
T.~Katori and M.~Martini, Journal of Physics G: Nuclear and Particle Physics
  \textbf{45}(1), 013001 (2017).
\newblock \doi{10.1088/1361-6471/aa8bf7}.
\newblock \urlprefix\url{https://doi.org/10.1088/1361-6471/aa8bf7}

\bibitem{pandey_nufact2022}
V.~Pandey, in \emph{23rd International Workshop on Neutrinos from Accelerators
  (NuFact2022), Salt Lake City (UT, USA), July 30, August 6} (2022)

\bibitem{Abi:2020wmh}
B.~Abi, R.~Acciarri, M.~Acero, G.~Adamov, D.~Adams, M.~Adinolfi, et~al., J. of
  Instrumentation \textbf{15}(08), T08008 (2020).
\newblock \doi{10.1088/1748-0221/15/08/t08008}.
\newblock \urlprefix\url{https://doi.org/10.1088/1748-0221/15/08/t08008}

\bibitem{Abe:2018uyc}
K.~Abe, K.~Abe, H.~Aihara, A.~Aimi, and R.~Akutsu.
\newblock Hyper-kamiokande design report (2018)

\bibitem{Alvarez-Ruso:2017oui}
L.~Alvarez-Ruso, M.~S. Athar, M.~B. Barbaro, D.~Cherdack, M.~E. Christy,
  P.~Coloma, et~al., Prog. Part. Nucl. Phys. \textbf{100}, 1 (2018).
\newblock \doi{10.1016/j.ppnp.2018.01.006}

\bibitem{Branca:2021vis}
A.~Branca, G.~Brunetti, A.~Longhin, M.~Martini, F.~Pupilli, and F.~Terranova,
  Symmetry \textbf{13}(9), 1625 (2021).
\newblock \doi{10.3390/sym13091625}

\bibitem{EUdeliberation2020}
Deliberation document on the 2020 update of the european strategy for particle
  physics (2020).
\newblock {The European Strategy Group, CERN-ESU-014. Available at
  \texttt{https://cds.cern.ch/record/2720131} }

\bibitem{ENUBET_proposal}
F.~Acerbi, G.~Ballerini, M.~Bonesini, C.~Brizzolari, G.~Brunetti, M.~Calviani,
  et~al., The {ENUBET} project.
\newblock Tech. Rep. CERN-SPSC-2018-034. SPSC-I-248, CERN, Geneva (2018).
\newblock \urlprefix\url{https://cds.cern.ch/record/2645532}

\bibitem{ENUBET_spsc_2020}
F.~Acerbi, M.~Bonesini, A.~Branca, C.~Brizzolari, G.~Brunetti, M.~Calviani,
  et~al., {NP06/ENUBET} annual report for the {CERN-SPSC}.
\newblock Tech. Rep. CERN-SPSC-2020-009. SPSC-SR-268, CERN, Geneva (2020).
\newblock \urlprefix\url{https://cds.cern.ch/record/2714046}

\bibitem{ENUBET_spsc_2021}
F.~Acerbi, I.~Angelis, , M.~Bonesini, A.~Branca, C.~Brizzolari, et~al.,
  {NP06/ENUBET} annual report for the {CERN-SPSC}.
\newblock Tech. Rep. CERN-SPSC-2021-013. SPSC-SR-290, CERN, Geneva (2021).
\newblock \urlprefix\url{https://cds.cern.ch/record/2759849}

\bibitem{Longhin:2014yta}
A.~Longhin, L.~Ludovici, and F.~Terranova, Eur. Phys. J. C \textbf{75}, 155
  (2015).
\newblock \doi{10.1140/epjc/s10052-015-3378-9}.
\newblock \urlprefix\url{https://doi.org/10.1140/epjc/s10052-015-3378-9}

\bibitem{app11041644}
N.~Charitonidis, A.~Longhin, M.~Pari, E.~G. Parozzi, and F.~Terranova, Applied
  Sciences \textbf{11}(4) (2021).
\newblock \doi{10.3390/app11041644}

\bibitem{ENUBET_ERC}
Enubet: Enhanced neutrino beams from kaon tagging (2016).
\newblock {ERC-CoG-2015 (PI A. Longhin), grant agreement n. 681647
  \texttt{https://www.pd.infn.it/eng/enubet/} }

\bibitem{GEANT4:2002zbu}
S.~Agostinelli et~al., Nucl. Instrum. Meth. A \textbf{506}, 250 (2003).
\newblock \doi{10.1016/S0168-9002(03)01368-8}

\bibitem{Allison:2006ve}
J.~Allison et~al., IEEE Trans. Nucl. Sci. \textbf{53}, 270 (2006).
\newblock \doi{10.1109/TNS.2006.869826}

\bibitem{Allison:2016lfl}
J.~Allison et~al., Nucl. Instrum. Meth. A \textbf{835}, 186 (2016).
\newblock \doi{10.1016/j.nima.2016.06.125}

\bibitem{Danby:1962nd}
G.~Danby, J.~M. Gaillard, K.~A. Goulianos, L.~M. Lederman, N.~B. Mistry,
  M.~Schwartz, and J.~Steinberger, Phys. Rev. Lett. \textbf{9}, 36 (1962).
\newblock \doi{10.1103/PhysRevLett.9.36}

\bibitem{kopp2006}
S.~Kopp, Physics Reports \textbf{439}(3), 101  (2007)

\bibitem{MINERvA:2022vmb}
L.~Akhter, Z.~{Ahmad Dar}, F.~Akbar, V.~Ansari, M.~V. Ascencio, M.~{Sajjad
  Athar}, A.~Bashyal, A.~Bercellie, M.~Betancourt, et~al.
\newblock {Improved constraint on the MINERVA medium energy neutrino flux using
  $\bar{\nu}e^{-} \!\rightarrow \bar{\nu}e^{-}$ data}.
\newblock {a}rXiv:2209.05540, 2022

\bibitem{Hand1969}
L.~Hand, in \emph{{Proceedings. Second NAL Summer Study, Jun 9 - Aug 3}} (1969)

\bibitem{Pontevcorvo1979}
B.~Pontecorvo, Lett. Nuovo Cim. \textbf{25}, 257 (1979)

\bibitem{Bernstein}
R.~H. Bernstein, F.~Borcherding, D.~Jovanovic, M.~J. Lamm, and F.~Vannucci, {A
  Proposal for a Neutrino Oscillation Experiment in a Tagged Neutrino Line}.
\newblock Tech. Rep. FERMILAB-PROPOSAL-0788, FERMILAB (1988)

\bibitem{Ludovici:1996sx}
L.~Ludovici and P.~Zucchelli,   (1996)

\bibitem{ammosov}
V.~Ammosov, A.~Belkov, A.~Bugorskij, and et~al., JINR-R-1-90-458  (1990)

\bibitem{Terranova:2015nsa}
F.~Terranova, A.~Longhin, and L.~Ludovici, PoS \textbf{NUFACT2014}, 037 (2015).
\newblock \doi{10.22323/1.226.0037}

\bibitem{Perrin-Terrin:2021jtl}
M.~Perrin-Terrin, Eur. Phys. J. C \textbf{82}(5), 465 (2022).
\newblock \doi{10.1140/epjc/s10052-022-10397-8}

\bibitem{NUTECH}
{NUTECH} (neutrino time-tagged beams with cherenkov detectors).
\newblock Grant MIUR – Bando FARE (2017-2022)

\bibitem{Longhin:2022tkk}
F.~Acerbi, I.~Angelis, M.~Bonesini, C.~B. A.~Branca~and, G.~Brunetti,
  M.~Calviani, S.~Capelli, S.~Carturan, M.~Catanesi, et~al.
\newblock {Enhanced NeUtrino BEams from kaon Tagging (ENUBET)} (2022).
\newblock \urlprefix\url{https://arxiv.org/abs/2203.08319}.
\newblock Snowmass 2022 ar{X}iv:2203.08319

\bibitem{protodune:tdr}
B.~Abi, R.~Acciarri, M.~Acero, M.~Adamowski, C.~Adams, D.~Adams, et~al., The
  single-phase protodune technical design report.
\newblock Tech. Rep. FERMILAB-DESIGN-2017-02 (2017).
\newblock \urlprefix\url{http://cds.cern.ch/record/2271524}

\bibitem{slawg:2019}
M.~Fraser, B.~Balhan, H.~Bartosik, and J.~Bernhard, in \emph{Proc. 10th
  International Particle Accelerator Conference (IPAC'19), Melbourne,
  Australia, 19-24 May 2019} (2019), no.~10 in International Particle
  Accelerator Conference, pp. 2391--2394.
\newblock \doi{doi:10.18429/JACoW-IPAC2019-WEPMP031}.
\newblock \urlprefix\url{http://jacow.org/ipac2019/papers/wepmp031.pdf}

\bibitem{Kain:2019qxl}
V.~Kain, F.~M. Velotti, M.~A. Fraser, B.~Goddard, J.~Prieto, L.~S. Stoel, and
  M.~Pari, Phys. Rev. Accel. Beams \textbf{22}(10), 101001 (2019).
\newblock \doi{10.1103/PhysRevAccelBeams.22.101001}

\bibitem{pari:tesi}
M.~Pari, Study and development of {SPS} slow extraction schemes and focusing of
  secondary particles for the {ENUBET} monitored neutrino beam.
\newblock Ph.D. thesis, {Universit\'a degli Studi Di Padova, Dipartimento di
  Fisica e Astronomia G. Galilei}, Padova (2021).
\newblock \urlprefix\url{https://hdl.handle.net/11577/3426254}

\bibitem{pari:ripples}
M.~Pari, F.~M. Velotti, M.~A. Fraser, V.~Kain, and O.~Michels, Phys. Rev.
  Accel. Beams \textbf{24}, 083501 (2021).
\newblock \doi{10.1103/PhysRevAccelBeams.24.083501}.
\newblock
  \urlprefix\url{https://link.aps.org/doi/10.1103/PhysRevAccelBeams.24.083501}

\bibitem{FLUKA1}
G.~Battistoni, T.~Boehlen, F.~Cerutti, P.~W. Chin, L.~S. Esposito,
  A.~Fass{\`o}, et~al., Annals of Nuclear Energy \textbf{82}, 10 (2015)

\bibitem{FLUKA2}
T.~B{\"o}hlen, F.~Cerutti, M.~Chin, A.~Fass{\`o}, A.~Ferrari, P.~G. Ortega,
  A.~Mairani, P.~R. Sala, G.~Smirnov, and V.~Vlachoudis, Nuclear data sheets
  \textbf{120}, 211 (2014)

\bibitem{g4beamline}
T.~J. Roberts, K.~Beard, S.~Ahmed, D.~Huang, and D.~M. Kaplan, in \emph{Proc.
  PAC}, vol. 2013 (2011), vol. 2013, pp. 373--375

\bibitem{Feynman1969}
R.~P. Feynman, Phys. Rev. Lett. \textbf{23}, 1415 (1969).
\newblock \doi{10.1103/PhysRevLett.23.1415}

\bibitem{parozzi_thesis}
E.~G. Parozzi.
\newblock Design and optimisation of a variable momentum secondary beamline for
  the {NP06/ENUBET} project (2022).
\newblock {Ph.D. Thesis, Univ. of Milano Bicocca }

\bibitem{T2K:2011qtm}
K.~Abe et~al., Nucl. Instrum. Meth. A \textbf{659}, 106 (2011).
\newblock \doi{10.1016/j.nima.2011.06.067}

\bibitem{T2K:2019eao}
K.~Abe et~al.,   (2019)

\bibitem{Adamson:2015dkw}
P.~Adamson et~al., Nucl. Instrum. Meth. A \textbf{806}, 279 (2016).
\newblock \doi{10.1016/j.nima.2015.08.063}

\bibitem{Brown:1973jce}
K.~L. Brown, F.~Rothacker, D.~C. Carey, and F.~C. Iselin,   (1973).
\newblock \doi{10.5170/CERN-1973-016}

\bibitem{MAD-X}
The {MAD-X} project.
\newblock \urlprefix\url{http://cern.ch/madxJPARC-MR}

\bibitem{Roberts:2007nte}
T.~J. Roberts and D.~M. Kaplan, Conf. Proc. C \textbf{070625}, 3468 (2007).
\newblock \doi{10.1109/PAC.2007.4440461}

\bibitem{NA62:2017rwk}
E.~Cortina~Gil et~al., JINST \textbf{12}(05), P05025 (2017).
\newblock \doi{10.1088/1748-0221/12/05/P05025}

\bibitem{inermet}
M.~Cauchi, R.~Assmann, A.~Bertarelli, F.~Carra, F.~Cerutti, L.~Lari,
  P.~Mollicone, S.~Redaelli, and N.~Sammut, Physical Review Special Topics -
  Accelerators and Beams \textbf{18} (2015).
\newblock \doi{10.1103/PhysRevSTAB.18.041002}

\bibitem{kochenderfer}
M.~J. Kochenderfer and T.~A. Wheeler, \emph{Algorithms for Optimization} (MIT
  press, 2019)

\bibitem{Liu:2015ylc}
A.~Liu, A.~Bross, and D.~Neuffer, Nucl. Instrum. Meth. A \textbf{794}, 200
  (2015).
\newblock \doi{10.1016/j.nima.2015.05.035}

\bibitem{in2p3}
Centre de calcule de l'in2p3.
\newblock \urlprefix\url{https://cc.in2p3.fr}.
\newblock Calcule scientific et traitement de donne\'es

\bibitem{Berra:2016thx}
A.~Berra et~al., Nucl. Instrum. Meth. A \textbf{830}, 345 (2016).
\newblock \doi{10.1016/j.nima.2016.05.123}

\bibitem{Berra:2017rsi}
A.~Berra, C.~Brizzolari, S.~Cecchini, and F.~Chignoli, IEEE Trans. Nucl. Sci.
  \textbf{64}(4), 1056 (2017).
\newblock \doi{10.1109/TNS.2017.2672500}

\bibitem{Ballerini:2018hus}
G.~Ballerini, A.~Berra, R.~Boanta, and C.~Brizzolari, J. of Instrumentation
  \textbf{13}(01), P01028 (2018).
\newblock \doi{10.1088/1748-0221/13/01/P01028}

\bibitem{Acerbi:2019wti}
F.~Acerbi, G.~Ballerini, A.~Berra, and C.~Brizzolari, J. of Instrumentation
  \textbf{14}(02), P02029 (2019).
\newblock \doi{10.1088/1748-0221/14/02/P02029}

\bibitem{Acerbi:2020itd}
F.~Acerbi, A.~Branca, C.~Brizzolari, and G.~Brunetti, Nucl. Instrum. Meth. A
  \textbf{956}, 163379 (2020).
\newblock \doi{10.1016/j.nima.2019.163379}

\bibitem{Acerbi:2020nwd}
F.~Acerbi, M.~Bonesini, F.~Bramati, and A.~Branca, J. of Instrumentation
  \textbf{15}(08), P08001 (2020).
\newblock \doi{10.1088/1748-0221/15/08/P08001}

\bibitem{Torti:2023hsc}
M.~Torti et~al., EPJ Web Conf. \textbf{282}, 01018 (2023).
\newblock \doi{10.1051/epjconf/202328201018}

\bibitem{demonstrator}
F.~Acerbi, I.~Angelis, M.~Bonesini, A.~Branca, C.~Brizzolari, G.~Brunetti,
  M.~Calviani, S.~Carturan, M.~Catanesi, et~al.
\newblock {The ENUBET Demostrator}.
\newblock In preparation

\bibitem{Eckert:2012yr}
P.~Eckert, R.~Stamen, and H.~C. Schultz-Coulon, JINST \textbf{7}, P08011
  (2012).
\newblock \doi{10.1088/1748-0221/7/08/P08011}

\bibitem{systematics}
F.~Acerbi, I.~Angelis, M.~Bonesini, A.~Branca, C.~Brizzolari, G.~Brunetti,
  M.~Calviani, S.~Carturan, M.~Catanesi, et~al.
\newblock {Assessment of the systematic budget at the ENUBET monitored neutrino
  beam}.
\newblock In preparation

\bibitem{Musienko:2017znn}
Y.~Musienko, A.~Heering, R.~Ruchti, M.~Wayne, Y.~Andreev, A.~Karneyeu, and
  V.~Postoev, JINST \textbf{12}(07), C07030 (2017).
\newblock \doi{10.1088/1748-0221/12/07/C07030}

\bibitem{Garutti:2018hfu}
E.~Garutti and Y.~Musienko, Nucl. Instrum. Meth. A \textbf{926}, 69 (2019).
\newblock \doi{10.1016/j.nima.2018.10.191}

\bibitem{Leroy:2011goz}
C.~Leroy and P.-G. Rancoita, \emph{{Principles of radiation interaction in
  matter and detection}} (World Scientific, Singapore, 2011).
\newblock \doi{10.1142/5578}

\bibitem{Hocker:2007ht}
A.~Hocker et~al.,   (2007)

\bibitem{PIMENT}
T.~Papaevangelou.
\newblock Développement d'un détecteur picosec-micromegas pour enubet –
  piment (2022).
\newblock {Available at \texttt{https://anr.fr/Projet-ANR-21-CE31-0027} }

\bibitem{Bonesini2001}
M.~Bonesini, A.~Marchionni, F.~Pietropaolo, and T.~T. de~Fatis, Eur. Phys. J. C
  \textbf{20}(1), 13 (2001)

\bibitem{Ambrosini:1999id}
G.~Ambrosini, R.~Arsenescu, K.~Bernier, C.~Biino, M.~Bonesini, et~al., Eur.
  Phys. J. C \textbf{10}(4), 605 (1999).
\newblock \doi{10.1007/s100520050601}.
\newblock \urlprefix\url{https://doi.org/10.1007/s100520050601}

\bibitem{Atherton:1980vj}
H.~W. Atherton, C.~Bovet, N.~Doble, G.~von Holtey, L.~Piemontese, A.~Placci,
  M.~Placidi, D.~E. Plane, M.~Reinharz, and E.~Rossa,   (1980).
\newblock \doi{10.5170/CERN-1980-007}

\bibitem{multiuniverse}
M.~Kordowsky.
\newblock Error bands from the many universes method.
\newblock Minerva note, n.7433

\bibitem{MINERvA:2021ddh}
B.~Messerly et~al., EPJ Web Conf. \textbf{251}, 03046 (2021).
\newblock \doi{10.1051/epjconf/202125103046}

\bibitem{Verkerke:2003ir}
W.~Verkerke and D.~P. Kirkby, eConf \textbf{C0303241}, MOLT007 (2003)

\end{thebibliography}

\end{document}